\documentclass[10pt, conference, letterpaper]{IEEEtran}
\usepackage{comment}
\usepackage{listings}
\usepackage{amssymb}
\usepackage{amsfonts}
\usepackage{amsmath}
\usepackage{graphicx}
\usepackage{nomencl}
\usepackage{multirow}
\usepackage{url}
\usepackage{booktabs}
\usepackage{subfiles}
\usepackage{amsthm}
\usepackage{todonotes}
\usepackage{subcaption}
\theoremstyle{definition}
\usepackage{tikz, pgfplots}
\usetikzlibrary{decorations.pathreplacing}
\usetikzlibrary{matrix,positioning}
\pgfplotsset{compat=newest}
\newenvironment{bluebox}{\color{blue}
}{}

\includecomment{bluebox}

\begin{document}

\author
{
    \IEEEauthorblockN
    {
	Belma Turkovic\IEEEauthorrefmark{1},
	Fernando A. Kuipers\IEEEauthorrefmark{2} and
	Steve Uhlig \IEEEauthorrefmark{3}
    }

    \IEEEauthorblockA
    {
        \IEEEauthorrefmark{1} \IEEEauthorrefmark{2} Delft University of Technology, Delft, The Netherlands \\
        \IEEEauthorrefmark{3} Queen Mary University of London, London, United Kingdom
    }   
    Email: 
    \{\IEEEauthorrefmark{1}B.Turkovic-2,
    \IEEEauthorrefmark{2}F.A.Kuipers\}@tudelft.nl,
     \IEEEauthorrefmark{3}Steve.Uhlig@qmul.ac.uk,

}

\title{Fifty Shades of Congestion Control: A Performance and Interactions Evaluation}

\maketitle

\begin{abstract}
Congestion control algorithms are crucial in achieving high utilization while preventing overloading the network. Over the years, many different congestion control algorithms have been developed, each trying to improve in specific situations. However, 
their interactions and co-existence has, to date, not been thoroughly evaluated, which is the focus of this paper. 
Through head-to-head comparisons of representatives from loss-based, delay-based and hybrid types of congestion control algorithms, we reveal that fairness in resources claimed is often not attained, especially when flows sharing a link have different RTTs.

\end{abstract}

\section{Introduction\label{Sec_intro}}
In the wake of the 
growing demand for higher bandwidth, higher reliability, and lower latency, novel congestion control algorithms have been developed. 
For example, in 2016, Google published its bottleneck bandwidth and round-trip time (BBR) congestion control algorithm \cite{3022184}, claiming it was able to operate without 
filling buffers. 
Around the same time, TCP LoLa \cite{lola} and TIMELY \cite{TIMELY} were proposed, focusing on low latency and bounding of the queuing delay. 
Moreover, new transport protocols such as QUIC allow the implementation of algorithms directly in user space, which facilitates quick development of new transport features. 
However, \emph{congestion control algorithms have been typically developed in isolation}, without thoroughly investigating their behaviour in the presence of other congestion control algorithms, which is the goal of this paper. 


In this paper, we first divide existing congestion control algorithms into three groups: loss-based, delay-based, and hybrid. Based on experiments in a testbed, we study the interactions over a bottleneck link among flows of the same group, across groups, as well as when flows have different RTTs. We find that flows using loss-based algorithms are over-powering flows using delay-based, as well as hybrid algorithms. Moreover, as flows using loss-based algorithms fill the queues, increase of queuing delay of all the flows sharing the bottleneck is determined by their presence. 
Non-loss-based groups thus cannot be used in a typical network, where flows typically rely on a loss-based algorithm. In addition, we observe that convergence times can be large (up to $60s$), which may surpass the flow duration of many applications. Finally, we find that hybrid algorithms, such as BBR, not only favour flows with a higher RTT, but they also cannot maintain a low queuing delay.

In Section \ref{Sec_Congestion_Control} and 
\ref{Sec_Classification}, we provide an overview of congestion control mechanisms. These algorithms are classified in 3 main groups, namely loss based, delay based, and hybrid. In Section \ref{Sec_eval}, we (1) identify a set of key performance metrics to compare them, (2) describe our measurement setup, and (3) present our measurement results. 

\section{Congestion control\label{Sec_Congestion_Control}}
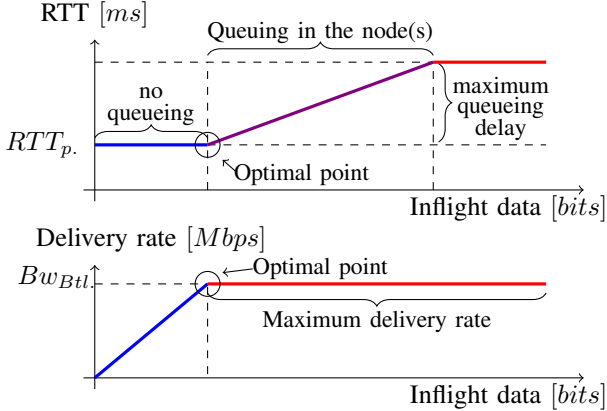
\begin{figure}[h!]
    \centering
    	\begin{tikzpicture}
	\draw[->] (0.35,0) -- (7,0); 
	\draw (6,-0.25) node[] {\begin{tabular}{c}Inflight data $\lbrack bits \rbrack$ \end{tabular}};
	
	\draw (0.5,2) node[above] {\begin{tabular}{c}RTT $\lbrack ms\rbrack$ \end{tabular}};
	\draw[->] (0.5,-0.15) -- (0.5,2);
	\draw[-, dashed] (2,0)  -- (2,1.7);
	\draw[-, dashed] (5,0) -- (5,1.7);
	\draw [decorate,decoration={brace,amplitude=5pt,raise=2pt,mirror}] (5,0.64) -- (5,1.68);
	\draw (5.9,1.4) node[] {\small maximum};
	\draw (5.9,1.1) node[] {\small queueing};
    \draw (5.9,0.8) node[] {\small delay};
    \draw (3.5,2.1) node[] {\small Queuing in the node(s)};
	\draw [decorate,decoration={brace,amplitude=5pt,raise=2pt}] (2,1.72) -- (5,1.72);
	\draw [decorate,decoration={brace,amplitude=5pt,raise=2pt}] (0.5,0.62) -- (2,0.62);
	\draw (1.25,1.3) node[] {\small no};
	\draw (1.25,1) node[] {\small queueing};

	\draw[-, dashed] (2,0.6) -- (6.5,0.6);
	\draw[-, dashed] (0.5,1.7) -- (5,1.7);
	\draw[-, color = blue, line width = 1.2] (0.5,0.6) -- (2,0.6);
	\draw[-, color = blue!50!red, line width = 1.2] (2,0.6) -- (5,1.7);
	\draw[-, color = red, line width = 1.2] (5,1.7) -- (6.5,1.7);
	\draw (-0.2,0.6) node[rectangle,minimum width=7mm] (ll) {$RTT_{p.}$};

	\draw[->] (0.35,-2.5) -- (7,-2.5); 
	\draw (6,-2.75) node[] {\begin{tabular}{c}Inflight data $\lbrack bits \rbrack$ \end{tabular}};
	
	\draw (1.25,-1) node[above] {\begin{tabular}{c}Delivery rate $\lbrack Mbps\rbrack$ \end{tabular}};
	\draw[->] (0.5,-2.65) -- (0.5,-1);
	\draw[-, dashed] (0.5,-1.25) -- (2,-1.25);
	\draw (0,-1.5) node[above] {\begin{tabular}{c}$Bw_{Btl.}$ \end{tabular}};
	\draw[-, color = blue, line width = 1.2] (0.5,-2.5) -- (2,-1.25);
	\draw[-, color = red, line width = 1.2] (2,-1.25) -- (6.5,-1.25);
	\draw [decorate,decoration={brace,amplitude=5pt,raise=2pt, mirror}] (2,-1.3) -- (6.5,-1.3);
    \draw (4.25,-1.75) node[] {\small Maximum delivery rate};
     \draw (2,-1.25) node[draw, circle,radius=0.1cm] {};
     \draw (2,0.6) node[draw, circle,radius=0.1cm] {};
     
    \draw (3.25,0.2) node[] {\small Optimal point};
    \draw (3.5,-1.05) node[] {\small Optimal point};
	\draw[->] (2.37,0.25)  -- (2.2,0.4);
	\draw[->] (2.6,-1.05)  -- (2.2,-1.1);

	\draw[-, dashed] (2,-2.45)  -- (2,-1.25);
	\end{tikzpicture}    
    \caption{Effect of the amount of packets sent on the RTT (top) and
delivery rate (bottom). Based on \cite{3022184,scholztowards}.}\label{fig:optimalpoint}
\end{figure}

When a packet arrives at a switch, it is processed based on the installed forwarding rules and forwarded to an output link. Output links have a fixed bandwidth and, when packets arrive too fast, queues can form and congestion may occur. Network buffers are added to absorb short-term bursts in network traffic and to prevent packet loss, but they add delay to every packet passing through the buffer (as shown in Fig. \ref{fig:optimalpoint}). 

As network nodes process thousands of flows every second, bandwidth is often shared among flows. Hence, the maximum rate of a connection is limited by the so-called bottleneck link, i.e., the link with the least amount of available resources to process that flow on the path.

Congestion occurs when a network node needs to process more traffic than it is capable of sending further along the network.

\begin{figure}[h!]
    \centering
    	\begin{tikzpicture}[>=latex]
	\node (controler) at (0.25,0) [rectangle, draw=black!150,rounded corners = .25ex, minimum height=2.75cm, minimum width=6.2cm] {};
	\node at (-2.5,1.6) {Switch};	
	
	\node (controler) at (0.25,0) [rectangle, draw=black!150,rounded corners = .25ex, minimum height=2.5cm, minimum width=1.5cm] {};
	\node at (0.25,0) {\begin{tabular}{c} Switch\\  Fabric\end{tabular}};
	\node (DUT2) at (2.25,0.65) [rectangle, draw=black!150, fill=black!10,,rounded corners = .25ex, minimum height=1cm, minimum width=1.75cm] {};
	\node (DUT) at (2.25,-0.65) [rectangle, draw=black!150, fill=black!10,,rounded corners = .25ex, minimum height=1cm, minimum width=1.75cm] {};
	\node at (2.2,0.98) {\begin{tabular}{c} {\footnotesize Out. Queue 0}\end{tabular}};	
	\node at (2.2,-0.33) {\begin{tabular}{c} {\footnotesize Out. Queue 1}\end{tabular}};		
	
	\draw (1.75,0.3) -- ++(1cm,0) -- ++(0,0.5cm) -- ++(-1cm,0);
	\foreach \i in {1,...,4}
	\draw (2.75cm-\i*6pt,0.3) -- +(0,0.5cm);
	\draw (1.75,-0.5) -- ++(1cm,0) -- ++(0,-0.5cm) -- ++(-1cm,0);
	\foreach \i in {1,...,4}
	\draw (2.75cm-\i*6pt,-0.5) -- +(0,-0.5cm);
	\draw[->, line width=1pt] (3.4,0.65) -- (4.3,0.65);
	\draw[->, line width=1pt] (3.4,-0.65) -- (4.3,-0.65);
	\node at (3.8,0.85) {\begin{tabular}{c} \footnotesize Link0 \end{tabular}};	
	\node at (3.9,1.1) {\begin{tabular}{c} \footnotesize Out.\end{tabular}};		
	\node at (3.8,-0.45) {\begin{tabular}{c} \footnotesize Link1\end{tabular}};	
	\node at (3.9,-0.2) {\begin{tabular}{c} \footnotesize Out.\end{tabular}};

	\node (DUT2) at (-1.75,0.65) [rectangle, draw=black!150, fill=black!10,,rounded corners = .25ex, minimum height=1cm, minimum width=1.75cm] {};
	\node (DUT) at (-1.75,-0.65) [rectangle, draw=black!150, fill=black!10,,rounded corners = .25ex, minimum height=1cm, minimum width=1.75cm] {};
	\node at (-1.85,0.98) {\begin{tabular}{c} {\footnotesize In. Queue 0}\end{tabular}};	
	\node at (-1.85,-0.33) {\begin{tabular}{c} {\footnotesize In. Queue 1}\end{tabular}};		
	
	\draw (-2.25,0.3) -- ++(1cm,0) -- ++(0,0.5cm) -- ++(-1cm,0);
	\foreach \i in {1,...,4}
	\draw (-1.25cm-\i*6pt,0.3) -- +(0,0.5cm);
	\draw (-2.25,-0.5) -- ++(1cm,0) -- ++(0,-0.5cm) -- ++(-1cm,0);
	\foreach \i in {1,...,4}
	\draw (-1.25cm-\i*6pt,-0.5) -- +(0,-0.5cm);
	\draw[->, line width=1pt] (-3.85,0.65) -- (-2.95,0.65);
	\draw[->, line width=1pt] (-3.85,-0.65) -- (-2.95,-0.65);
	\node at (-3.35,0.85) {\begin{tabular}{c} \footnotesize Link0 \end{tabular}};	
	\node at (-3.45,1.1) {\begin{tabular}{c} \footnotesize Out.\end{tabular}};		
	\node at (-3.35,-0.45) {\begin{tabular}{c} \footnotesize Link1\end{tabular}};	
	\node at (-3.45,-0.2) {\begin{tabular}{c} \footnotesize Out.\end{tabular}};

	\end{tikzpicture}
    \caption{Buffers in a network node.}\label{fig:buffers}
\end{figure}
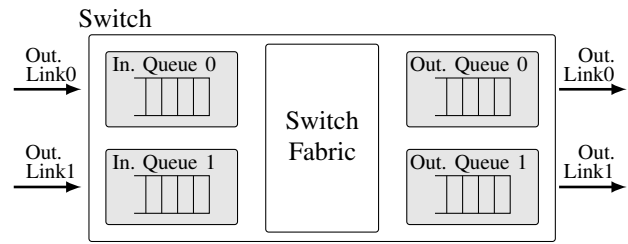

If a TCP connection sends less data than the bottleneck link bandwidth ($Bw_{Btl.}$), and no other flow shares the link, there is no congestion and the RTT equals the propagation and processing delay ($RTT_{p.}$). In this case, the delivery rate corresponds to the sending rate. When it reaches the bottleneck link bandwidth, the TCP connection is at its optimal operating point, because the sender sends as much data as possible without filling the buffers in the intermediate nodes. 

By increasing the sending rate further, buffers in the network nodes start to fill and queues might form. Packets arrive at the bottleneck faster than they can be forwarded causing increased delay, while the delivery rate remains the same. Finally, when buffers are full, the network node has to drop packets. Increasing buffer size will not improve the performance of the network and instead will lead to bufferbloat, i.e., the formation of queues in the network devices that unnecessarily add delay to every packet passing through \cite{gettys2011bufferbloat}.

Congestion control algorithms exploit the fact that packets arrive at the receiver at a rate the bottleneck can support (maximum delivery rate). Upon reception of a packet, the receiver informs the sender by sending an ACK. The congestion control algorithm of the sender based on the spacing and/or the reception of these ACKs, estimates the current state of the network. If the algorithm detects that the network is congested, it will back-off, and switch to a more conservative approach. Otherwise, if a congestion-free state is detected, the algorithm will increase the sending rate to probe for more resources.



\section{Background\label{Sec_Classification}}

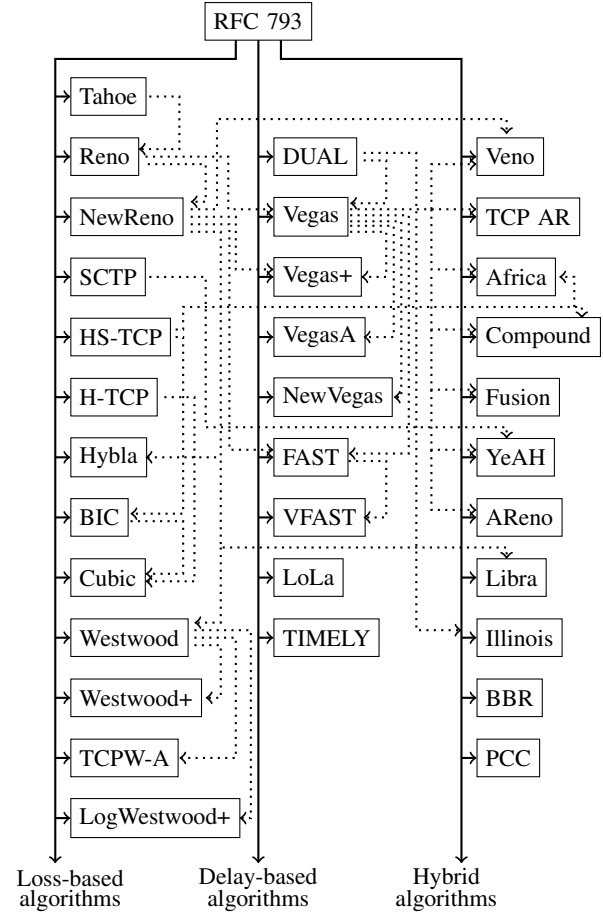
\begin{figure}[t!]
	\centering
	\begin{tikzpicture}
\draw (0,0) node[draw, minimum height = 0.5cm] (s1) {\small RFC 793};

\draw[->, thick, color=black] (-0.3,-0.25) -- (-0.3, -0.5) -- (-2.7,-0.5) -- (-2.7,-11.2);
\draw[->, thick, color=black] (0, -0.25) -- (0, -11.2);
\draw[->, thick, color=black] (0.3, -0.25) -- (0.3, -0.5) -- (2.7,-0.5) -- (2.7,-11.2);
\node at (-2.5,-11.4) {\small Loss-based};
\node at (-2.5,-11.7) {\small algorithms};
\node at (0,-11.4) {\small Delay-based};
\node at (0,-11.7) {\small algorithms};
\node at (2.5,-11.4) {\small Hybrid};
\node at (2.5,-11.7) {\small algorithms};

\draw (-2.5,-1) node[draw,  anchor=west, minimum height = 0.5cm] (s1) {\small Tahoe};
\draw (-2.5,-1.8) node[draw,  anchor=west, minimum height = 0.5cm] (s1) {\small Reno};
\draw (-2.5,-2.6) node[draw,  anchor=west, minimum height = 0.5cm] (s1) {\small NewReno};
\draw (-2.5,-3.4) node[draw,  anchor=west, minimum height = 0.5cm] (s1) {\small SCTP};
\draw (-2.5,-4.2) node[draw,  anchor=west, minimum height = 0.5cm] (s1) {\small HS-TCP};
\draw (-2.5,-5) node[draw,  anchor=west, minimum height = 0.5cm] (s1) {\small H-TCP};
\draw (-2.5,-5.8) node[draw,  anchor=west, minimum height = 0.5cm] (s1) {\small Hybla};
\draw (-2.5,-6.6) node[draw,  anchor=west, minimum height = 0.5cm] (s1) {\small BIC};
\draw (-2.5,-7.4) node[draw,  anchor=west, minimum height = 0.5cm] (s1) {\small Cubic};
\draw (-2.5,-8.2) node[draw,  anchor=west, minimum height = 0.5cm] (s1) {\small Westwood
};
\draw (-2.5,-9) node[draw,  anchor=west, minimum height = 0.5cm] (s1) {\small Westwood+
};
\draw (-2.5,-9.8) node[draw,  anchor=west, minimum height = 0.5cm] (s1) {\small TCPW-A};
\draw (-2.5,-10.6) node[draw,  anchor=west, minimum height = 0.5cm] (s1) {\small LogWestwood+
};

\draw[->, thick, color=black] (-2.7, -1) --   (-2.5, -1);
\draw[->, thick, color=black] (-2.7, -1.8) -- (-2.5, -1.8);
\draw[->, thick, color=black] (-2.7, -2.6) -- (-2.5, -2.6);
\draw[->, thick, color=black] (-2.7, -4.2) -- (-2.5, -4.2);
\draw[->, thick, color=black] (-2.7, -5) -- (-2.5, -5);
\draw[->, thick, color=black] (-2.7, -3.4) -- (-2.5, -3.4);
\draw[->, thick, color=black] (-2.7, -5.8) -- (-2.5, -5.8);
\draw[->, thick, color=black] (-2.7, -6.6) -- (-2.5, -6.6);
\draw[->, thick, color=black] (-2.7, -7.4) -- (-2.5, -7.4);
\draw[->, thick, color=black] (-2.7, -8.2) -- (-2.5, -8.2);
\draw[->, thick, color=black] (-2.7, -9) -- (-2.5, -9);
\draw[->, thick, color=black] (-2.7, -9.8) -- (-2.5, -9.8);
\draw[->, thick, color=black] (-2.7, -10.6) -- (-2.5, -10.6);

\draw[->, dotted, color=black, thick] (-1.45, -1) -- (-1.05, -1) -- (-1.05, -1.7) -- (-1.59,-1.7);

\draw[<-, dotted, color=black, thick] (-1.5, -5.8) -- (-0.5, -5.8);

\draw[->, dotted, color=black, thick] (-1.25, -5) -- (-0.85, -5) -- (-0.85, -7.45) -- (-1.5,-7.45);
\draw[->, dotted, color=black, thick] (-1.7, -6.65) -- (-1, -6.65) -- (-1, -7.35) -- (-1.5,-7.35);
\draw[->, dotted, color=black, thick] (-1.1,- 4.2) -- (-1, -4.2) -- (-1, -6.55) -- (-1.7,-6.55);
\draw[->, dotted, color=black, thick] (-1.47, -1.9) -- (-0.7, -1.9) -- (-0.7, -2.4) -- (-0.9,-2.4);
\draw[->, dotted, color=black, thick] (-0.5, -7) -- (-0.5,-8) -- (-0.85,-8);

\draw[->, dotted, color=black, thick] (-0.85,-8.3) -- (-0.5,-8.3) --(-0.5,-9) -- (-0.7,-9);
\draw[->, dotted, color=black, thick] (-0.85,-8.2) -- (-0.3,-8.2) --(-0.3,-9.8) -- (-1.05,-9.8);
\draw[->, dotted, color=black, thick] (-0.85,-8.1) -- (-0.1,-8.1) --(-0.1,-10.6) -- (-0.25,-10.6);

\draw (0.2,-1.8) node[draw, anchor=west, minimum height = 0.5cm] (s1) {\small DUAL};
\draw (0.2,-2.6) node[draw, anchor=west, minimum height = 0.5cm] (s1) {\small Vegas};
\draw (0.2,-3.4) node[draw,  anchor=west, minimum height = 0.5cm] (s1) {\small Vegas+};
\draw (0.2,-4.2) node[draw,  anchor=west, minimum height = 0.5cm] (s1) {\small VegasA};
\draw (0.2,-5) node[draw,  anchor=west, minimum height = 0.5cm] (s1) {\small NewVegas};
\draw (0.2,-5.8) node[draw,  anchor=west, minimum height = 0.5cm] (s1) {\small FAST};
\draw (0.2,-6.6) node[draw,  anchor=west, minimum height = 0.5cm] (s1) {\small VFAST};
\draw (0.2,-7.4) node[draw, anchor=west, minimum height = 0.5cm] (s1) {\small LoLa};
\draw (0.2,-8.2) node[draw,  anchor=west, minimum height = 0.5cm] (s1) {\small TIMELY};
\draw[->, thick, color=black] (0, -1.8) -- (0.2, -1.8);
\draw[->, thick, color=black] (0, -2.6) -- (0.2, -2.6);
\draw[->, thick, color=black] (0, -3.4) -- (0.2, -3.4);
\draw[->, thick, color=black] (0, -4.2) -- (0.2, -4.2);
\draw[->, thick, color=black] (0, -5) -- (0.2, -5);
\draw[->, thick, color=black] (0, -5.8) -- (0.2, -5.8);
\draw[->, thick, color=black] (0, -7.4) -- (0.2, -7.4);
\draw[->, thick, color=black] (0, -6.6) -- (0.2, -6.6);
\draw[->, thick, color=black] (0, -8.2) -- (0.2, -8.2);

\draw[->, dotted, color=black, thick] (1.4, -1.85) -- (1.7, -1.85) -- (1.7, -2.42) -- (1.22,-2.42);
\draw[->, dotted, color=black, thick] (1.22, -2.8) -- (1.7, -2.8) -- (1.7, -3.4) -- (1.37,-3.4);

\draw[->, dotted, color=black, thick] (1.22, -2.5) -- (2.3, -2.5) --  (2.9,-2.5);
\draw[->, dotted, color=black, thick] (2.3, -2.5) -- (2.3, -3.3) -- (2.9,-3.3);

\draw[->, dotted, color=black, thick] (2.3,-3.4) -- (2.3,-4.1) -- (2.9,-4.1);
\draw[->, dotted, color=black, thick] (2.3,-4.1) -- (2.3,-4.9) -- (2.9,-4.9);
\draw[->, dotted, color=black, thick] (2.3,-4.9) -- (2.3,-5.7) --(2.9,-5.7);
\draw[->, dotted, color=black, thick] (2.3,-5.7) -- (2.3,-6.5) --(2.9,-6.5);
\draw[->, dotted, color=black, thick] (2.3,-2.5) -- (2.3,-1.9) --(2.9,-1.9);

\draw[->, dotted, color=black, thick] (1.22, -2.72) -- (1.8, -2.72) -- (1.8, -4.2) -- (1.4,-4.2);

\draw[->, dotted, color=black, thick] (1.22, -2.58) -- (2, -2.58) -- (2.0, -5.75) -- (1.2,-5.75);
\draw[->, dotted, color=black, thick]  (1.2, -5.85) -- (1.7,-5.85) -- (1.7,-6.6) --(1.4,-6.6);
\draw[->, dotted, color=black, thick] (1.4, -1.75) -- (2.1, -1.75) -- (2.1, -8.1) -- (2.7,-8.1);
\draw[->, dotted, color=black, thick] (1.22, -2.66) -- (1.9, -2.66) -- (1.9, -5) -- (1.8,-5);

\draw (2.9,-1.8) node[draw,  anchor=west, minimum height = 0.5cm] (s1) {\small Veno};
\draw (2.9,-2.6) node[draw,  anchor=west, minimum height = 0.5cm] (s1) {\small TCP AR};
\draw (2.9,-3.4) node[draw,  anchor=west, minimum height = 0.5cm] (s1) {\small Africa};
\draw (2.9,-4.2) node[draw,  anchor=west, minimum height = 0.5cm] (s1) {\small Compound};
\draw (2.9,-5) node[draw,  anchor=west, minimum height = 0.5cm] (s1) {\small Fusion};
\draw (2.9,-6.6) node[draw,  anchor=west, minimum height = 0.5cm] (s1) {\small  AReno};
\draw (2.9,-5.8) node[draw,  anchor=west, minimum height = 0.5cm] (s1) {\small YeAH};
\draw (2.9,-7.4) node[draw,  anchor=west, minimum height = 0.5cm] (s1) {\small Libra};
\draw (2.9,-8.2) node[draw,  anchor=west, minimum height = 0.5cm] (s1) {\small Illinois};

\draw (2.9,-9) node[draw,  anchor=west, minimum height = 0.5cm] (s1) {\small BBR};
\draw (2.9,-9.8) node[draw,  anchor=west, minimum height = 0.5cm] (s1) {\small PCC};
\draw[->, thick, color=black] (2.7, -1.8) -- (2.9, -1.8);
\draw[->, thick, color=black] (2.7, -2.6) -- (2.9, -2.6);
\draw[->, thick, color=black] (2.7, -3.4) -- (2.9, -3.4);
\draw[->, thick, color=black] (2.7, -4.2) -- (2.9, -4.2);
\draw[->, thick, color=black] (2.7, -5) -- (2.9, -5);
\draw[->, thick, color=black] (2.7, -5.8) -- (2.9, -5.8);
\draw[->, thick, color=black] (2.7, -6.6) -- (2.9, -6.6);
\draw[->, thick, color=black] (2.7, -7.4) -- (2.9, -7.4);
\draw[->, thick, color=black] (2.7, -8.2) -- (2.9, -8.2);
\draw[->, thick, color=black] (2.7, -9) --   (2.9, -9);
\draw[->, thick, color=black] (2.7, -9.8) --   (2.9, -9.8);

\draw[->, dotted, color=black, thick] (-1.5, -1.8) -- (-0.4, -1.8) -- (-0.4, -2.5) -- (0.2,-2.5);
\draw[->, dotted, color=black, thick] (-0.9, -2.6) -- (-0.3, -2.6) -- (-0.3, -3.3) -- (0.2,-3.3);
\draw[->, dotted, color=black, thick] (-0.9, -2.7) -- (-0.385, -2.7) -- (-0.385, -5.7) -- (0.2,-5.7);
\draw[->, dotted, color=black, thick] (-1.45, -3.4) -- (-0.7, -3.4) -- (-0.7, -5.4) -- (3.3,-5.4) -- (3.3,-5.55);
\draw[->, dotted, color=black, thick] (-1.1,-4.1) -- (-1, -4.1) -- (-1,-3.8) -- (-0.9, -3.8) --  (4.3, -3.8) -- (4.3,-3.95);
\draw[->, dotted, color=black, thick] (4.2, -3.8) -- (4.2,-3.4) -- (4,-3.4);
\draw[->, dotted, color=black, thick] (-0.9, -2.5) -- (-0.55, -2.5) -- (-0.55, -1.3) -- (3.3,-1.3)-- (3.3,-1.5);
\draw[->, dotted, color=black, thick] (-0.9, -2.8) -- (-0.5, -2.8) -- (-0.5, -7) -- (3.3,-7) -- (3.3,-7.15);

\end{tikzpicture}
	\caption{Classification of different congestion control algorithms. Dotted arrows indicate that one was based on the other.}\label{fig:classification}
	\vspace{-0.3cm}
\end{figure}

Since the original TCP specification (RFC 793 \cite{rfc793}), numerous congestion control algorithms 
have been developed. In this paper, we focus mostly on the algorithms designed for wired networks. They can be used by both QUIC and TCP and they can be divided into three main groups (see Fig. \ref{fig:classification}): (1) loss-based algorithms detect congestion when buffers are already full and packets are dropped, (2) delay-based algorithms rely on Round Trip Time (RTT) measurements and detect congestion by an increase in RTT, indicating buffering, and (3) hybrid algorithms use some combination of the other two methods. 

\subsection{Loss-based algorithms}

The original congestion control algorithms from \cite{rfc793} were loss-based algorithms. TCP Reno was the first that was widely deployed.
With the increase in network speeds, Reno's conservative approach of halving the congestion window became an issue. TCP connections were unable to fully utilize the available bandwidth, so that other loss-based algorithms were proposed, such as NewReno \cite{rfc5681}, Highspeed-TCP (HS-TCP \cite{floyd2003highspeed}), Hamilton-TCP (H-TCP \cite{leith2004h}), Scalable TCP (STCP \cite{stcp}), Westwood (TCPW \cite{mascolo2001tcp}), TCPW+ (TCP Westwood+ \cite{grieco2004performance}), TCPW-A \cite{1312665}, and LogWestwood+ \cite{kliazovich2008logarithmic}. They all improved upon Reno by including additional mechanisms to probe for network resources more aggressively. They also react more conservatively to loss detection events, and discriminate between different causes of packet loss. 

However, these improvements also came with RTT-fairness issues \cite{xu2004binary,caini2004tcp}. Indeed, when two flows with different RTTs share the same bottleneck link, the flow with the smaller RTT is likely to obtain more resources than other flows. This is due to the algorithm used to discover resources, i.e., the congestion window size function. If it depends on RTT, flows with smaller RTTs probe for resources more often, 
and thus claim more resources. For example, calculations showed that an HS-TCP flow with $x$ times smaller RTT will get a network share that is $x^{4.56}$ times larger than the network share received by the flow with a higher RTT \cite{afanasyev2010host,xu2004binary}. 

To address this issue, BIC \cite{xu2004binary} and Hybla \cite{caini2004tcp} were proposed. Hybla modified NewReno's Slow Start and Congestion Avoidance phases and made them semi-independent of RTT. However, the achieved RTT-fairness meant that flows with higher RTTs behaved more aggressively. As loss detection time is proportional to RTT, these aggressive flows congested the network easily. Xu et al. addressed the RTT-fairness problem by proposing the BIC (Binary Increase Congestion control) algorithm \cite{xu2004binary}. 
The main idea of BIC was to use a binary search algorithm to approach the optimal congestion window size. As a consequence, the closer the algorithm got to the optimum value of the congestion window, 
the less aggressive it became, thereby improving RTT-fairness. 
However, later evaluations \cite{1400105} showed that BIC still has poor fairness, 
as well as a complex implementation. 
In response, Cubic was proposed in \cite{1400105}. 
\begin{bluebox}
Since Cubic is the current default algorithm in the Linux kernel, and thus widely used, we will describe it in more detail and use it as a reference for loss-based algorithms throughout this paper. 

\textbf{Cubic's} main difference compared to traditional algorithms is the use of a cubic function (see Fig.~\ref{fig:cubic}) for the congestion window size, defined as \cite{afanasyev2010host}:  
\begin{equation}
   cwnd = C \cdot \left( \Delta - \sqrt{\beta \cdot \frac{cwnd_{max}}{C}} \right)  ^3 + cwnd_{max}
\end{equation}
where $\Delta$ is the time elapsed since the last congestion event, $\beta$ is a coefficient of the multiplicative decrease in Fast Recovery, and $cwnd_{max}$ is the observed congestion window size just before the last registered loss.

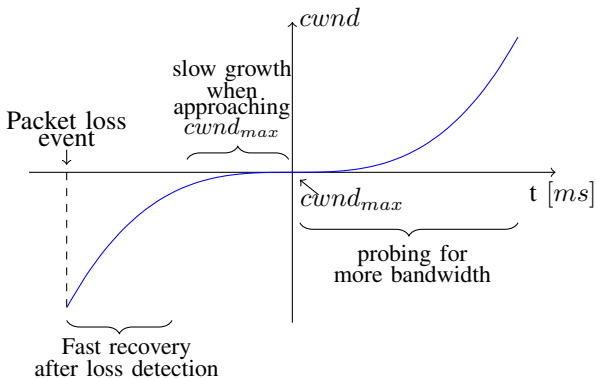
\begin{figure}[h!]
    \centering
    	\begin{tikzpicture}
									
	\draw[->] (-3.5,0) -- (3.5,0); 
	\draw (3.6,-0.3) node[] {\begin{tabular}{c} t $\left[ ms \right]$ \end{tabular}};
	\draw[->] (0,-2) -- (0,2);
	\draw (0.5,1.7) node[above] {\begin{tabular}{c} $cwnd$ \end{tabular}};
	\draw (0.8,-0.7) node[above] {\begin{tabular}{c} $cwnd_{max}$ \end{tabular}};
    \draw[->] (0.35,-0.3) -- (0.1,-0.1);

	\draw[-, dashed] (-3,0)  -- (-3,-1.8);
	\draw (-3,0.7) node[] {Packet loss};
	\draw (-3,0.45) node[] {event};
    \draw[->] (-3,0.3) -- (-3,0.1);

	\draw [decorate,decoration={brace,amplitude=5pt,raise=2pt, mirror}] (-3,-1.9) -- (-1.6,-1.9);
	\draw (1.6,-1.1) node[] {\small probing for};
	\draw (1.6,-1.35) node[] {\small more bandwidth};
	\draw [decorate,decoration={brace,amplitude=5pt,raise=2pt}] (-1.4,0.1) -- (-0.05,0.1);
	\draw [decorate,decoration={brace,amplitude=5pt,raise=2pt, mirror}] (0.1,-0.6) -- (3,-0.6);
	\draw (-0.8,1.35) node[] {\small slow growth};
	\draw (-0.8,1.1) node[] {\small when};
	\draw (-0.8,0.85) node[] {\small approaching};
	\draw (-0.8,0.6) node[] {\small $cwnd_{max}$};
	\draw (-2.2,-2.35) node[] {\small Fast recovery};
	\draw (-2.2,-2.6) node[] {\small after loss detection};

	
	\draw[domain=-3:3,smooth,variable=\x,blue] plot ({\x},  {1/15*(\x )^3});	
	
	

	\end{tikzpicture}    
    \caption{Cubic function used for the congested window size (cwnd).}\label{fig:cubic}
\end{figure}

This choice for the congestion window function has two main advantages. First, the algorithm becomes less aggressive the closer it gets to the target congestion window ($cwnd_{max}$). Second, when the current congestion window $cwnd$ is far from the estimated target $cwnd_{max}$, the algorithm adopts a very fast growth rate, making is especially suitable for flows that need high bandwidth. 

At the start of a flow, since the target window $cwnd_{max}$ is unknown, a conservative approach using the right side of the cubic function is applied. This discovery is more conservative than the exponential one used by Reno. If no new packet loss is detected and the algorithm has reached $cwnd_{max}$, it will continue to probe for more bandwidth according to the same right branch of the cubic function and the congestion window will continue to grow slowly, as shown in Fig. \ref{fig:cubic}.

Cubic has multiple advantages compared to the other algorithms in its group. First, since only one cubic function is used to compute cwnd, it is less complex than BIC, H-TCP or HS-TCP. Second, it enforces RTT-fairness through an RTT-independent congestion window growth function. However, \textbf{Cubic} cannot achieve 100\% resource utilization and requires packet drops since loss is its indicator for congestion.
\end{bluebox}

\subsection{Delay-based algorithms}

In contrast to loss-based algorithms, delay-based algorithms are proactive. They try to find the point where the queues in the network start to fill, by monitoring the variations in RTT. An increase in RTT, or a packet drop, causes them to reduce their sending rate, while a steady RTT indicates a congestion-free state. 

The use of delay as a congestion indicator has multiple advantages. First, these algorithms try to prevent queue buildup. This minimizes the RTT experienced by packets making them best suitable for low-latency applications. Second, they avoid the oscillations in throughput inherent in loss-based algorithms. 
Unfortunately, RTT estimates can be inaccurate due to delayed ACKs, cross traffic, routing dynamics, and queues in the network \cite{TIMELY, afanasyev2010host}. 

The first algorithm that used queuing delay as a congestion indicator was TCP Dual. It maintains the minimum and maximum RTT values observed by the sender, and uses them to compute the maximum queuing delay. Finally, a fraction of the estimated maximum queuing delay is used as a threshold to detect congestion. This approach has multiple drawbacks. First, if any other Dual flow was present in the network at the start of the flow and the minimum RTT got overestimated, unfairness between different Dual flows is possible. Second, due to its conservative nature, network resources are rarely fully utilized. Third, when competing with existing loss-based algorithms, Dual flows suffer from a huge decrease in performance. 
The first improvement to this algorithm was Vegas \cite{brakmo1994tcp}. It focused on estimating the number of packets in the queues, instead of the queuing delay, and keeping it under a certain threshold. 

However, after noticing several problems, such as the inability to coexist with other loss-based flows as well as the inability to respond correctly to routing changes, several modifications to Vegas were proposed, including VegasA, Vegas+  \cite{896302, srijith2005tcp}. For example, VegasA uses an adaptive algorithm to detect route changes, while Vegas+ transitions to a more aggressive approach after it detects a loss-based environment \cite{896302, srijith2005tcp}. 

To fully utilize the available bandwidth in high-speed long-distance networks, two modifications of Vegas were proposed: FAST \cite{1383434, 1354670} and NewVegas \cite{sing2005tcp}. FAST defines a periodic fixed-rate congestion window update (e.g., every $x$ ms) and extends the Vegas congestion window function by including a scaling parameter $\alpha$ providing a trade-off between stability and high throughput. Using a high value of $\alpha$, FAST can easily achieve high throughput and efficiently utilize the existing high capacity infrastructure. Conversely, for lower values of $\alpha$, the algorithm behaves in the same way as the original Vegas algorithm. 
VFAST, an extension to FAST, was proposed to minimize throughput and queue oscillations \cite{belhaj2008vfast}. NewVegas introduces new mechanisms that extend the slow start phase, but with a slower cwnd growth than the typical doubling used during slow start. This allows the window to grow faster at the start of the connection and therefore to claim more resources faster.

Recently, as low latency became important, several new algorithms have been proposed. Hock et al. designed LoLa \cite{lola}, focusing on low latency and convergence to a fair share between flows. To improve performance in datacenter networks, Google proposed TIMELY \cite{TIMELY}, which relies on very precise RTT measurements. 
\begin{bluebox}
Since Vegas is used as the base algorithm by many other delay or hybrid algorithms (see Fig. \ref{fig:classification}), 
we will describe it in more details and use it as a reference for delay-based algorithms.

\textbf{Vegas} continuously computes an estimate of the used buffer size at the bottleneck router (based on the observed RTT measurements), and attempts to keep it under a predefined threshold. Similar to Dual, the minimum RTT value is used as a baseline measurement for a congestion-free network state. However, unlike Dual, Vegas tries to quantify, not a relative, but an absolute number of packets in the queue at the bottleneck router as a function of the expected and actual transmission rate \cite{afanasyev2010host}.

The expected transmission rate is computed using the $RTT_{min}$ as $cwnd/RTT_{min}$ and represents the theoretical rate of a TCP flow in a congestion-free network state. A Vegas flow will achieve this rate if all the transmitted packets are acknowledged within the minimum RTT, i.e., if $RTT_{i} = RTT_{min}$, where $RTT_i$ is the RTT of the $i^{th}$ packet in the flow. Similarly, the actual transmission rate is computed using the current observed RTT as $cwnd/RTT_i$. The number of packets queued $\Delta$ at the bottleneck is: 
\begin{equation}
\Delta = cwnd \cdot \frac{RTT_i - RTT_{min}}{RTT_i}
\end{equation}
For every RTT, Vegas calculates the difference $\Delta$ and tries to keep it between a set of predefined thresholds $\alpha$ and $\beta$ (e.g., in the Linux implementation 2 and 4). If $\Delta$ is higher than $\beta$, it detects congestion and decreases the congestion window by one. If $\Delta$ is lower than $\alpha$, the congestion window is increased by one. 

This approach minimizes the queuing delay, and unlike the loss-based algorithms, can keep the sending rate stable. Oscillations in the network are therefore reduced and the overall throughput of a flow improved \cite{afanasyev2010host}. However, Vegas has several issues. First, due to its conservative nature, the growth of the congestion window is very slow and the algorithm may not fully utilize all the available bandwidth in high-speed networks. Second, any change in RTT (e.g., due to a path change in the network) is interpreted as congestion, resulting in a significant reduction of the sending rate. Finally, and most importantly, it suffers from a huge decrease in performance if used in a network that also has loss-based flows present. In order to counteract the last issue, it switches to a loss-based algorithm upon detecting an ``unfriendly environment'', thereby  losing all its benefits \cite{hock2016}.   

\end{bluebox}

\subsection{Hybrid algorithms}

Hybrid algorithms use both loss and delay as congestion indicators. 
In a network that is congested, or has a really high utilization a conservative approach, such as the one used by the delay-based algorithms, is desirable. However, in a high-speed network that has a low resource utilization, an aggressive approach with a fast cwnd value is needed. 
 
Thus, hybrid algorithms were developed to improve on loss-based congestion control by detecting congestion before the queues are completely full and packets need to be dropped, while keeping the throughput high in presence of other variants of TCP. 
Some of the best known algorithms from this group are TCP Compound \cite{INFOCOM2006}, used as a default congestion control algorithm for the Microsoft Windows operating systems, and TCP BBR \cite{3022184}, recently developed and used by Google. 

The first hybrid algorithm was Veno \cite{fu2003tcp}. It is a modification of Reno that extends the additive increase and multiplicative decrease functions by also using queuing delay as the secondary metric. 
When Veno determines (using this additional delay estimate) that the network is most likely congested, it adjusts its additive increase parameter to probe for network resources more conservatively. If it determines that the network is most likely not congested, but at the same time detects a loss, it adjusts sshtresh to 80\% of its current value. 

To efficiently utilize the available bandwidth in high-speed networks, many algorithms use similar modifications based on the Vegas or Dual network state estimations. Some of the most important ones are Africa \cite{king2005tcp}, 
Compound \cite{INFOCOM2006}, 
and YeAH \cite{baiocchi2007yeah}. 
Other algorithms modify the congestion window increase function to follow a function of both the RTT and the bottleneck link capacity, such as Illinois \cite{liu2008tcp}, AR \cite{shimonishi2005improving}, Fusion \cite{kaneko2007tcp}, TCP-Adaptive Reno (AReno) \cite{shimonishi2006tcp}, and Libra \cite{marfia2007tcp}.

In 2016, Google developed the bottleneck bandwidth and round-trip time (BBR) algorithm. 
At the same time, a new approach to congestion control using online learning was proposed in PCC \cite{dong2015pcc}. We use BBR as our representative for hybrid algorithms, since it is actually deployed (in Google's network) and implemented in the Linux kernel (since v4.9).
\begin{bluebox}

\textbf{Bottleneck bandwidth and round-trip time (BBR)} periodically estimates the available bandwidth ($Bw_{btl}$) and the propagation round-trip time ($RTT_{p}$). $RTT_{p}$ is computed as the minimum of all observed RTT measurements (similar to delay-based algorithms), while the $Bw_{btl}$ is the maximum data delivery rate to the receiver, measured at the sender using the received ACKs (similar to loss-based algorithms). In theory, it can operate at Kleinrock's optimal operating point  \cite{Kleinrock:1979:Power-and-Deterministic} of maximum delivery rate with minimal congestion. 
This maximizes the throughput and prevents the creation of queues, keeping delay minimal. 

BBR uses four different phases: (1) Startup, (2) Drain, (3) Probe Bandwidth, and (4) Probe RTT \cite{cardwell2017bbr, 3022184}. The Startup phase uses the exponential startup function from NewReno (Slow Start), doubling the sending rate each RTT. For each received ACK, the current delivery rate is estimated. Once the measured delivery rate stops increasing for at least 3 consecutive RTTs, BBR assumes to have reached the bottleneck bandwidth. Consequently, it enters the Drain phase to drain the queue formed in the previous RTTs, by reducing the sending rate in the next RTT to 0.75 of the estimated bandwidth delay product ($RTT_{p} \times Bw_{btl}$). 

At the same time, for every data packet sent, BBR calculates an RTT sample. $RTT_{p}$ is set to the minimum recent RTT sample calculated by the sender over the past 10 seconds. After these two values are estimated ($RTT_{p}$ and  $Bw_{btl}$), the cwnd value is set to the measured bandwidth delay product ($BDP = RTT_{p} \times Bw_{btl}$). 

To adapt to changing network conditions, BBR periodically probes for more bandwidth by deliberately sending at a rate 1.25 times higher than the measured bandwidth delay product (from the Probe Bandwidth Phase), for one $RTT_{p}$ interval. This is followed by a new Drain phase in which the rate is set 0.75 times lower than the measured BDP. This is performed every eight cycle, each lasting $RTT_{p}$. 

Additionally, if the $RTT_{p}$ did not change for ten seconds, BBR will stop probing for bandwidth and enter the Probe RTT phase. To measure the $RTT_{p}$ as accurately as possible, the algorithm  quickly reduces the volume of in-flight data to drain the bottleneck queue. To this end, the amount of in-flight data is reduced to four packets for the next 200 ms plus one RTT. To maximize the throughput, BBR is designed to spend the vast majority of time ($\approx 98 \%$) in Probe Bandwidth, and the rest in Probe RTT (the Probe RTT phase lasts $\approx$ 200ms every 10s) \cite{cardwell2017bbr}. 

Since BRR was published in 2016, several problems, mostly related to the Probe RTT phase, were discovered: (1) bandwidth can be shared unfairly depending on the timing of new flows and their RTT, 
(2) the time until a bandwidth equilibrium is regained can last up to 30s, which is bad for short-lived flows, and (3) unfairness towards other protocols, especially Cubic \cite{cardwell2017bbr, scholztowards, 8117540, bbrissues}. 
\end{bluebox}

\section{Evaluation}\label{Sec_eval}

In this section, by using the metrics described in Sec. \ref{Sec:metrics} and via the set-up described in Sec. \ref{Sec:experimentSetup}, we evaluated the algorithms implemented in the Linux kernel and available in the Chromium project. 

\subsection{Performance metrics}\label{Sec:metrics}

\textbf{Sending rate} represents the bit-rate (incl. data-link layer overhead) of a flow generated by the source, per time unit. 

\textbf{Throughput} measures the number of bits (incl. the data-link layer overhead) received at the receiver, per time unit. 

\textbf{Goodput} measures the amount of useful data (i.e., excl. overhead) delivered by the network between specific hosts, per time unit. This value is an indicator of the application-level QoS experienced by the end-users. 
\begin{bluebox}
\begin{equation}
    Goodput = \frac{(D_{s} - D_{r} - D_{o})}{\Delta t}
\end{equation}
where $D_s$ is the number of useful bits transmitted, $D_r$ the number of bits retransmitted and $D_{o}$ the number of overhead bits in time interval $\Delta t$. 
\end{bluebox}
Additionally, we use the \textbf{goodput ratio}, i.e., the amount of useful data transmitted divided by the total amount of data transmitted. 
\begin{bluebox}
\begin{equation}
    Goodput\_ratio = \frac{(D_{s} - D_{r} - D_{o})}{D_{s}}
\end{equation}
\end{bluebox}
 
\begin{bluebox} 
\end{bluebox}

\textbf{Fairness} describes how the available bandwidth is shared among multiple users. 
We consider three different types of fairness: (1) \textbf{intra-fairness} describes the resource distribution between flows running the same congestion control algorithm; (2) \textbf{inter-fairness} describes the resource distribution between flows running different congestion control algorithms, and (3) \textbf{RTT-fairness} describes the resource distribution between flows having different RTTs. Fairness is represented by Jain's index \cite{jain1984quantitative}. This index is based on the throughput and indicates how fair the available bandwidth at the bottleneck is shared between all flows present. 
This fairness index ranges from 0 to 1, with 1 corresponding to all users receiving an equal share. 




\subsection{Experiment setup}\label{Sec:experimentSetup}
\begin{figure}[htb!]
\begin{center}
\begin{tikzpicture}
\draw (-0.5,0) node[circle, draw] (s1) {1};
\draw (-2.5,0.6) node[circle, draw] (h2) {Cn};
\draw (-2.5,-0.6) node[circle, draw] (h1) {C1};
\draw (1.5,0) node[circle, draw] (s2) {2};
\draw (3.5,-0.6) node[circle, draw] (h3) {S1};
\draw (3.5,0.6) node[circle, draw] (h4) {Sn};

\draw (-2.5,0) node[circle, ] (dot) {.};
\draw (3.5,0) node[circle, ] (dot2) {.};
\draw (-2.5,0.1) node[circle, ] (dot) {.};
\draw (3.5,0.1) node[circle, ] (dot2) {.};
\draw (-2.5,-0.1) node[circle, ] (dot) {.};
\draw (3.5,-0.1) node[circle, ] (dot2) {.};

\draw (s1) -- (s2);
\draw (s2) -- (h3);
\draw (s2) -- (h4);
\draw (h2) -- (s1);
\draw (h1) -- (s1);
\draw[->, thick, color=black] (0.7,0.7) -- (1.1,0.2);
\draw[->, thick, color=black] (0.3,0.7) -- (-0.1,0.2);
\node at (0.5,0.9) {{\footnotesize Bandwidth of the bottleneck}};	
\node (controler) at (-2.5,0.1) [rectangle, draw=black!150,rounded corners = .25ex, minimum height=2.5cm, minimum width=1cm] {};
\node at (-2.5,1.17) {\footnotesize Clients};	
\node (controler) at (3.5,0.1) [rectangle, draw=black!150,rounded corners = .25ex, minimum height=2.5cm, minimum width=1cm] {};
\node at (3.5,1.17) {\footnotesize Servers};	
\end{tikzpicture}
\caption{Experiment topology.}\label{topo}
\vspace{-0.3cm}
\end{center}
\end{figure}
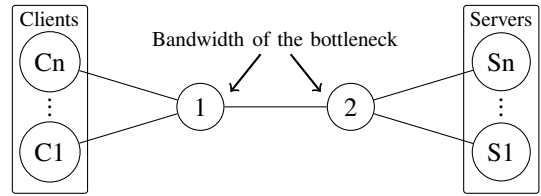

Each server in our testbed has a 64-bit Quad-Core Intel Xeon CPU running at 3GHz with 4GB of main memory and has 6 independent 1 Gbps NICs. Each server can play the role of a 6-degree networking node. All nodes run Linux with kernel version 4.13 with the txqueuelen set to 1000, and were connected as shown in Fig. \ref{topo}. Since the performance of congestion control algorithms is affected by the bottleneck link on the path, it suffices to use such a simple topology.
The maximum bandwidth and the bottleneck (between s1 and s2) was limited to a pre-configured value ($100Mbps$ in case of TCP and $10Mbps$ in case of QUIC) with the use of ethtool. 
To perform measurements, we rely on tshark, iperf, QUIC client and server (available in the Chromium project \cite{chromium}) and socket statistics. From traffic traces (before and after the bottleneck), we calculate the metrics described in Sec.~\ref{Sec:metrics}. All the values are averaged per flow, using a configurable time interval. We consider the following three scenarios: 

\begin{bluebox}
\textbf{Base-Line scenario.}
The purpose of this scenario is to isolate the characteristics of each algorithm. Client $C1$ generates TCP flows (using iperf3) towards a server $S1$, using different congestion control algorithms. To test the sensitivity of the algorithm to ACK-compression~\cite{zhang1991observations}, UDP traffic (between 0\% and 90\% of the available bandwidth) is sent in the opposite direction of the TCP flow (between h2 and h4).

\begin{table*}[!htb]
	\begin{center}
		\caption{Base-Line scenario using just one TCP flow.}
		\label{tab:base}
		\begin{tabular}{l | l | l | c | c | c | c | c | c}
		\toprule                                                                                          Protocol & Group & Algorithm & Average & Average &  Average & Average & Average & Average\\              
          & & & goodput & goodput ratio   & cwnd & RTT  & sending rate & throughput\\                     & &           & $\lbrack Mbps \rbrack$ & $\lbrack \% \rbrack$  & $\lbrack \#packets \rbrack$ &  $\lbrack ms \rbrack$ & $\lbrack Mbps \rbrack$ & $\lbrack Mbps \rbrack$ \\                                                                                    
						\midrule                                                              
            \multirow{13}{1cm}{TCP} & \multirow{6}{2cm}{Loss-based}                                         
             &Reno     & 69.415 & 94.85 &  898.90 & 137.57 & 73.182  & 72.591  \\  
            & &BIC      & 68.580 & 94.88 & 1607.80 & 244.91 & 72.284  & 71.717  \\  
            & &Cubic    & 67.656 & 94.75 & 1216.19 & 203.56 & 71.406  & 70.751  \\  
            & &HS-TCP   & 69.377 & 94.88 & 1553.56 & 245.61 & 73.122  & 72.551  \\  
            & &H-TCP    & 69.040 & 94.87 & 1231.47 & 193.10 & 72.774  & 72.199  \\  
            & &Hybla    & 69.433 & 94.86 &  968.71 & 156.56 & 73.192  & 72.609  \\  
            & &Westwood & 69.543 & 94.84 &  631.16 &  94.20 & 73.324  & 72.725  \\  
            \cline{2-9}                                                                
            & \multirow{2}{2cm}{Delay-based}                                        
             &Vegas    & 52.271 & 95.48 &    7.54 &   1.48 & 54.746 & 54.662   \\  
            & &LoLA     & 63.953 & 95.51 &   17.62 &   2.70 & 66.963 & 66.879   \\  
            \cline{2-9}                                                                
            & \multirow{5}{2cm}{Hybrid}                                             
             &Veno     & 69.291 & 94.83  &  715.98 & 106.79 & 73.067 & 72.461  \\  
            & &Illinois & 69.299 & 94.89  & 1399.04 & 221.61 & 73.034 & 72.470  \\  
            & &YeAH     & 68.501 & 94.82  &  339.77 &  54.73 & 72.243 & 71.636  \\  
            & &BBR      & 67.442 & 95.28  &   31.22 &   4.37 & 70.779 & 70.527  \\  
            \cline{1-9}                                                                
            \multirow{2}{1cm}{QUIC} & \multirow{1}{2cm}{Loss-based}  
            & Cubic & 9.47 & 95.23 & / & / & 9.84 & 9.75 \\
            \cline{2-9}                                                                
            
             & \multirow{1}{2cm}{Hybrid}  
            & BBR & 9.41 & 95.90 & / & / & 9.87 & 9.69 \\            
            
		\end{tabular}
	\end{center}
\end{table*}
Queues in the network fill up quickly, and, as a consequence, the algorithms have to drop packets periodically 
even when no other flow is present in the network.

\end{bluebox}

\textbf{BW scenario.}
Each analyzed algorithm is compared to itself and all others. Host $C_i$ generates TCP flows (using iperf3) towards servers running at $S_i$ using different congestion control algorithms. The number of flows varies between 2 and 4. 

\textbf{RTT scenario with flows having different RTTs.}
The purpose of this scenario is to test the RTT-fairness of different congestion control algorithms. In addition to the setup of the previous scenario, 
the delay at links between $S_i$ and node 1 is artificially increased using Linux TC (adding $0-400ms$). The number of flows varies between 2 and 4. 



\begin{bluebox}
\subsection{Results: Base-Line scenario}\label{Sec:base1}

\textbf{Throughput \& Sending rate.} 
None of the evaluated congestion control algorithms is able to fully utilize the available bandwidth, even when no additional traffic is present on the link. When the bandwidth on the bottleneck link (between nodes $1$ and $2$) is set to $100 Mbps$, the highest measured average throughput is $\approx 74 Mbps$.

Delay-based algorithms, such as Vegas, have the lowest sending rate and throughput, because they are conservative. Their averaged measured sending rate was $\approx 1.1 - 1.4$ times lower than the sending rate of the other evaluated loss-based or hybrid algorithms. 
Loss-based algorithms on the other hand, being very aggressive, have the highest number of retransmissions (between $14 - 223$). 
\end{bluebox}
\begin{bluebox}
\begin{figure}[!htb]
\begin{center}
	\begin{tikzpicture}
	\begin{axis}[
            title style={at={(0.5,1.1)},anchor=north},
	        width = 8.5cm, height = 4cm,
		    xlabel={t $\lbrack s \rbrack$},
            xlabel style={at={(0.5,-0.37)},anchor=south},
		    ylabel={\small{\# Retransmissions}},
    		legend columns=2,
    		legend style={at={(0.75,0.75)},anchor=south east},	
		    xmin=0, xmax=60, ymin=-0.1, ymax=35, grid,]
		    \addplot [color=red, mark=none, line width=1.2pt] 
		    table [col sep=semicolon, x index=0, y index=1,] {csv/retransmissionsCubic.csv};
		    \addplot [color=green, mark=none, line width=1.2pt] 
		    table [col sep=semicolon, x index=0, y index=1,] {csv/retransmissionsVeno.csv};
		    \legend{\small Cubic, Veno}
		\end{axis}
	\end{tikzpicture}
\vspace{-0.3cm}
\caption{Baseline scenario: Number of retransmissions.}\label{fig:losscubic}
\vspace{-0.3cm}
\end{center}
\end{figure}
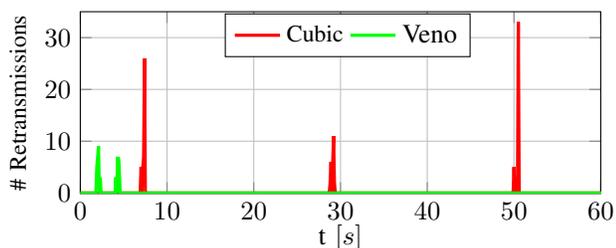
\begin{figure*}[tbh!]
\begin{center}
\begin{subfigure}{0.34\textwidth}
	\begin{tikzpicture}
	\begin{axis}[
            title style={at={(0.5,1.1)},anchor=north},
	        width = 6cm, height = 4cm,
		    xlabel={t $\lbrack s \rbrack$},
            xlabel style={at={(0.5,-0.3)},anchor=south},
    		legend style={at={(0.4,0.73)},anchor=south east},	
		    ylabel={\small{Avg. RTT $\lbrack ms \rbrack$}},
		    xmin=0, xmax=60, ymin=0, ymax=550, grid,]
		    \addplot [color=red, mark=none, line width=1.2pt] 
		    table [ col sep=semicolon, x index=0, y index=1,] {csv/Cubicrtt.csv};
		    \legend{\small Cubic}
		\end{axis}
	\end{tikzpicture}
	\caption{Loss-based algorithms: Cubic.\label{fig:rttcubic}}
\end{subfigure}%
\begin{subfigure}{0.315\textwidth}
	\begin{tikzpicture}
	\begin{axis}[
            title style={at={(0.5,1.1)},anchor=north},
	        width = 6cm, height = 4cm,
		    xlabel={t $\lbrack s \rbrack$},
            xlabel style={at={(0.5,-0.3)},anchor=south},
    		legend style={at={(0.98,0.73)},anchor=south east},	
		    ylabel={\small{Avg. RTT $\lbrack ms \rbrack$}},
		    xmin=0, xmax=60, ymin=0, ymax=5, grid,]
		    \addplot [color=green, mark=none, line width=1.2pt] 
		    table [ col sep=semicolon, x index=0, y index=1,] {csv/Vegasrtt.csv};
		    \legend{\small Vegas}
		\end{axis}
	\end{tikzpicture}
	\caption{Delay-based algorithms: Vegas.}
	\label{fig:rttvegas}
\end{subfigure}%
\begin{subfigure}{0.315\textwidth}
	\begin{tikzpicture}
	\begin{axis}[
            title style={at={(0.5,1.1)},anchor=north},
	        width = 6cm, height = 4cm, 
		    xlabel={t $\lbrack s \rbrack$},
            xlabel style={at={(0.5,-0.3)},anchor=south},
    		legend style={at={(0.37,0.73)},anchor=south east},	
		    ylabel={\small{Avg. RTT $\lbrack ms \rbrack$}},
		    xmin=0, xmax=60, ymin=0, ymax=8, grid,]
		    \addplot [color=blue, mark=none, line width=1.2pt] 
		    table [ col sep=semicolon, x index=0, y index=1,] {csv/Bbrrtt.csv};
		    \legend{\small BBR}
		\end{axis}
	\end{tikzpicture}
	\caption{Hybrid algorithms: BBR.}
	\label{fig:rttbbr}
\end{subfigure}%
\caption{Comparison of the average RTT (time unit 100ms) for reference flavours of the three groups of congestion control algorithms, baseline scenario using just one TCP flow.}\label{fig:rtt}
\end{center}
\end{figure*}
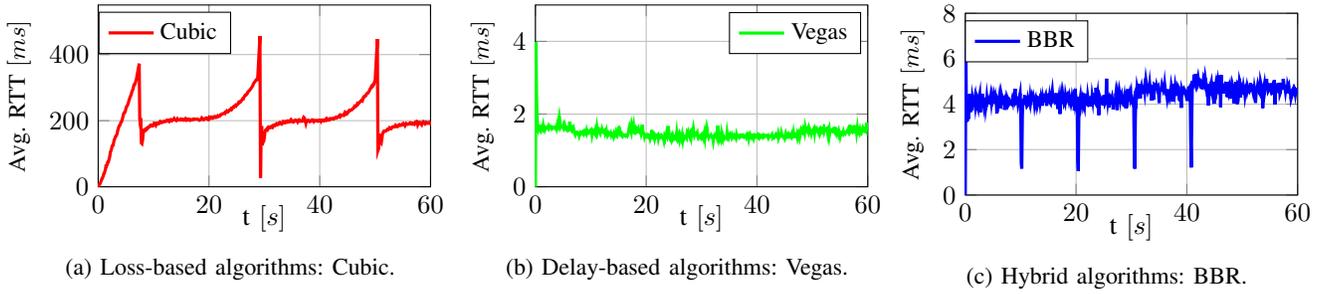
\end{bluebox}
\begin{bluebox}
Similar to delay-based algorithms that have no retransmissions, hybrid algorithms have a very small number of retransmissions at the start of the connection. After they estimate the bandwidth, they are able to send traffic without filling the buffers.

\textbf{Congestion window (cwnd).} Algorithms intended for high-speed communications (HS-TCP, H-TCP, BIC, Cubic, Illinois) see, as expected, a really fast growth of the congestion window. However, a higher cwnd value does not translate into a higher sending rate (Table \ref{tab:base}). The congestion window depends on the estimated RTT. For example, Cubic can send $\approx 1220$ packets every $\approx 200\;ms$. This equals to $\approx$ 6 packets every $1 \; ms$, in contrast to Westwood that can send $\approx 6.7$, despite a smaller cwnd.

\textbf{Delay.}
Due to their conservative nature, delay-based algorithms have the lowest average RTT, with a maximum RTT of $\approx 2ms$. Their measured average RTT is $\approx 35 - 166$ times lower than the one of loss-based algorithms and at least twice as low as that of hybrid algorithms (Table \ref{tab:base}). In comparison, loss-based algorithms (Cubic) have a maximum delay in the order of $500\;ms$, which happens when the queues are filled and packets need to be dropped. These algorithms have a very aggressive approach: the queues get filled very quickly and only during short time intervals immediately after loss detection does the RTT get below $200\;ms$.

Hybrid algorithms, such as BBR, have a slightly higher maximum RTT (around $4\;ms$) than delay-based ones. During the ProbeRTT phase, the cwnd is reduced to 4 packets, and, as no queues are present in the network, the RTT drops to $\leq 1.6\;ms$, comparable to delay-based algorithms. However, during other phases, a smaller queuing delay is always present ($\approx 4-5\;ms$). 

\textbf{Goodput.}
Due to their conservative nature, proactive algorithms, such as Vegas, obtain the highest goodput ratio. However, due to their lower sending rate, the average goodput is lower than the one achieved by loss-based algorithms (similar to the sending rate).

\begin{table*}[!htb]
	\begin{center}
		\caption{Base-Line scenario: Different metrics for different congestion control algorithm groups for the baseline scenario with two flows, one TCP flow between and one UDP flow sent in the opposite direction to the TCP flow.}
		\label{tab:base2}
		\begin{tabular}{l | l | l | c | c | c | c | c | c}
			\toprule                                                                Protocol & Group & Algorithm & Average & Average &  Average & Average & Average & Average\\              
			& & & goodput & goodput ratio   & cwnd & RTT  & sending rate & throughput\\                      &           & $\lbrack Mbps \rbrack$ & & $\lbrack \% \rbrack$  & $\lbrack \#packets \rbrack$ &  $\lbrack ms \rbrack$ & $\lbrack Mbps \rbrack$ & $\lbrack Mbps \rbrack$ \\                                                                                          
			\midrule                                                                \multirow{13}{0.5cm}{TCP}                   &\multirow{6}{2cm}{Loss-based}                                      
			&Reno      & 29.34  & 96.39 & 1426.44 & 349.98 & 30.44 & 30.01 \\  
			& &Bic       & 38.36  & 95.76 & 2300.95 & 642.30 & 40.06 & 39.17 \\  
			& &Cubic     & 51.45  & 96.15 & 1723.35 & 322.45 & 53.51 & 52.32 \\  
			& &TCP-HS    & 57.56  & 95.75 & 2366.46 & 352.45 & 60.11 & 57.53 \\  
			& &H-TCP     & 45.53  & 96.26 & 2934.63 & 358.95 & 47.30 & 46.45 \\  
			& &Hybla     & 41.61  & 93.16 & 1993.45 & 333.87 & 44.67 & 40.93 \\  
			& &Westwood  & 57.87  & 96.12 & 4148.01 & 358.51 & 60.20 & 58.87 \\  
			\cline{2-9}                                                             
			& \multirow{2}{2cm}{Delay-based}                                     
			&Vegas     & 2.856  & 96.23 & 5.29252 & 23.023 & 2.968 & 2.945 \\  
			& &LoLa      & 2.921  & 96.86 & 8.10688 & 34.288 & 3.016 & 2.997 \\  
			\cline{2-9}                                                             
			& \multirow{5}{2cm}{Hybrid}                                          
			&Veno      & 6.461  & 97.22 & 69.1582 & 124.56 & 6.645 & 6.589 \\  
			& &Illinois  & 57.26  & 96.01 & 3733.65 & 354.37 & 59.64 & 58.15 \\  
			& &YeAH      & 6.696  & 97.93 & 82.5048 & 141.98 & 6.838 & 6.744 \\  
			& &BBR       & 10.35  & 98.29 & 207.796 & 127.73 & 10.53 & 10.46 \\  
			\hline
		\end{tabular}
		\vspace{-0.5cm}
	\end{center}
\end{table*}

\textbf{Sensitivity to ACK-compression.} TCP congestion control exploits the fact that packets arrive at the receiver at a rate the bottleneck can support. Upon reception, the receiver informs the sender by sending an ACK. The sender, consequently, sends new data packets at the same rate (or higher depending on the current phase and the congestion algorithm used) and with the same spacing, to avoid overloading the bottleneck. This property is called self-clocking. However, to correctly exploit it, ACKs need to arrive with the same spacing with which the receiver generated them. If ACKs spend any time in queues or get lost, the sender might be misled into sending more data than the network can accept (if ACKs arrive in bursts), or will detect congestion even if bandwidth is available on its path to the sender. This effect can be observed in Table~\ref{tab:base2}. 

We observe that ACKs were either lost or received at a different rate than they were sent. Algorithms were not able to correctly estimate the bandwidth and detected congestion prematurely. The sending rate of all the evaluated algorithms experiences a drop of at least 10Mbps (when compared to the scenario with no cross-traffic). Delay-based algorithms are particularly vulnerable to this effect, with their rate dropping with a factor of $18-22$ to $\approx 3\%$ of the available bandwidth on the link between h2 and h4 (Table \ref{tab:base2}). 

The average delay detected at the sender side increased, due to ACK packets being queued at node s2. This causes an increase in cwnd, although less packets were sent between hosts h2 and h4. This has negative effects for these flows (that measure higher RTT) as they are more vulnerable to changes in the network as well as packet loss (explained further in Sec. \ref{Sec:dual3}).

Illinois is less affected than other hybrid algorithms (BBR, Veno and YeAh), thanks to the way it controls the cwnd growth. Parameters of additive increase and multiplicative decrease are defined as functions of the average queuing delay. Thus, measured delay is used to pace the growth of cwnd. However, as the measured queuing delay does not vary too much in this scenario (traffic in the opposite direction is constant), the algorithm behaves similarly to other loss-based algorithms. 
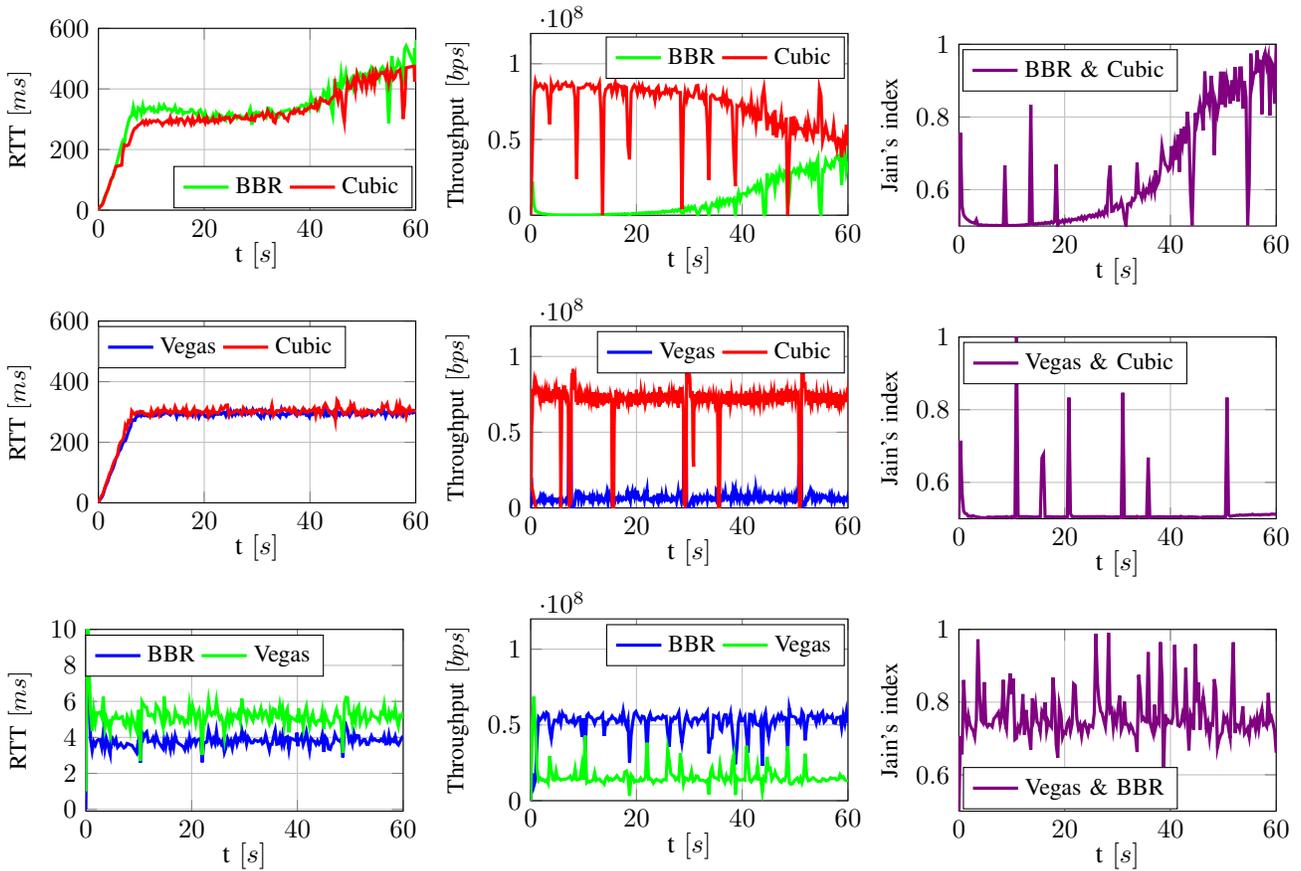
\begin{figure*}[!htb]
	\begin{center}
		\begin{subfigure}{0.32\textwidth}
    \centering
	\begin{tikzpicture}
	\begin{axis}[
            title style={at={(0.5,1.1)},anchor=north},
	        width = 5.8cm, height = 4cm,
		    xlabel={t $\lbrack s \rbrack$},
            xlabel style={at={(0.5,-0.37)},anchor=south},
		    ylabel={\small{RTT $\lbrack ms \rbrack$}},
    		legend style={at={(0.99,0.01)},anchor=south east},	
    		legend columns=2,
		    xmin=0, xmax=60, ymin=-0.1, ymax=600, grid,]
		    \addplot [color=green, mark=none, line width=1.2pt] 
		    table [col sep=semicolon, x index=0, y index=1,] {csv/rttBBRCubic1sec.csv};
		    \addplot [color=red, mark=none, line width=1.2pt] 
		    table [col sep=semicolon, x index=2, y index=3,] {csv/rttBBRCubic1sec.csv};
		    \legend{\small BBR, \small Cubic}
		\end{axis}
	\end{tikzpicture}
	\label{fig:rttbbcubic}
\end{subfigure}%
\begin{subfigure}{0.32\textwidth}
    \centering
	\begin{tikzpicture}
	\begin{axis}[
            title style={at={(0.5,1.1)},anchor=north},
	        width = 5.8cm, height = 4cm,
		    xlabel={t $\lbrack s \rbrack$},
            xlabel style={at={(0.5,-0.37)},anchor=south},
		    ylabel={\small{Throughput $\lbrack bps \rbrack$}},
    		legend style={at={(0.99,0.75)},anchor=south east},	
    		legend columns=2,
		    xmin=0, xmax=60, ymin=-0.1, ymax=120000000, grid,]
		    \addplot [color=green, mark=none, line width=1.2pt] 
		    table [col sep=semicolon, x index=0, y index=1,] {csv/throughputBBRCubic1sec.csv};
		    \addplot [color=red, mark=none, line width=1.2pt] 
		    table [col sep=semicolon, x index=2, y index=3,] {csv/throughputBBRCubic1sec.csv};
		    \legend{\small BBR, \small Cubic}
		\end{axis}
	\end{tikzpicture}
	\label{fig:throughputbbcubic}
\end{subfigure}%
\begin{subfigure}{0.32\textwidth}
    \centering
	\begin{tikzpicture}
	\begin{axis}[
            title style={at={(0.5,1.1)},anchor=north},
	        width = 5.8cm, height = 4cm,
		    xlabel={t $\lbrack s \rbrack$},
            xlabel style={at={(0.5,-0.37)},anchor=south},
		    ylabel={\small{Jain's index}},
    		legend style={at={(0.69,0.74)},anchor=south east},	
		    xmin=0, xmax=60, ymin=0.5, ymax=1, grid,]
		    \addplot [color=blue!50!red, mark=none, line width=1.2pt] 
		    table [col sep=semicolon, x index=0, y index=1,] {csv/fairnessbbrcubic1sec.csv};
		  \legend{\small BBR \& Cubic}
		\end{axis}
	\end{tikzpicture}
	\label{fig:fairnesbbrcubic}
\end{subfigure}
\vspace{-0.2cm}

\begin{subfigure}{0.32\textwidth}
    \centering
	\begin{tikzpicture}
	\begin{axis}[
            title style={at={(0.5,1.1)},anchor=north},
	        width = 5.8cm, height = 4cm,
		    xlabel={t $\lbrack s \rbrack$},
            xlabel style={at={(0.5,-0.37)},anchor=south},
		    ylabel={\small{RTT $\lbrack ms \rbrack$}},
    		legend style={at={(0.78,0.74)},anchor=south east},	
    		legend columns=2,
		    xmin=0, xmax=60, ymin=-0.1, ymax=600, grid,]
		    \addplot [color=blue, mark=none, line width=1.2pt] 
		    table [col sep=semicolon, x index=0, y index=1,] {csv/cubicvegasrttnew1sec.csv};
		    \addplot [color=red, mark=none, line width=1.2pt] 
		    table [col sep=semicolon, x index=2, y index=3,] {csv/cubicvegasrttnew1sec.csv};
		    \legend{\small Vegas, \small Cubic}
		\end{axis}
	\end{tikzpicture}
	\label{fig:rttvegascubic}
\end{subfigure}%
\begin{subfigure}{0.32\textwidth}
    \centering
	\begin{tikzpicture}
	\begin{axis}[
            title style={at={(0.5,1.1)},anchor=north},
	        width = 5.8cm, height = 4cm,
		    xlabel={t $\lbrack s \rbrack$},
            xlabel style={at={(0.5,-0.37)},anchor=south},
		    ylabel={\small{Throughput $\lbrack bps \rbrack$}},
    		legend style={at={(0.99,0.74)},anchor=south east},	
    		legend columns=2,
		    xmin=0, xmax=60, ymin=-0.1, ymax=120000000, grid,]
		    \addplot [color=blue, mark=none, line width=1.2pt] 
		    table [col sep=semicolon, x index=0, y index=1,] {csv/CubicVegasthroughput.csv};
		    \addplot [color=red, mark=none, line width=1.2pt] 
		    table [col sep=semicolon, x index=2, y index=3,] {csv/CubicVegasthroughput.csv};
		    \legend{\small Vegas, \small Cubic}
		\end{axis}
	\end{tikzpicture}
	\label{fig:throughputvegascubic}
\end{subfigure}%
\begin{subfigure}{0.32\textwidth}
    \centering
	\begin{tikzpicture}
	\begin{axis}[
            title style={at={(0.5,1.1)},anchor=north},
	        width = 5.8cm, height = 4cm,
		    xlabel={t $\lbrack s \rbrack$},
            xlabel style={at={(0.5,-0.37)},anchor=south},
		    ylabel={\small{Jain's index}},
    		legend style={at={(0.72,0.74)},anchor=south east},	
		    xmin=0, xmax=60, ymin=0.5, ymax=1, grid,]
		    \addplot [color=blue!50!red, mark=none, line width=1.2pt] 
		    table [col sep=semicolon, x index=0, y index=1,] {csv/fairnessvegascubic1sec.csv};
		    \legend{\small Vegas \& Cubic}
		\end{axis}
	\end{tikzpicture}
	\label{fig:fairnessvegascubic}
\end{subfigure}
\vspace{-0.2cm}

\begin{subfigure}{0.32\textwidth}
    \centering
	\begin{tikzpicture}
	\begin{axis}[
            title style={at={(0.5,1.1)},anchor=north},
	        width = 5.8cm, height = 4cm,
		    xlabel={t $\lbrack s \rbrack$},
            xlabel style={at={(0.5,-0.37)},anchor=south},
		    ylabel={\small{RTT $\lbrack ms \rbrack$}},
    		legend style={at={(0.75,0.74)},anchor=south east},	
    		legend columns=2,
		    xmin=0, xmax=60, ymin=-0.1, ymax=10, grid,]
		    \addplot [color=blue, mark=none, line width=1.2pt] 
		    table [col sep=semicolon, x index=0, y index=1,] {csv/VegasBbrrttnew1sec.csv};	
		    \addplot [color=green, mark=none, line width=1.2pt] 
		    table [col sep=semicolon, x index=2, y index=3,] {csv/VegasBbrrttnew1sec.csv};		    
		    \legend{\small BBR, \small Vegas}
		\end{axis}
	\end{tikzpicture}
	\label{fig:rttvegasbbr}
\end{subfigure}%
\begin{subfigure}{0.32\textwidth}
    \centering
	\begin{tikzpicture}
	\begin{axis}[
            title style={at={(0.5,1.1)},anchor=north},
	        width = 5.8cm, height = 4cm,
		    xlabel={t $\lbrack s \rbrack$},
            xlabel style={at={(0.5,-0.37)},anchor=south},
		    ylabel={\small{Throughput $\lbrack bps \rbrack$}},
    		legend style={at={(0.99,0.74)},anchor=south east},	
    		legend columns=2,
		    xmin=0, xmax=60, ymin=-0.1, ymax=120000000, grid,]
		    \addplot [color=blue, mark=none, line width=1.2pt] 
		    table [col sep=semicolon, x index=0, y index=1,] {csv/VegasBbrthroughput1sec.csv};
		    \addplot [color=green, mark=none, line width=1.2pt] 
		    table [col sep=semicolon, x index=2, y index=3,] {csv/VegasBbrthroughput1sec.csv};
		    \legend{\small BBR, \small Vegas}
		\end{axis}
	\end{tikzpicture}
	\label{fig:throughputvegasbbr}
\end{subfigure}%
\begin{subfigure}{0.32\textwidth}
    \centering
	\begin{tikzpicture}
	\begin{axis}[
            title style={at={(0.5,1.1)},anchor=north},
	        width = 5.8cm, height = 4cm,
		    xlabel={t $\lbrack s \rbrack$},
            xlabel style={at={(0.5,-0.37)},anchor=south},
		    ylabel={\small{Jain's index}},
    		legend style={at={(0.69,0.01)},anchor=south east},	
		    xmin=0, xmax=60, ymin=0.5, ymax=1, grid,]
		    \addplot [color=blue!50!red, mark=none, line width=1.2pt] 
		    table [col sep=semicolon, x index=0, y index=1,] {csv/fairnessvegasbbr1sec.csv};
		    \legend{\small Vegas \&  BBR}
		\end{axis}
	\end{tikzpicture}
	\label{fig:fairnessvegasbbr}
\end{subfigure}
		\caption{BW scenario: Comparison of average RTT, average throughput and fairness index for representatives of the congestion control algorithm groups in case the link is shared by 2 flows (time unit 300ms).}\label{fig:dual1}
	\end{center}
\end{figure*}

\textbf{Summary.}
From our baseline scenario, we observe that no TCP flavour is able to use the full capacity of a high-speed link, despite no cross-traffic. Further, as expected, we observe that aggressive loss-based algorithms trigger significant retransmissions, while delay-based and hybrid TCP flavours function without creating losses and retransmissions. We also observe the impact of aggressive behaviors on RTT, with loss-based algorithms leading to higher RTTs, while the others are capable of using resources without filling buffers and therefore increasing the RTT. As expected, we observe a high sensitivity of all algorithms to ACK compression leading to a drop in throughput, especially for delay-based algorithms. Delay-based algorithms are particularly sensitive to this, while hybrid algorithms behave differently, either like loss-based ones, or more like the delay-based ones.

\end{bluebox}

\subsection{Results: BW scenario}\label{Sec:dual1}

\begin{figure*}[!ht]
\begin{center}
\begin{subfigure}{0.32\textwidth}
    \centering
	\begin{tikzpicture}
	\begin{axis}[
            title style={at={(0.5,1.1)},anchor=north},
	        width = 5.8cm, height = 4cm,
		    xlabel={t $\lbrack s \rbrack$},
            xlabel style={at={(0.5,-0.37)},anchor=south},
		    ylabel={\small{RTT $\lbrack ms \rbrack$}},
    		legend style={at={(0.79,0.74)},anchor=south east},	
    		legend columns=2,
		    xmin=0, xmax=60, ymin=-0.1, ymax=800, grid,]
		    \addplot [color=green, mark=none, line width=1.2pt] 
		    table [col sep=semicolon, x index=0, y index=1,] {csv/cubiccubicrtt1sec.csv};
		    \addplot [color=red, mark=none, line width=1.2pt] 
		    table [col sep=semicolon, x index=2, y index=3,] {csv/cubiccubicrtt1sec.csv};
		    \legend{\small Cubic, \small Cubic}
		\end{axis}
	\end{tikzpicture}
	\label{fig:rttbbcubic2}
\end{subfigure}%
\begin{subfigure}{0.32\textwidth}
    \centering
	\begin{tikzpicture}
	\begin{axis}[
            title style={at={(0.5,1.1)},anchor=north},
	        width = 5.8cm, height = 4cm,
		    xlabel={t $\lbrack s \rbrack$},
            xlabel style={at={(0.5,-0.37)},anchor=south},
		    ylabel={\small{Throughput $\lbrack bps \rbrack$}},
    		legend style={at={(0.98,0.74)},anchor=south east},	
    		legend columns=2,
		    xmin=0, xmax=60, ymin=-0.1, ymax=120000000, grid,]
		    \addplot [color=green, mark=none, line width=1.2pt] 
		    table [col sep=semicolon, dotted, x index=0, y index=1,] {csv/cubiccubicthroughput1sec.csv};
		    \addplot [color=red,mark=none, line width=1.2pt] 
		    table [col sep=semicolon, x index=2, y index=3,] {csv/cubiccubicthroughput1sec.csv};
		    \legend{\small Cubic, \small Cubic}
		\end{axis}
	\end{tikzpicture}
	\label{fig:throughputbbcubic2}
\end{subfigure}%
\begin{subfigure}{0.32\textwidth}
    \centering
	\begin{tikzpicture}
	\begin{axis}[
            title style={at={(0.5,1.1)},anchor=north},
	        width = 5.8cm, height = 4cm,
		    xlabel={t $\lbrack s \rbrack$},
            xlabel style={at={(0.5,-0.37)},anchor=south},
		    ylabel={\small{Jain's index}},
    		legend style={at={(0.51,0.02)},anchor=south east},	
		    xmin=0, xmax=60, ymin=0.5, ymax=1, grid,]
		    \addplot [color=blue!50!red, mark=none, line width=1.2pt] 
		    table [col sep=semicolon, x index=0, y index=1,] {csv/fairnescubicubic1sec.csv};
		    \legend{\small 2xCubic}
		\end{axis}
	\end{tikzpicture}
	\label{fig:fairnesbbrcubic2}
\end{subfigure}

\begin{subfigure}{0.32\textwidth}
    \centering
	\begin{tikzpicture}
	\begin{axis}[
            title style={at={(0.5,1.1)},anchor=north},
	        width = 5.8cm, height = 4cm,
		    xlabel={t $\lbrack s \rbrack$},
            xlabel style={at={(0.5,-0.37)},anchor=south},
		    ylabel={\small{RTT $\lbrack ms \rbrack$}},
    		legend style={at={(0.775,0.74)},anchor=south east},	
    		legend columns=2,
		    xmin=0, xmax=60, ymin=-0.1, ymax=4, grid,]
		    \addplot [color=blue, mark=none, line width=1.2pt] 
		    table [col sep=semicolon, x index=0, y index=1,] {csv/vegasvegasrtt1sec.csv};
		    \addplot [color=red, mark=none, line width=1.2pt] 
		    table [col sep=semicolon, x index=2, y index=3,] {csv/vegasvegasrtt1sec.csv};
		    \legend{\small Vegas, \small Vegas}
		\end{axis}
	\end{tikzpicture}
	\label{fig:rttvegascubic2}
\end{subfigure}%
\begin{subfigure}{0.32\textwidth}
    \centering
	\begin{tikzpicture}
	\begin{axis}[
            title style={at={(0.5,1.1)},anchor=north},
	        width = 5.8cm, height = 4cm,
		    xlabel={t $\lbrack s \rbrack$},
            xlabel style={at={(0.5,-0.37)},anchor=south},
		    ylabel={\small{Throughput $\lbrack bps \rbrack$}},
    		legend style={at={(0.98,0.74)},anchor=south east},	
    		legend columns=2,
		    xmin=0, xmax=60, ymin=-0.1, ymax=120000000, grid,]
		    \addplot [color=blue, mark=none, line width=1.2pt] 
		    table [col sep=semicolon, x index=0, y index=1,] {csv/vegasvegasthroughput1sec.csv};
		    \addplot [color=red, mark=none, line width=1.2pt] 
		    table [col sep=semicolon, x index=2, y index=3,] {csv/vegasvegasthroughput1sec.csv};
		    \legend{\small Vegas, \small Vegas}
		\end{axis}
	\end{tikzpicture}
	\label{fig:throughputvegascubic2}
\end{subfigure}%
\begin{subfigure}{0.32\textwidth}
    \centering
	\begin{tikzpicture}
	\begin{axis}[
            title style={at={(0.5,1.1)},anchor=north},
	        width = 5.8cm, height = 4cm,
		    xlabel={t $\lbrack s \rbrack$},
            xlabel style={at={(0.5,-0.37)},anchor=south},
		    ylabel={\small{Jain's index}},
    		legend style={at={(0.5,0.02)},anchor=south east},	
		    xmin=0, xmax=60, ymin=0.4, ymax=1, grid,]
		    \addplot [color=blue!50!red, mark=none, line width=1.2pt] 
		    table [col sep=semicolon, x index=0, y index=1,] {csv/fairnevegasvegas1sec.csv};
		    \legend{\small 2xVegas}
		\end{axis}
	\end{tikzpicture}
	\label{fig:fairnessvegascubic2}
\end{subfigure}

\begin{subfigure}{0.32\textwidth}
    \centering
	\begin{tikzpicture}
	\begin{axis}[
            title style={at={(0.5,1.1)},anchor=north},
	        width = 5.8cm, height = 4cm,
		    xlabel={t $\lbrack s \rbrack$},
            xlabel style={at={(0.5,-0.37)},anchor=south},
		    ylabel={\small{RTT $\lbrack ms \rbrack$}},
    		legend style={at={(0.72,0.74)},anchor=south east},	
    		legend columns=2,
		    xmin=0, xmax=60, ymin=-0.1, ymax=10, grid,]
		    \addplot [color=blue, mark=none, line width=1.2pt] 
		    table [col sep=semicolon, x index=0, y index=1,] {csv/bbrbbrrtt1sec.csv};	
		    \addplot [color=green, mark=none, line width=1.2pt] 
		    table [col sep=semicolon, x index=2, y index=3,] {csv/bbrbbrrtt1sec.csv};		    
		    \legend{\small BBR, \small BBR}
		\end{axis}
	\end{tikzpicture}
	\label{fig:rttvegasbbr2}
\end{subfigure}%
\begin{subfigure}{0.32\textwidth}
    \centering
	\begin{tikzpicture}
	\begin{axis}[
            title style={at={(0.5,1.1)},anchor=north},
	        width = 5.8cm, height = 4cm,
		    xlabel={t $\lbrack s \rbrack$},
            xlabel style={at={(0.5,-0.37)},anchor=south},
		    ylabel={\small{Throughput $\lbrack bps \rbrack$}},
    		legend style={at={(0.98,0.74)},anchor=south east},	
    		legend columns=2,
		    xmin=0, xmax=60, ymin=-0.1, ymax=120000000, grid,]
		    \addplot [color=blue, mark=none, line width=1.2pt] 
		    table [col sep=semicolon, x index=2, y index=3,] {csv/bbrbbrthroughput1sec.csv};
		    \addplot [color=green, mark=none, line width=1.2pt] 
		    table [col sep=semicolon, x index=0, y index=1,] {csv/bbrbbrthroughput1sec.csv};
		    \legend{\small BBR, \small BBR}
		\end{axis}
	\end{tikzpicture}
	\label{fig:throughputvegasbbr2}
\end{subfigure}%
\begin{subfigure}{0.32\textwidth}
    \centering
	\begin{tikzpicture}
	\begin{axis}[
            title style={at={(0.5,1.1)},anchor=north},
	        width = 5.8cm, height = 4cm,
		    xlabel={t $\lbrack s \rbrack$},
            xlabel style={at={(0.5,-0.37)},anchor=south},
		    ylabel={\small{Jain's index}},
    		legend style={at={(0.48,0.02)},anchor=south east},	
		    xmin=0, xmax=60, ymin=0.4, ymax=1, grid,]
		    \addplot [color=blue!50!red, mark=none, line width=1.2pt] 
		    table [col sep=semicolon, x index=0, y index=1,] {csv/fairnesbbrbbr1sec.csv};
		    \legend{\small 2xBBR}
		\end{axis}
	\end{tikzpicture}
	\label{fig:fairnessvegasbbr2}
\end{subfigure}
\caption{BW scenario: Comparison of average RTT, average throughput, and fairness index for representatives of the congestion control algorithm classes groups in case the link is shared by 2 flows (time unit 300ms).}\label{fig:dual2}
\end{center}
\end{figure*}
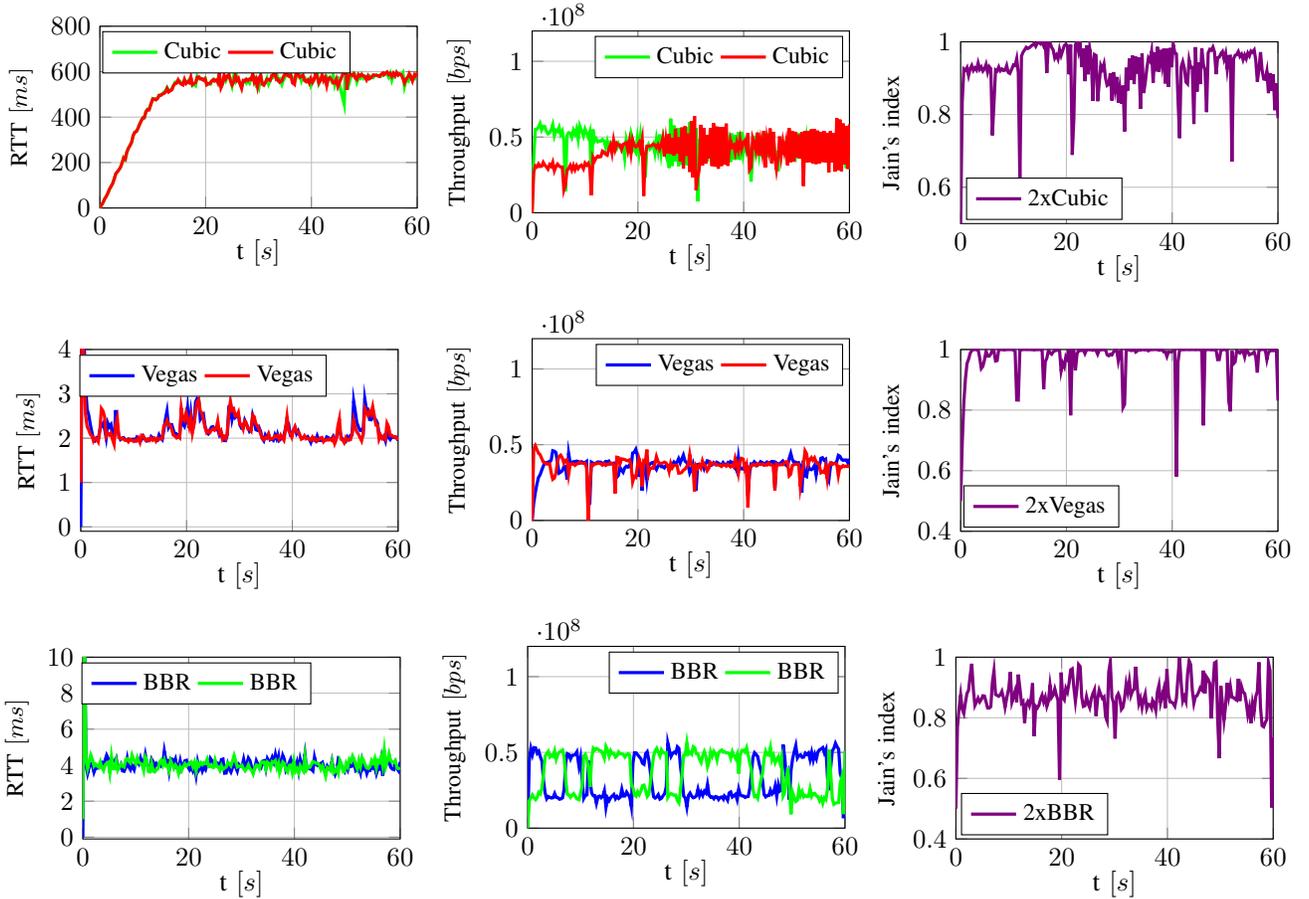

\textbf{Intra-Fairness.} As expected, flows that use delay-based algorithms experience a huge decrease in throughput if they share the bottleneck with loss-based or hybrid algorithms (Fig.~\ref{fig:dual1}, Tab.~\ref{tab:fairness}). This is because they detect congestion earlier, at the point when the queues start to fill. Loss-based algorithms on the other hand continue to increase their sending rate as no loss is detected. This increases the observed RTT (Fig.~\ref{fig:dual2}) of all flows, triggering the delay-based flow to back-off. As a consequence, only a few hundred milliseconds after the start of the connections, delay-based algorithms reduced their sending rate to $1/10-1/15$ of the sending rate of the loss-based algorithms. This process continued until almost no resources were available for the delay-based algorithm.

\begin{table*}[!htb]
	\begin{center}
		\caption{BW scenario with 2 flows: Different metrics for representatives of the three congestion control algorithm groups.}
		\label{tab:dual1}
		\begin{tabular}{l | l | l | c | c | c | c | c | c | c}
			\toprule
			Protocol & Group & Algorithm & Average & Average & Average & Average & Average & Average & Average\\
			& & & goodput & goodput ratio  & cwnd & RTT  & sending rate & throughput & Jain's index\\ 
			&   &           & $\lbrack Mbps \rbrack$ & $\lbrack \% \rbrack$ & $\lbrack \#packets \rbrack$ &  $\lbrack ms \rbrack$ & $\lbrack Mbps \rbrack$ & $\lbrack Mbps \rbrack$ \\            \midrule
			\multirow{12}{0.5cm}{TCP} & \multirow{2}{2.7cm}{Loss-based vs. Hybrid}                                                
			&Cubic   & 71.96 & 95.35 & 1937.96 & 379.90 & 75.47 & 73.22 &  \multirow{2}{1cm}{0.62} \\ 
			& &BBR     &  9.84 & 90.69 & 451.64  & 308.34 & 10.85 & 10.53 &                          \\ 
			\cline{2-10}                                                            
			& \multirow{2}{2.7cm}{Loss- vs. Delay-based}                                           
			&Cubic  &  81.38 & 95.36 & 2005.91 &  228.79 & 85.34 & 82.63 & \multirow{2}{1cm}{0.51}  \\
			& &Vegas  &  0.62  & 89.86 & 11.96   &  279.39 & 0.72  &  0.71 &                          \\
			\cline{2-10}                                                            
			& \multirow{2}{2.7cm}{Delay-based vs. Hybrid}                                               
			&Vegas  & 15.77 & 94.49 & 6.86  & 5.18 & 16.69 & 16.26  & \multirow{2}{1cm}{0.76}   \\    
			& &BBR    & 50.42 & 95.33 & 23.82 & 3.82 & 52.89 & 51.62  &                           \\    
			\cline{2-10}          
			&\multirow{2}{2.7cm}{Loss- vs. Loss-based}                                             
			&Cubic  & 39.74 & 95.62 & 1806.66 & 481.54 & 41.56 & 40.58 & \multirow{2}{1cm}{0.93 }  \\ 
			& &Cubic  & 44.32 & 95.58 & 2149.88 & 497.35 & 46.37 & 45.52 &                           \\ 
			\cline{2-10}                                                                                   
			& \multirow{2}{2.7cm}{Delay- vs. Delay-based}                                           
			&Vegas    & 35.52 & 95.46 & 6.98 & 2.18 & 37.21 & 36.52 &  \multirow{2}{1cm}{0.98 }  \\  
			& &Vegas    & 35.53 & 95.51 & 6.73 & 2.18 & 37.20 & 36.48 &                            \\  
			\cline{2-10}                                                                                   
			& \multirow{2}{2.7cm}{Hybrid vs. Hybrid}                                                     
			&BBR    & 31.62 & 94.81 & 16.03 & 4.05 & 33.35 & 32.64  &  \multirow{2}{1cm}{0.87}  \\   
			& &BBR    & 35.99 & 94.99 & 17.68 & 4.08 & 37.89 & 37.15  &                           \\   
			\hline
		\end{tabular}
	\end{center}
\end{table*}

A similar behaviour is observed when a bottleneck is shared between a hybrid and a delay-based algorithm. The average fairness index is always low (Tab.~\ref{tab:fairness}) and highest in case of BBR. When we increase the number of Vegas or BBR flows at the bottleneck to four, the new flows increase their bandwidth at the expense of the BBR flow, reducing its share from $50Mbps$ to $20Mbps$, and increasing the fairness index to $0.9-0.94$. The reason for this is that BBR tries to operate without filling the queues, allowing the delay-based algorithm to grow and claim more bandwidth. However, due to Vegas' conservative nature, the increase is slow, allowing BBR to always claim more bandwidth than the corresponding Vegas flow.

When the bottleneck is shared between a hybrid (BBR) and a loss-based algorithm (Cubic), the two flows oscillate, confirming results from related work \cite{scholztowards,hock2017experimental}. 
The fairness index at the start of the connection is very low. Either Cubic or BBR takes the whole available bandwidth at the expense of the other flow. After a Cubic flow fills the buffers, BBR measures an increased RTT and adopts, as a consequence, a more aggressive approach (Fig.~\ref{fig:dual2}). As RTT keeps increasing, BBR will increase the rate until the buffers are drained and BBR measures a lower RTT estimate. However, when the number of Cubic flows was increased to three, the throughout of the BBR flow dropped close to zero. 


Even if one loss-based algorithm is present at the bottleneck, the observed RTT is determined by it, nullifying the advantages of delay-based and hybrid algorithms, namely the prevention of the queue buildup. Even though BBR, as well as Vegas, claim to be able to operate with a small RTT, they suffer from a huge increase in average RTT (by more than $200\;ms$) when competing with Cubic (compared to $2-6ms$ without Cubic). However, when a link is shared between a hybrid and a delay-based flow, both of them are able to maintain a low RTT. Moreover, hybrid algorithms, such as BBR, even outperform the delay-based algorithms, by maintaining a lower RTT value.  

\begin{bluebox}
\begin{figure*}[!htb]
	\centering
	\input{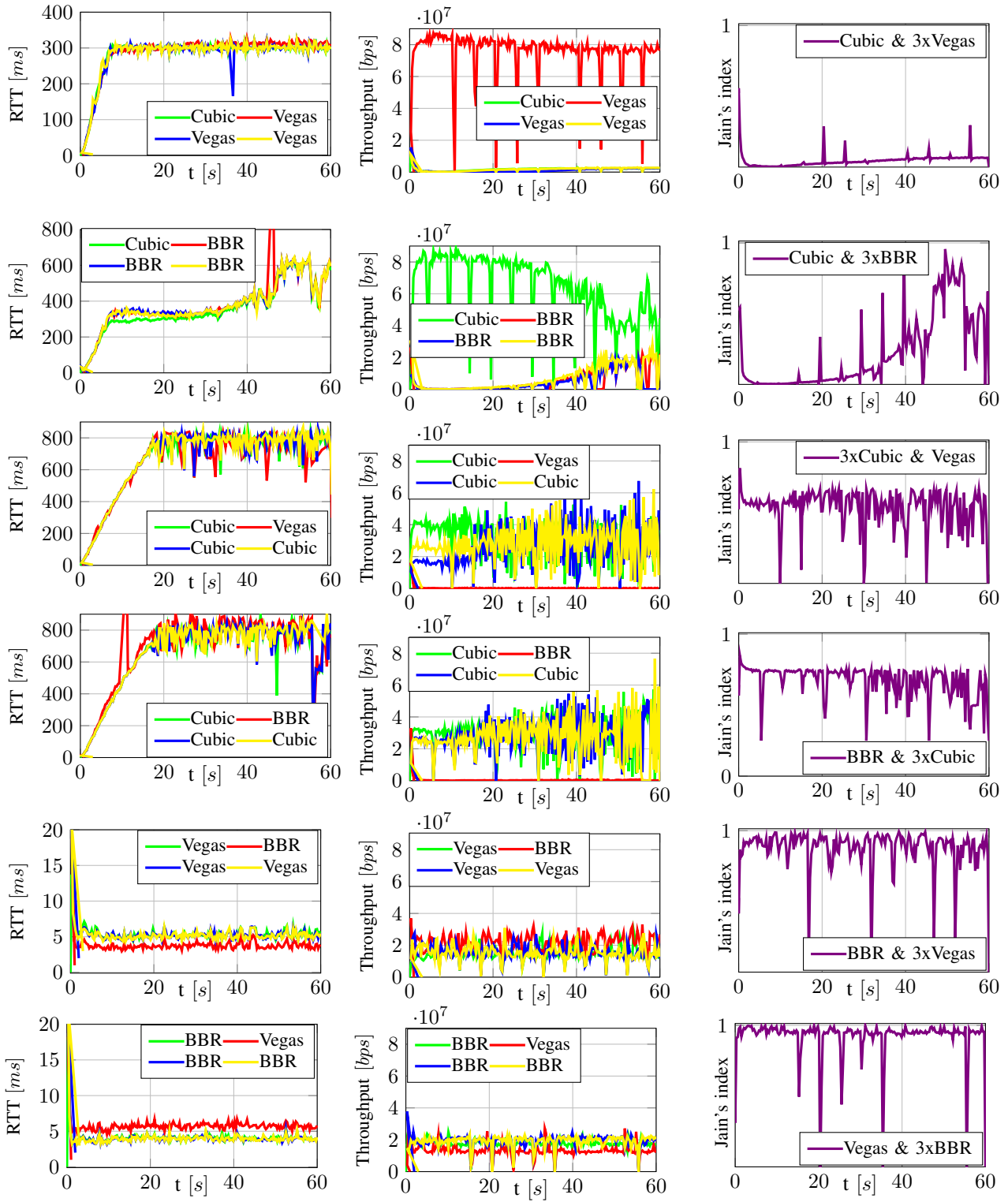}
	\caption{BW scenario: Comparison of average RTT, average throughput, and fairness index for representatives of the congestion control algorithm classes groups in case the link is shared by 4 flows (time unit 300ms).}\label{fig:dual6}
\end{figure*}
\begin{table*}[!htb]
	\begin{center}
		\caption{BW scenario with 4 flows: Different metrics for representatives of the three congestion control algorithm groups.}
		\label{tab:dual6}
		\begin{tabular}{l | l | l | c | c | c | c | c | c | c}
			\toprule
			Protocol & Group & Algorithm & Average & Average & Average & Average & Average & Average & Average\\
			& & & goodput & goodput ratio  &  cwnd & RTT  & sending rate & throughput & Jain's index\\ 
			& &           & $\lbrack Mbps \rbrack$ & $\lbrack \% \rbrack$ & $\lbrack \#packets \rbrack$ &  $\lbrack ms \rbrack$ & $\lbrack Mbps \rbrack$ & $\lbrack Mbps \rbrack$ \\            
			\midrule    
			\multirow{24}{0.5cm}{TCP}   
			&\multirow{4}{2.7cm}{Loss-based vs. Delay-based}             
			 &Cubic  & 72.97 & 94.34 & 2043.07 & 287.69 & 80.70 & 75.30 &  \multirow{4}{1cm}{0.28}  \\   
			& &Vegas  &  1.62 & 91.60 & 39.94 & 282.98 & 1.85 & 1.70 &   \\     
			& &Vegas  &  1.04 & 91.23 & 40.13 & 281.92 & 1.09 & 1.19 & \\     
			& &Vegas  &  1.65 & 89.73 & 23.37 & 283.50 & 1.73 & 1.87 & \\     
			\cline{2-10}                                                                                          
			& \multirow{4}{2.7cm}{Loss-based vs. Hybrid}
			 &Cubic  & 61.14 & 94.37 & 2014.79 & 358.66 & 69.20 & 63.14 &  \multirow{4}{1cm}{0.42} \\		
			& &BBR    &  5.12 & 91.15 & 306.414 & 379.34 & 6.07 & 5.41 &     \\       
			& &BBR    &  4.00 & 91.63 & 368.983 & 368.90 & 4.99 & 4.21 &     \\       
			& &BBR    &  6.46 & 83.63 & 229.312 & 375.20 & 6.80 & 7.686 &    \\       
			
			\cline{2-10}                                                                                            
			& \multirow{4}{2.7cm}{Delay-based vs. Loss-based}                                                             
			 &Vegas    &  0.28 & 81.90 & 7.21 & 576.67 & 0.34 & 0.29 &   \multirow{4}{1cm}{0.65} \\ 
			& &Cubic    & 29.80 & 94.06 & 1840.35 & 665.12 & 33.04 & 30.82 &  \\      
			& &Cubic    & 24.87 & 94.06 & 1671.9 & 668.63 & 27.47 & 26.28 &  \\      
			& &Cubic    & 26.54 & 94.35 & 1750.23 & 669.35 & 29.79 & 28.02 &  \\

			\cline{2-10}                                                                                            
			& \multirow{4}{2.7cm}{Hybrid vs.Loss-based}                                                             
			 &BBR      &  0.32 & 62.19 & 23.74 & 775.87 & 0.75 & 0.34 &   \multirow{4}{1cm}{0.67} \\ 
			& &Cubic    & 30.03 & 93.90 & 1793.47 & 747.01 & 31.57 & 31.04 &  \\      
			& &Cubic    & 29.22 & 94.43 & 1771.46 & 749.91 & 29.77 & 30.83 &  \\      
			& &Cubic    & 24.35 & 94.28 & 1743.42 & 790.97 & 28.15 & 25.74 &  \\      

						\cline{2-10}                                                                                            
			& \multirow{4}{2.7cm}{Hybrid vs. Delay-based}                                                             
			&BBR      & 20.38 & 93.45 & 9.86 & 3.85 & 22.940 & 21.35 &   \multirow{4}{1cm}{0.90} \\ 
			& &Vegas    & 13.34 & 94.25 & 6.25 & 5.39 & 15.08 & 13.99 &  \\      
			& &Vegas    & 14.45 & 94.17 & 6.78 & 5.32 & 16.09 & 15.22 &  \\      
			& &Vegas    & 14.52 & 94.14 & 6.87 & 5.26 & 16.03 & 15.29 &  \\

			\cline{2-10}                                                                                          
			& \multirow{4}{2.7cm}{Delay-based vs. Hybrid}                                                             
			 &Vegas    & 12.70 & 94.21 & 6.34 & 5.75 & 13.83 & 13.30 & \multirow{4}{1cm}{0.94} \\ 
			& &BBR      & 16.29 & 93.23 & 8.26 & 4.12 & 18.16 & 17.21 & \\      
			& &BBR      & 18.052 & 93.29 & 9.09 & 4.04 & 19.91 & 19.01 & \\      
			& &BBR      & 18.10 & 93.26 & 9.22 & 4.04 & 20.00 & 19.07 & \\  			
			\hline                                                                                           
		\end{tabular}
	\end{center}
\end{table*}
\end{bluebox}

\textbf{Inter-Fairness.} Delay-based algorithms have the best inter-fairness properties, see Fig.~\ref{fig:dual2}. Jain's index is always close to 1, even when the number of flows increases (Fig. \ref{fig:dual5}), indicating that all present flows receive an equal share of the resources. 


\begin{bluebox}

\end{bluebox}

\begin{bluebox}
\begin{figure*}[!htb]
	\begin{center}
		\begin{subfigure}{0.32\textwidth}
    \centering
	\begin{tikzpicture}
	\begin{axis}[
            title style={at={(0.5,1.1)},anchor=north},
	        width = 5.8cm, height = 4cm,
		    xlabel={t $\lbrack s \rbrack$},
		    title={4 Cubic flows},
            xlabel style={at={(0.5,-0.37)},anchor=south},
		    ylabel={\small{RTT $\lbrack ms \rbrack$}},
    		legend style={at={(0.99,0.0)},anchor=south east, column sep=-0.08cm, legend cell align=left, row sep=-0.05cm },	
    		legend columns=2,
		    xmin=0, xmax=60, ymin=-0.1, ymax=1500, grid,]
		    \addplot [color=green, mark=none, line width=1.2pt] 
		    table [col sep=semicolon, x index=0, y index=1,] {csv/4cubicrtt.csv};
		    \addplot [color=red, mark=none, line width=1.2pt] 
		    table [col sep=semicolon, x index=2, y index=3,] {csv/4cubicrtt.csv};
		    \addplot [color=blue, mark=none, line width=1.2pt] 
		    table [col sep=semicolon, x index=4, y index=5,] {csv/4cubicrtt.csv};
		    \addplot [color=yellow, mark=none, line width=1.2pt] 
		    table [col sep=semicolon, x index=6, y index=7,] {csv/4cubicrtt.csv};
		\end{axis}
	\end{tikzpicture}
	\label{fig:4cubicrtt}
\end{subfigure}%
\begin{subfigure}{0.32\textwidth}
    \centering
	\begin{tikzpicture}
	\begin{axis}[
            title style={at={(0.5,1.1)},anchor=north},
	        width = 5.8cm, height = 4cm,
		    xlabel={t $\lbrack s \rbrack$},
		    title={4 Cubic flows},
            xlabel style={at={(0.5,-0.37)},anchor=south},
		    ylabel={\small{Throughput $\lbrack bps \rbrack$}},
    		legend style={at={(0.99,0.62)},anchor=south east, column sep=-0.08cm, legend cell align=left, row sep=-0.05cm},	
    		legend columns=2,
		    xmin=0, xmax=60, ymin=-0.1, ymax=60000000, grid,]
		    \addplot [color=green, mark=none, line width=1.2pt] 
		    table [col sep=semicolon, x index=0, y index=1,] {csv/4cubicthroughput.csv};
		    \addplot [color=red,mark=none, line width=1.2pt] 
		    table [col sep=semicolon, x index=2, y index=3,] {csv/4cubicthroughput.csv};
		    \addplot [color=blue, mark=none, line width=1.2pt] 
		    table [col sep=semicolon, x index=4, y index=5,] {csv/4cubicthroughput.csv};
		    \addplot [color=yellow, mark=none, line width=1.2pt] 
		    table [col sep=semicolon, x index=6, y index=7,] {csv/4cubicthroughput.csv};
		\end{axis}
	\end{tikzpicture}
	\label{fig:4cubicth}
\end{subfigure}%
\begin{subfigure}{0.32\textwidth}
    \centering
	\begin{tikzpicture}
	\begin{axis}[
            title style={at={(0.5,1.1)},anchor=north},
	        width = 5.8cm, height = 4cm,
		    xlabel={t $\lbrack s \rbrack$},
		    title={4 Cubic flows},
            xlabel style={at={(0.5,-0.37)},anchor=south},
		    ylabel={\small{Jain's index}},
    		legend style={at={(0.99,0.76)},anchor=south east,column sep=-0.08cm, legend cell align=left, row sep=-0.05cm},	
		    xmin=0, xmax=60, ymin=0.25, ymax=1.01, grid,]
		    \addplot [color=blue!50!red, mark=none, line width=1.2pt] 
		    table [col sep=semicolon, x index=0, y index=1,] {csv/4fairnescubiccubic.csv};
		\end{axis}
	\end{tikzpicture}
	\label{fig:4cubicfairness}
\end{subfigure}

\begin{subfigure}{0.32\textwidth}
    \centering
	\begin{tikzpicture}
	\begin{axis}[
            title style={at={(0.5,1.1)},anchor=north},
	        width = 5.8cm, height = 4cm,
		    xlabel={t $\lbrack s \rbrack$},
		    title={4 Vegas flows},
            xlabel style={at={(0.5,-0.37)},anchor=south},
		    ylabel={\small{RTT $\lbrack ms \rbrack$}},
    		legend style={at={(0.99,0.0)},anchor=south east, column sep=-0.08cm, legend cell align=left, row sep=-0.05cm },	
    		legend columns=2,
		    xmin=0, xmax=60, ymin=-0.1, ymax=20, grid,]
		    \addplot [color=green, mark=none, line width=1.2pt] 
		    table [col sep=semicolon, x index=0, y index=1,] {csv/4vegasrttnew.csv};
		    \addplot [color=red, mark=none, line width=1.2pt] 
		    table [col sep=semicolon, x index=2, y index=3,] {csv/4vegasrttnew.csv};
		    \addplot [color=blue, mark=none, line width=1.2pt] 
		    table [col sep=semicolon, x index=4, y index=5,] {csv/4vegasrttnew.csv};
		    \addplot [color=yellow, mark=none, line width=1.2pt] 
		    table [col sep=semicolon, x index=6, y index=7,] {csv/4vegasrttnew.csv};
		\end{axis}
	\end{tikzpicture}
	\label{fig:4vegasrtt}
\end{subfigure}%
\begin{subfigure}{0.32\textwidth}
    \centering
	\begin{tikzpicture}
	\begin{axis}[
            title style={at={(0.5,1.1)},anchor=north},
	        width = 5.8cm, height = 4cm,
	        title={4 Vegas flows},
		    xlabel={t $\lbrack s \rbrack$},
            xlabel style={at={(0.5,-0.37)},anchor=south},
		    ylabel={\small{Throughput $\lbrack bps \rbrack$}},
    		legend style={at={(0.99,0.25)},anchor=south east, column sep=-0.08cm, legend cell align=left, row sep=-0.05cm},	
    		legend columns=2,
		    xmin=0, xmax=60, ymin=-0.1, ymax=40000000, grid,]
		    \addplot [color=green, mark=none, line width=1.2pt] 
		    table [col sep=semicolon, x index=0, y index=1,] {csv/4vegasthroughputnew.csv};
		    \addplot [color=red, mark=none, line width=1.2pt] 
		    table [col sep=semicolon, x index=2, y index=3,] {csv/4vegasthroughputnew.csv};
		    \addplot [color=blue, mark=none, line width=1.2pt] 
		    table [col sep=semicolon, x index=4, y index=5,] {csv/4vegasthroughputnew.csv};
		    \addplot [color=yellow, mark=none, line width=1.2pt] 
		    table [col sep=semicolon, x index=6, y index=7,] {csv/4vegasthroughputnew.csv};
		\end{axis}
	\end{tikzpicture}
	\label{fig:4vegasth}
\end{subfigure}%
\begin{subfigure}{0.32\textwidth}
    \centering
	\begin{tikzpicture}
	\begin{axis}[
            title style={at={(0.5,1.1)},anchor=north},
	        width = 5.8cm, height = 4cm,
		    xlabel={t $\lbrack s \rbrack$},
		    title={4 Vegas flows},
            xlabel style={at={(0.5,-0.37)},anchor=south},
		    ylabel={\small{Jain's index}},
    		legend style={at={(0.99,0.76)},anchor=south east,column sep=-0.08cm, legend cell align=left, row sep=-0.05cm},	
		    xmin=0, xmax=60, ymin=0.25, ymax=1.01, grid,]
		    \addplot [color=blue!50!red, mark=none, line width=1.2pt] 
		    table [col sep=semicolon, x index=0, y index=1,] {csv/4fairnesvegas.csv};
		\end{axis}
	\end{tikzpicture}
	\label{fig:4vegasfairnes}
\end{subfigure}

\begin{subfigure}{0.32\textwidth}
    \centering
	\begin{tikzpicture}
	\begin{axis}[
		    title={4 BBR flows},
            title style={at={(0.5,1.1)},anchor=north},
	        width = 5.8cm, height = 4cm,
		    xlabel={t $\lbrack s \rbrack$},
            xlabel style={at={(0.5,-0.37)},anchor=south},
		    ylabel={\small{RTT $\lbrack ms \rbrack$}},
    		legend style={at={(0.99,0.225)},anchor=south east, column sep=-0.08cm, legend cell align=left, row sep=-0.05cm},	
    		legend columns=2,
		    xmin=0, xmax=60, ymin=-0.1, ymax=10, grid,]
		    \addplot [color=green, mark=none, line width=1.2pt] 
		    table [col sep=semicolon, x index=0, y index=1,] {csv/4bbrrtt.csv};	
		    \addplot [color=red, mark=none, line width=1.2pt] 
		    table [col sep=semicolon, x index=2, y index=3,] {csv/4bbrrtt.csv};	
		    \addplot [color=blue, mark=none, line width=1.2pt] 
		    table [col sep=semicolon, x index=4, y index=5,] {csv/4bbrrtt.csv};		    
		    \addplot [color=yellow, mark=none, line width=1.2pt] 
		    table [col sep=semicolon, x index=6, y index=7,] {csv/4bbrrtt.csv};		    
		\end{axis}
	\end{tikzpicture}
	\label{fig:4bbrrtt}
\end{subfigure}%
\begin{subfigure}{0.32\textwidth}
    \centering
	\begin{tikzpicture}
	\begin{axis}[
		    title={4 BBR flows},
            title style={at={(0.5,1.1)},anchor=north},
	        width = 5.8cm, height = 4cm,
		    xlabel={t $\lbrack s \rbrack$},
            xlabel style={at={(0.5,-0.37)},anchor=south},
		    ylabel={\small{Throughput $\lbrack bps \rbrack$}},
    		legend style={at={(0.99,0.62)},anchor=south east, column sep=-0.08cm, legend cell align=left, row sep=-0.05cm},	
    		legend columns=2,
		    xmin=0, xmax=60, ymin=-0.1, ymax=50000000, grid,]
		    \addplot [color=green, mark=none, line width=1.2pt] 
		    table [col sep=semicolon, x index=0, y index=1,] {csv/4bbrthroughput.csv};
		    \addplot [color=red, mark=none, line width=1.2pt] 
		    table [col sep=semicolon, x index=2, y index=3,] {csv/4bbrthroughput.csv};
		    \addplot [color=blue, mark=none, line width=1.2pt] 
		    table [col sep=semicolon, x index=4, y index=5,] {csv/4bbrthroughput.csv};	
		    \addplot [color=yellow, mark=none, line width=1.2pt] 
		    table [col sep=semicolon, x index=6, y index=7,] {csv/4bbrthroughput.csv};	  
		\end{axis}
	\end{tikzpicture}
	\label{fig:4bbrbbrth}
\end{subfigure}%
\begin{subfigure}{0.32\textwidth}
    \centering
	\begin{tikzpicture}
	\begin{axis}[
		    title={4 BBR flows},
            title style={at={(0.5,1.1)},anchor=north},
	        width = 5.8cm, height = 4cm,
		    xlabel={t $\lbrack s \rbrack$},
            xlabel style={at={(0.5,-0.37)},anchor=south},
		    ylabel={\small{Jain's index}},
    		legend style={at={(0.01,0.76)},anchor=south west,column sep=-0.08cm, legend cell align=left, row sep=-0.05cm},	
    		legend columns=2,
		    xmin=0, xmax=60, ymin=0.25, ymax=1.01, grid,]
		    \addplot [color=blue!50!red, mark=none, line width=1.2pt] 
		    table [col sep=semicolon, x index=0, y index=1,] {csv/4bbrfairnes.csv};
		\end{axis}
	\end{tikzpicture}
	\label{fig:4bbrfairness}
\end{subfigure}
		\caption{BW scenario: Comparison of average RTT, average throughput, and fairness index for representatives of the congestion control algorithm classes groups in case the link is shared by 4 flows (time unit 600ms).}\label{fig:dual5}
	\end{center}
\end{figure*}
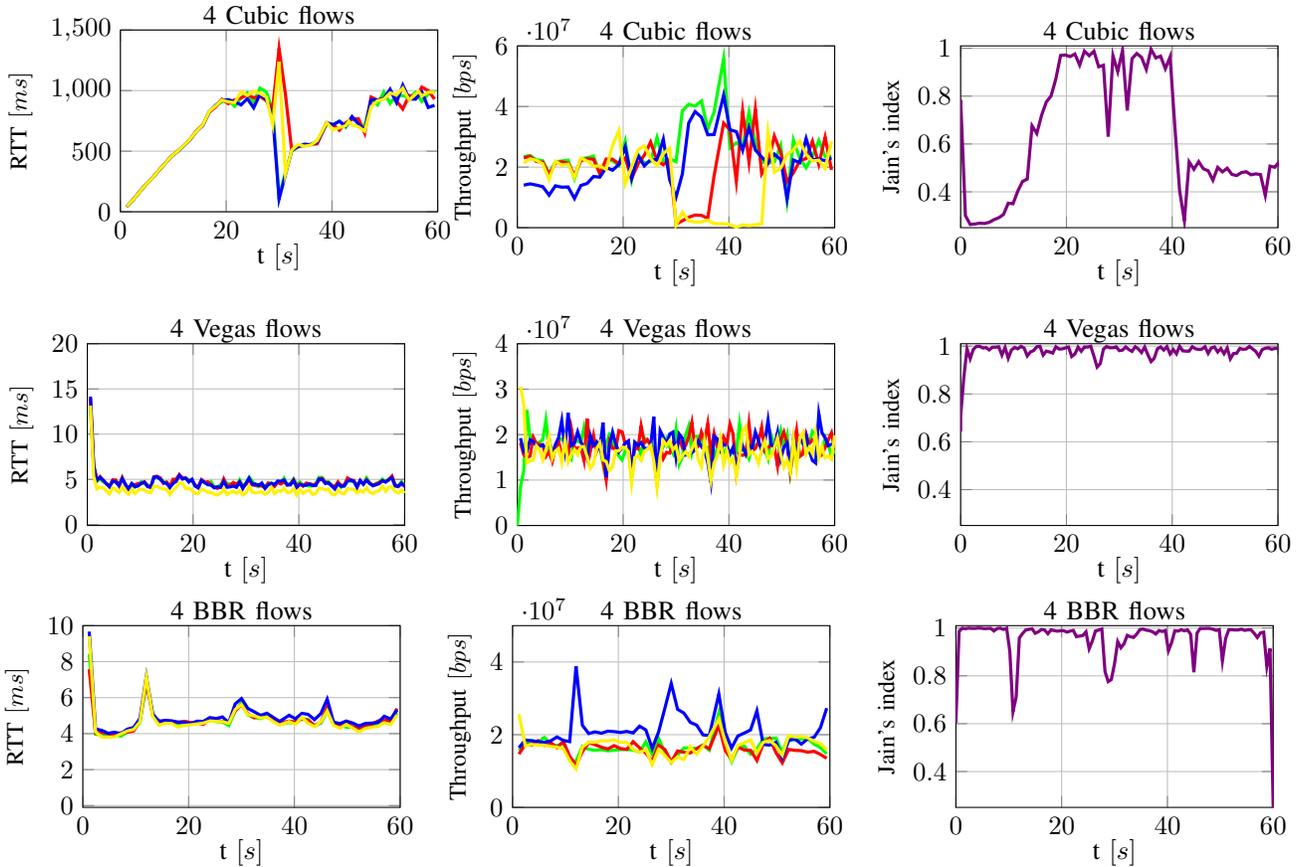
\begin{table*}[!htb]
	\begin{center}
		\caption{BW scenario with 4 flows: Different metrics for representatives of the three congestion control algorithm groups.}
		\label{tab:dual5}
		\begin{tabular}{l | l | l | c | c | c | c | c | c | c}
			\toprule
           Protocol & Group & Algorithm & Average & Average & Average & Average & Average & Average & Average\\
        &    & & goodput & goodput ratio  &  cwnd & RTT  & sending rate & throughput & Jain's index\\ 
          &        &           & $\lbrack Mbps \rbrack$ & $\lbrack \% \rbrack$ & $\lbrack \#packets \rbrack$ &  $\lbrack ms \rbrack$ & $\lbrack Mbps \rbrack$ & $\lbrack Mbps \rbrack$ \\            
						\midrule    
     \multirow{12}{0.5cm}{TCP} &
     \multirow{4}{2.7cm}{Loss- vs. Loss-based}             
    &Cubic  &  24.32 & 93.87 & 1644.7 & 697.807 & 27.29 & 25.16 &   \multirow{4}{1cm}{0.82}  \\   
    & &Cubic  &  19.67 & 93.93 & 1420.02  & 720.881 & 21.62 & 20.35 &   \\     
    & &Cubic  &  20.66 & 93.92 & 1573.35  & 689.727 & 22.49 & 21.84 &   \\     
    & &Cubic  &  15.51 & 93.35 & 1100.81  & 705.501 & 16.78 & 16.37 &   \\     
			\cline{2-10}                                                                                           
	&		\multirow{4}{2.7cm}{Delay- vs. Delay-based}
	&Vegas  & 16.30  & 93.72 & 6.69  & 4.73 & 17.79 & 17.08 &   \multirow{4}{1cm}{0.97} \\		
    & &Vegas  & 16.63 & 93.49 & 6.40 & 4.72 & 17.58 & 17.42 &     \\       
    & &Vegas  & 16.64 & 93.83 & 6.40 & 4.66 & 17.91 & 17.66 &     \\       
    & &Vegas  & 15.24 & 93.72 & 6.68  & 3.85 & 16.03 & 16.04 &     \\       
                                  
			\cline{2-10}                                                                                           
	&		\multirow{4}{2.7cm}{Hybrid vs. Hybrid}                                                             
    &BBR    & 15.47 & 92.73 & 8.51 & 4.72 & 16.82 & 16.31 &   \multirow{4}{1cm}{0.95} \\ 
    & &BBR    & 14.94 & 92.99 & 8.45 & 4.75 & 16.03 & 15.76 &  \\      
    & &BBR    & 19.92 & 92.90 & 10.83 & 4.94 & 21.46 & 20.95 &  \\      
    & &BBR    & 15.86 & 92.60 & 8.67 & 4.70 & 17.08 & 16.69 &  \\

      \hline                                                                                


		\end{tabular}
	\end{center}
\end{table*}
\end{bluebox}
\begin{table*}[!htb]
	\begin{center}
		\caption{BW scenario with 2 flows: Comparison of Jain's index for different congestion control algorithms.}
		\label{tab:fairness}
		\begin{tabular}{ l | l | c  c  c  c  c  c  c | c  c | c  c  c  c }
			
			Group &  & \multicolumn{7}{|c|}{Loss-based}    &   \multicolumn{2}{|c|}{Delay-based} & \multicolumn{4}{|c}{Hybrid} \\
			\hline
			& Algorithm & Reno & BIC & Cubic & HS-TCP & H-TCP & Hybla & Westwood & Vegas & LoLA & Veno & Illinois & YeAH & BBR \\            
			\hline
			\multirow{6}{1cm}{Loss-based}                                                     
			&Reno       & 0.94 & 0.81 & 0.85 & 0.93 & 0.72 & 0.93 & 0.57 & 0.51 & 0.50 & 0.82 & 0.91 & 0.56 & 0.68 \\ 
			&BIC        & 0.81 & 0.96 & 0.86 & 0.79 & 0.76 & 0.73 & 0.58 & 0.51 & 0.50 & 0.79 & 0.75 & 0.55 & 0.63 \\  
			&Cubic      & 0.85 & 0.86 & 0.93 & 0.70 & 0.86 & 0.68 & 0.68 & 0.51 & 0.50 & 0.67 & 0.69 & 0.68 & 0.62 \\  
			&HS-TCP     & 0.93 & 0.79 & 0.70 & 0.95 & 0.95 & 0.80 & 0.69 & 0.52 & 0.50 & 0.80 & 0.85 & 0.55 & 0.62 \\  
			&H-TCP      & 0.72 & 0.76 & 0.86 & 0.95 & 0.96 & 0.78 & 0.59 & 0.52 & 0.50 & 0.82 & 0.79 & 0.55 & 0.68 \\  
			&Hybla      & 0.93 & 0.73 & 0.68 & 0.80 & 0.78 & 0.90 & 0.75 & 0.52 & 0.50 & 0.78 &  0.72 & 0.56 & 0.62 \\  
			&Westwood   & 0.57 & 0.58 & 0.68 & 0.69 & 0.59 & 0.75 & 0.96 & 0.52 & 0.50 & 0.61 & 0.69 & 0.55 & 0.58 \\  
			
			\hline
			\multirow{2}{1cm}{Delay-based}                                        
			&Vegas    & 0.51 & 0.51 & 0.51 & 0.52 & 0.52 & 0.52 & 0.52 & 0.98 & 0.77 & 0.52 & 0.52 & 0.52 & 0.76  \\  
			&LoLA     & 0.50 & 0.50 & 0.50 & 0.50 & 0.50 & 0.50 & 0.50 & 0.77 & 0.86 & 0.50 & 0.50 & 0.51 & 0.63  \\  
			\hline 
			\multirow{5}{1cm}{Hybrid}                                             
			&Veno     & 0.82 & 0.79 & 0.67 & 0.80 & 0.82 & 0.78 & 0.61 & 0.52 & 0.50 & 0.89 & 0.95 & 0.55 & 0.65 \\  
			&Illinois & 0.91 & 0.75 & 0.69 & 0.85 & 0.79 & 0.72 & 0.69 & 0.52 & 0.50  & 0.95 & 0.94 & 0.55 & 0.60 \\  
			&YeAH     & 0.56 & 0.55 & 0.68 & 0.55 & 0.55 & 0.56 & 0.55 & 0.52 & 0.51 & 0.55 & 0.55 & 0.95 & 0.54 \\  
			&BBR      & 0.68 & 0.63 & 0.62 & 0.62 & 0.68 & 0.62 & 0.58 & 0.76 & 0.63 & 0.65 & 0.60 & 0.54 & 0.87 \\  
			
		\end{tabular}
	\end{center}
\end{table*}

Similarly, hybrid-based algorithms have in general good inter-fairness properties, with both of them claiming a similar amount of the available resources on average (Tab.~\ref{tab:fairness}). Similarly, they have no significant increase in average RTT when compared to the Base-Line scenario, and a lower value when compared to the loss-based algorithms. Thus, even when multiple flows (using the same algorithm or a different hybrid algorithm) are present they are able to operate without completely filling the queues. In addition, Fig.~\ref{fig:dual2} shows that two BBR flows never converge to the same bandwidth, but oscillate between $\approx 50\;Mbps$ and $\approx 20\;Mbps$, hence they are not particularly stable. 

Loss-based algorithms have good inter-fairness properties, with an average fairness index between $0.82-0.96$ (Tab.~\ref{tab:dual5}, ~\ref{tab:fairness}). Fig.~\ref{fig:dual2} shows that, when two Cubic flows compete, they do converge to the same bandwidth after $ \approx 15\;s$ (almost 10 seconds longer than Vegas flows). However, while they claim the same amount of bandwidth on average, their throughput oscillates the most from all the evaluated approaches. The queues are constantly filled, and packets needed to be dropped (Fig.~\ref{fig:dual2}). When the number of Cubic flows increases to 4, bandwidth oscillations increase as well, and fairness decreases to $0.82$. 

\textbf{Summary.} In terms of intra-fairness, the only combination that works well together is delay algorithms and BBR. In such a scenario, delay is low and the throughput fairly shared, the more flows the fairer the distribution of resources. We observed that the most popular TCP flavour, Cubic, is prone to oscillation. Further, the convergence time of Cubic flows is high ($\approx 20s$), and when a loss is detected, the flows need to synchronize all over again (again $20s$). We also observed that BBR is not stable, which was not reported in the literature.

\subsection{Results: RTT scenario}\label{Sec:dual3}

\begin{figure*}[!htb]
\begin{center}
\begin{subfigure}{0.32\textwidth}
    \centering
	\begin{tikzpicture}
	\begin{axis}[
            title style={at={(0.5,1.1)},anchor=north},
	        width = 5.8cm, height = 4cm,
		    xlabel={t $\lbrack s \rbrack$},
            xlabel style={at={(0.5,-0.37)},anchor=south},
		    ylabel={\small{RTT $\lbrack ms \rbrack$}},
    		legend style={at={(0.99,0.0)},anchor=south east, column sep=-0.08cm, legend cell align=left, row sep=-0.05cm },	
		    xmin=0, xmax=60, ymin=-0.1, ymax=800, grid,]
		    \addplot [color=green, mark=none, line width=1.2pt] 
		    table [col sep=semicolon, x index=0, y index=1,] {csv/RTTcubiccubicrtt1sec.csv};
		    \addplot [color=red, mark=none, line width=1.2pt] 
		    table [col sep=semicolon, x index=2, y index=3,] {csv/RTTcubiccubicrtt1sec.csv};
		    \legend{\small Cubic(0ms), \small Cubic(200ms)}
		\end{axis}
	\end{tikzpicture}
	\label{fig:rttbbcubic3}
\end{subfigure}%
\begin{subfigure}{0.32\textwidth}
    \centering
	\begin{tikzpicture}
	\begin{axis}[
            title style={at={(0.5,1.1)},anchor=north},
	        width = 5.8cm, height = 4cm,
		    xlabel={t $\lbrack s \rbrack$},
            xlabel style={at={(0.5,-0.37)},anchor=south},
		    ylabel={\small{Throughput $\lbrack bps \rbrack$}},
    		legend style={at={(0.99,0.62)},anchor=south east, column sep=-0.08cm, legend cell align=left, row sep=-0.05cm},	
		    xmin=0, xmax=60, ymin=-0.1, ymax=120000000, grid,]
		    \addplot [color=green, mark=none, line width=1.2pt] 
		    table [col sep=semicolon, dotted, x index=0, y index=1,] {csv/RTTcubiccubicthroughput1sec.csv};
		    \addplot [color=red,mark=none, line width=1.2pt] 
		    table [col sep=semicolon, x index=2, y index=3,] {csv/RTTcubiccubicthroughput1sec.csv};
		    \legend{\small Cubic(0ms), \small Cubic(200ms)}
		\end{axis}
	\end{tikzpicture}
	\label{fig:throughputbbcubic3}
\end{subfigure}%
\begin{subfigure}{0.32\textwidth}
    \centering
	\begin{tikzpicture}
	\begin{axis}[
            title style={at={(0.5,1.1)},anchor=north},
	        width = 5.8cm, height = 4cm,
		    xlabel={t $\lbrack s \rbrack$},
            xlabel style={at={(0.5,-0.37)},anchor=south},
		    ylabel={\small{Jain's index}},
    		legend style={at={(0.01,0.76)},anchor=south west,column sep=-0.08cm, legend cell align=left, row sep=-0.05cm},	
		    xmin=0, xmax=60, ymin=0.5, ymax=1, grid,]
		    \addplot [color=blue!50!red, mark=none, line width=1.2pt] 
		    table [col sep=semicolon, x index=0, y index=1,] {csv/RTTfairnescubiccubic1sec.csv};
		    \legend{\small 2xCubic}
		\end{axis}
	\end{tikzpicture}
	\label{fig:fairnesbbrcubic3}
\end{subfigure}

\begin{subfigure}{0.32\textwidth}
    \centering
	\begin{tikzpicture}
	\begin{axis}[
            title style={at={(0.5,1.1)},anchor=north},
	        width = 5.8cm, height = 4cm,
		    xlabel={t $\lbrack s \rbrack$},
            xlabel style={at={(0.5,-0.37)},anchor=south},
		    ylabel={\small{RTT $\lbrack ms \rbrack$}},
    		legend style={at={(0.99,0.225)},anchor=south east, column sep=-0.08cm, legend cell align=left, row sep=-0.05cm},	
		    xmin=0, xmax=60, ymin=-0.1, ymax=260, grid,]
		    \addplot [color=blue, mark=none, line width=1.2pt] 
		    table [col sep=semicolon, x index=0, y index=1,] {csv/RTTvegasvegasrtt.csv};
		    \addplot [color=red, mark=none, line width=1.2pt] 
		    table [col sep=semicolon, x index=2, y index=3,] {csv/RTTvegasvegasrtt.csv};
		    \legend{\small Vegas(0ms), \small Vegas(200ms)}
		\end{axis}
	\end{tikzpicture}
	\label{fig:rttvegascubic3}
\end{subfigure}%
\begin{subfigure}{0.32\textwidth}
    \centering
	\begin{tikzpicture}
	\begin{axis}[
            title style={at={(0.5,1.1)},anchor=north},
	        width = 5.8cm, height = 4cm,
		    xlabel={t $\lbrack s \rbrack$},
            xlabel style={at={(0.5,-0.37)},anchor=south},
		    ylabel={\small{Throughput $\lbrack bps \rbrack$}},
    		legend style={at={(0.99,0.62)},anchor=south east, column sep=-0.08cm, legend cell align=left, row sep=-0.05cm},	
		    xmin=0, xmax=60, ymin=-0.1, ymax=120000000, grid,]
		    \addplot [color=blue, mark=none, line width=1.2pt] 
		    table [col sep=semicolon, x index=0, y index=1,] {csv/RTTvegasvegasthroughput.csv};
		    \addplot [color=red, mark=none, line width=1.2pt] 
		    table [col sep=semicolon, x index=2, y index=3,] {csv/RTTvegasvegasthroughput.csv};
		    \legend{\small Vegas(0ms), \small Vegas(200ms)}
		\end{axis}
	\end{tikzpicture}
	\label{fig:throughputvegascubic3}
\end{subfigure}%
\begin{subfigure}{0.32\textwidth}
    \centering
	\begin{tikzpicture}
	\begin{axis}[
            title style={at={(0.5,1.1)},anchor=north},
	        width = 5.8cm, height = 4cm,
		    xlabel={t $\lbrack s \rbrack$},
            xlabel style={at={(0.5,-0.37)},anchor=south},
		    ylabel={\small{Jain's index}},
    		legend style={at={(0.99,0.02)},anchor=south east,column sep=-0.08cm, legend cell align=left, row sep=-0.05cm},	
		    xmin=0, xmax=60, ymin=0.5, ymax=1, grid,]
		    \addplot [color=blue!50!red, mark=none, line width=1.2pt] 
		    table [col sep=semicolon, x index=0, y index=1,] {csv/RTTfairnevegasvegas.csv};
		    \legend{\small 2xVegas}
		\end{axis}
	\end{tikzpicture}
	\label{fig:fairnessvegascubic3}
\end{subfigure}

\begin{subfigure}{0.32\textwidth}
    \centering
	\begin{tikzpicture}
	\begin{axis}[
            title style={at={(0.5,1.1)},anchor=north},
	        width = 5.8cm, height = 4cm,
		    xlabel={t $\lbrack s \rbrack$},
            xlabel style={at={(0.5,-0.37)},anchor=south},
		    ylabel={\small{RTT $\lbrack ms \rbrack$}},
    		legend style={at={(0.99,0.225)},anchor=south east, column sep=-0.08cm, legend cell align=left, row sep=-0.05cm},	
		    xmin=0, xmax=60, ymin=-0.1, ymax=300, grid,]
		    \addplot [color=blue, mark=none, line width=1.2pt] 
		    table [col sep=semicolon, x index=0, y index=1,] {csv/RTTbbrbbrrtt1sec.csv};	
		    \addplot [color=green, mark=none, line width=1.2pt] 
		    table [col sep=semicolon, x index=2, y index=3,] {csv/RTTbbrbbrrtt1sec.csv};		    
		    \legend{\small BBR(0ms), \small BBR(200ms)}
		\end{axis}
	\end{tikzpicture}
	\label{fig:rttvegasbbr3}
\end{subfigure}%
\begin{subfigure}{0.32\textwidth}
    \centering
	\begin{tikzpicture}
	\begin{axis}[
            title style={at={(0.5,1.1)},anchor=north},
	        width = 5.8cm, height = 4cm,
		    xlabel={t $\lbrack s \rbrack$},
            xlabel style={at={(0.5,-0.37)},anchor=south},
		    ylabel={\small{Throughput $\lbrack bps \rbrack$}},
    		legend style={at={(0.99,0.62)},anchor=south east, column sep=-0.08cm, legend cell align=left, row sep=-0.05cm},	
		    xmin=0, xmax=60, ymin=-0.1, ymax=120000000, grid,]
		    \addplot [color=blue, mark=none, line width=1.2pt] 
		    table [col sep=semicolon, x index=0, y index=1,] {csv/RTTbbrbbrthroughput1sec.csv};
		    \addplot [color=green, mark=none, line width=1.2pt] 
		    table [col sep=semicolon, x index=2, y index=3,] {csv/RTTbbrbbrthroughput1sec.csv};
		    \legend{\small BBR(0ms), \small BBR(200ms)}
		\end{axis}
	\end{tikzpicture}
	\label{fig:throughputvegasbbr3}
\end{subfigure}%
\begin{subfigure}{0.32\textwidth}
    \centering
	\begin{tikzpicture}
	\begin{axis}[
            title style={at={(0.5,1.1)},anchor=north},
	        width = 5.8cm, height = 4cm,
		    xlabel={t $\lbrack s \rbrack$},
            xlabel style={at={(0.5,-0.37)},anchor=south},
		    ylabel={\small{Jain's index}},
    		legend style={at={(0.99,0.02)},anchor=south east,column sep=-0.08cm, legend cell align=left, row sep=-0.05cm},	
		    xmin=0, xmax=60, ymin=0.5, ymax=1, grid,]
		    \addplot [color=blue!50!red, mark=none, line width=1.2pt] 
		    table [col sep=semicolon, x index=0, y index=1,] {csv/RTTfairnesbbrbbr1sec.csv};
		    \legend{\small 2xBBR}
		\end{axis}
	\end{tikzpicture}
	\label{fig:fairnessvegasbbr3}
\end{subfigure}
\caption{RTT scenario: Comparison of average RTT, average throughput, and fairness index for representatives of the congestion control algorithm groups in the case the link is shared by 2 flows using the same algorithm but different RTTs (time unit 300ms).}\label{fig:dual3}
\end{center}
\end{figure*}
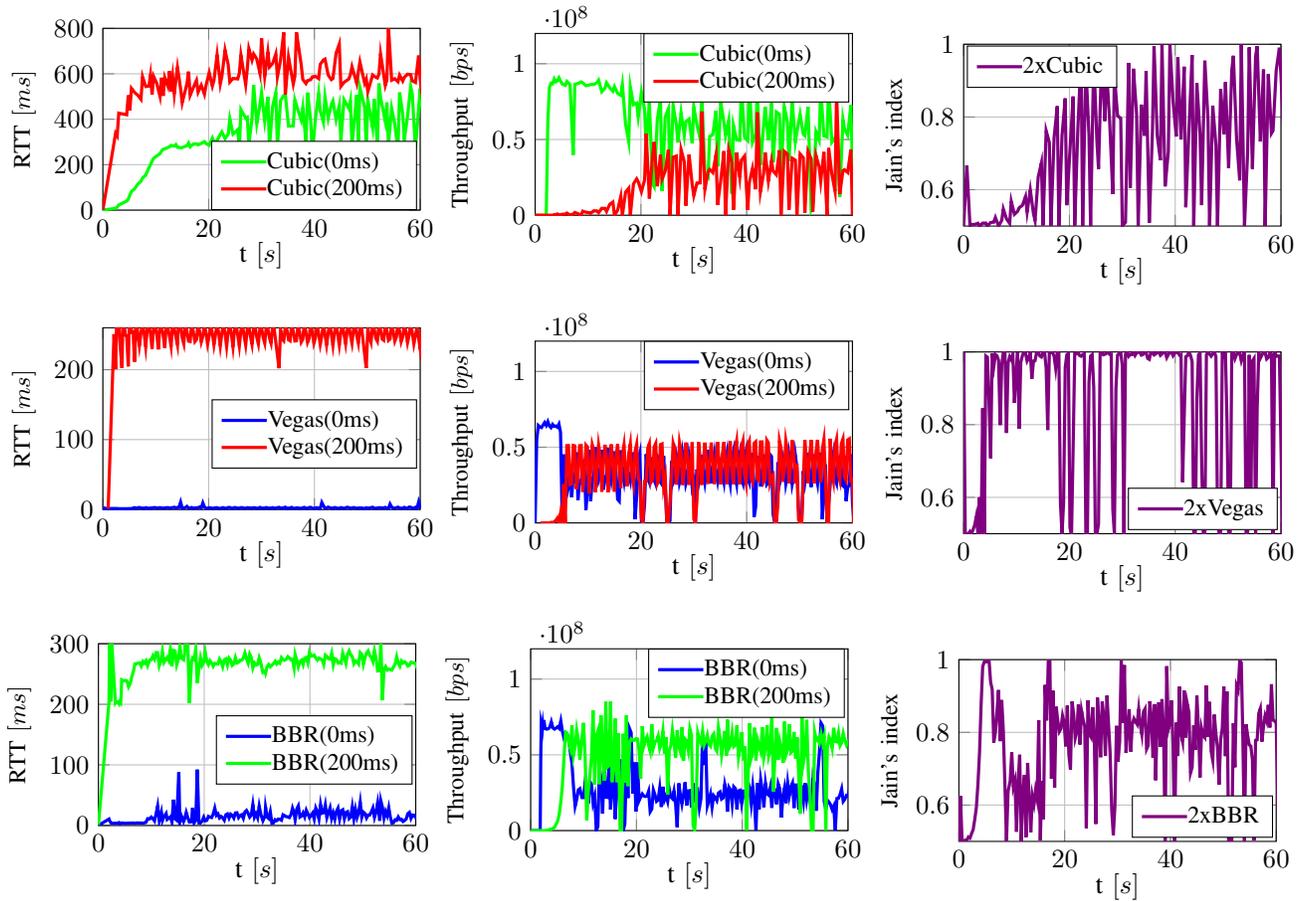
\begin{table*}[!htb]
	\begin{center}
		\caption{RTT scenario with 2 flows: Different metrics for representatives of the congestion control algorithm groups in case the link is shared by two flows using the same algorithm and having different RTTs.}
		\label{tab:dual3}
		\begin{tabular}{l| l | l | c | c | c | c | c | c | c}
			\toprule
           Protocol & Group & Algorithm & Average & Average & Average & Average & Average & Average & Average\\
            & & & goodput & goodput ratio  &  cwnd & RTT  & sending rate & throughput & Jain's index\\ 
               &   &           & $\lbrack Mbps \rbrack$ & $\lbrack \% \rbrack$ & $\lbrack \#packets \rbrack$ &  $\lbrack ms \rbrack$ & $\lbrack Mbps \rbrack$ & $\lbrack Mbps \rbrack$ \\            
						\midrule  
	  \multirow{6}{0.5cm}{TCP} & 
      \multirow{2}{2.7cm}{Loss- vs. Loss-based}             
      &Cubic(0ms)  &   64.01 &  95.01 & 2005.51 & 368.25 & 67.37 & 64.20 &   \multirow{2}{1cm}{0.72}  \\   
     & &Cubic(200ms)  &   23.68 &  85.12 & 1667.61 & 570.07 & 27.82 & 25.05 &   \\     
                               
			\cline{2-10}                                                                                          
	 &\multirow{2}{2.7cm}{Delay- vs. Delay-based}
	  &Vegas(0ms)  &   34.54 & 95.15 &    7.35 &  2.37 & 36.30 & 34.62 &   \multirow{2}{1cm}{0.87} \\		
     & &Vegas(200ms)  &   36.68 & 94.71 & 1236.40 &249.33 & 38.73 & 36.72 &    \\       
                                   
      \cline{2-10}                                                                                           
	 &\multirow{2}{2.7cm}{Hybrid vs. Hybrid}                                                             
      &BBR(0ms)   &  28.87 & 94.84 &  29.17   & 15.91  &     30.44 & 30.21  &   \multirow{2}{1cm}{0.76} \\ 
     & &BBR(200ms)    &  50.50 & 94.82 &  4019.57 & 268.30 &     53.26 & 50.98  & \\      
      \hline                                                                                           

		\end{tabular}
	\end{center}
\end{table*}

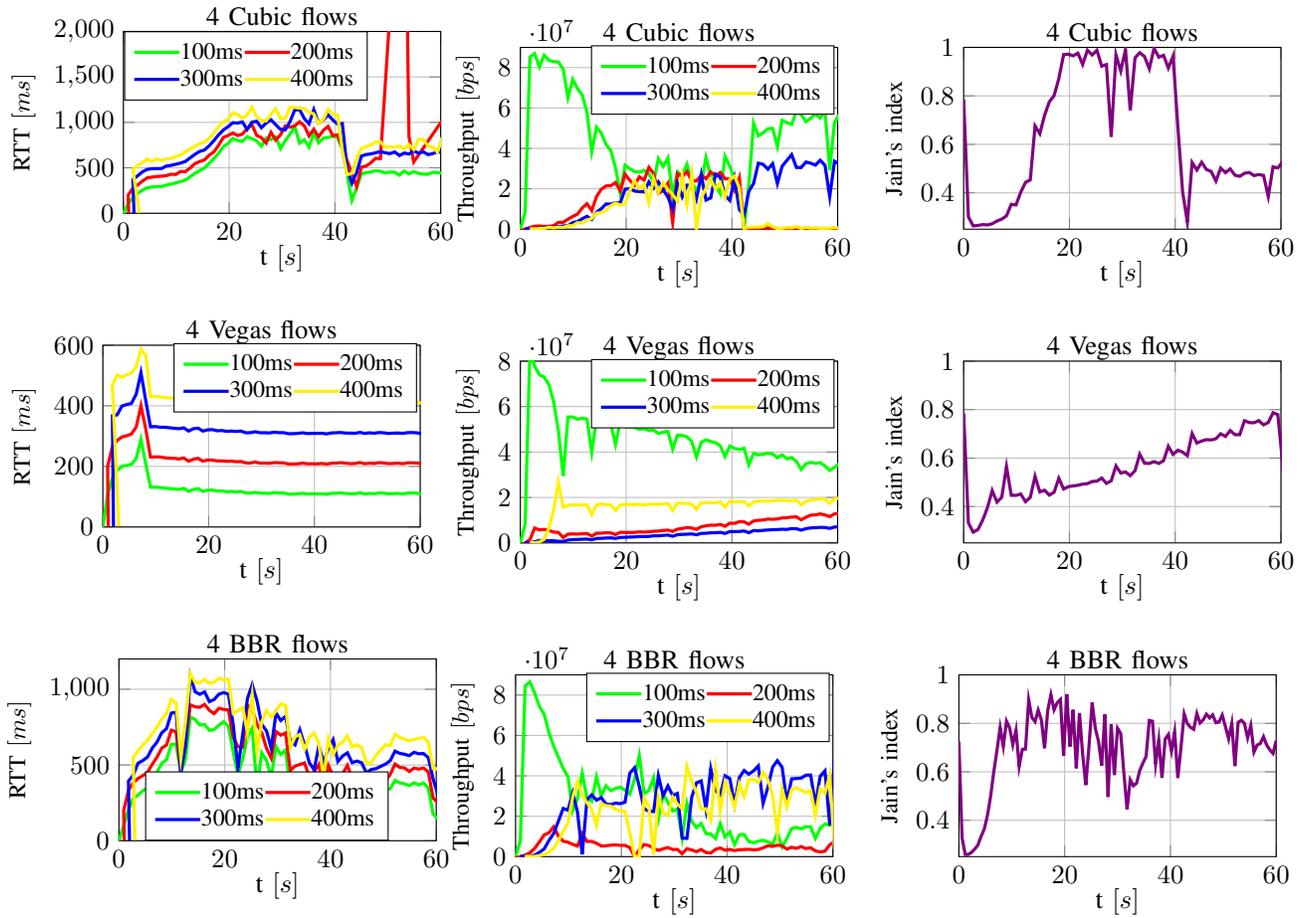
\begin{figure*}[!htb]
	\begin{center}
		\begin{subfigure}{0.32\textwidth}
    \centering
	\begin{tikzpicture}
	\begin{axis}[
            title style={at={(0.5,1.1)},anchor=north},
	        width = 5.8cm, height = 4cm,
		    xlabel={t $\lbrack s \rbrack$},
		    title={4 Cubic flows},
            xlabel style={at={(0.5,-0.37)},anchor=south},
		    ylabel={\small{RTT $\lbrack ms \rbrack$}},
    		legend style={at={(0.77,0.62)},anchor=south east, column sep=-0.08cm, legend cell align=left, row sep=-0.05cm },	
    		legend columns=2,
		    xmin=0, xmax=60, ymin=-0.1, ymax=2000, grid,]
		    \addplot [color=green, mark=none, line width=1.2pt] 
		    table [col sep=semicolon, x index=0, y index=1,] {csv/4cubiccubicrtt.csv};
		    \addplot [color=red, mark=none, line width=1.2pt] 
		    table [col sep=semicolon, x index=2, y index=3,] {csv/4cubiccubicrtt.csv};
		    \addplot [color=blue, mark=none, line width=1.2pt] 
		    table [col sep=semicolon, x index=4, y index=5,] {csv/4cubiccubicrtt.csv};
		    \addplot [color=yellow, mark=none, line width=1.2pt] 
		    table [col sep=semicolon, x index=6, y index=7,] {csv/4cubiccubicrtt.csv};
		    \legend{\small 100ms, \small 200ms, \small 300ms, \small 400ms}
		\end{axis}
	\end{tikzpicture}
	\label{fig:4cubiccubicrttrtt}
\end{subfigure}%
\begin{subfigure}{0.32\textwidth}
    \centering
	\begin{tikzpicture}
	\begin{axis}[
            title style={at={(0.5,1.1)},anchor=north},
	        width = 5.8cm, height = 4cm,
		    xlabel={t $\lbrack s \rbrack$},
		    title={4 Cubic flows},
            xlabel style={at={(0.5,-0.37)},anchor=south},
		    ylabel={\small{Throughput $\lbrack bps \rbrack$}},
    		legend style={at={(0.99,0.63)},anchor=south east, column sep=-0.08cm, legend cell align=left, row sep=-0.05cm},	
    		legend columns=2,
		    xmin=0, xmax=60, ymin=-0.1, ymax=90000000, grid,]
		    \addplot [color=green, mark=none, line width=1.2pt] 
		    table [col sep=semicolon, x index=0, y index=1,] {csv/4cubiccubicthroughput.csv};
		    \addplot [color=red,mark=none, line width=1.2pt] 
		    table [col sep=semicolon, x index=2, y index=3,] {csv/4cubiccubicthroughput.csv};
		    \addplot [color=blue, mark=none, line width=1.2pt] 
		    table [col sep=semicolon, x index=4, y index=5,] {csv/4cubiccubicthroughput.csv};
		    \addplot [color=yellow, mark=none, line width=1.2pt] 
		    table [col sep=semicolon, x index=6, y index=7,] {csv/4cubiccubicthroughput.csv};
		    \legend{\small 100ms, \small 200ms, \small 300ms, \small 400ms}
		\end{axis}
	\end{tikzpicture}
	\label{fig:4cubicthrtt}
\end{subfigure}%
\begin{subfigure}{0.32\textwidth}
    \centering
	\begin{tikzpicture}
	\begin{axis}[
            title style={at={(0.5,1.1)},anchor=north},
	        width = 5.8cm, height = 4cm,
		    xlabel={t $\lbrack s \rbrack$},
		    title={4 Cubic flows},
            xlabel style={at={(0.5,-0.37)},anchor=south},
		    ylabel={\small{Jain's index}},
    		legend style={at={(0.99,0.76)},anchor=south east,column sep=-0.08cm, legend cell align=left, row sep=-0.05cm},	
		    xmin=0, xmax=60, ymin=0.25, ymax=1, grid,]
		    \addplot [color=blue!50!red, mark=none, line width=1.2pt] 
		    table [col sep=semicolon, x index=0, y index=1,] {csv/4fairnescubiccubic.csv};
		\end{axis}
	\end{tikzpicture}
	\label{fig:4fairnescubicrtt}
\end{subfigure}

\begin{subfigure}{0.32\textwidth}
    \centering
	\begin{tikzpicture}
	\begin{axis}[
            title style={at={(0.5,1.1)},anchor=north},
	        width = 5.8cm, height = 4cm,
		    xlabel={t $\lbrack s \rbrack$},
		    title={4 Vegas flows},
            xlabel style={at={(0.5,-0.37)},anchor=south},
		    ylabel={\small{RTT $\lbrack ms \rbrack$}},
    		legend style={at={(0.99,0.63)},anchor=south east, column sep=-0.08cm, legend cell align=left, row sep=-0.05cm },	
    		legend columns=2,
		    xmin=0, xmax=60, ymin=-0.1, ymax=600, grid,]
		    \addplot [color=green, mark=none, line width=1.2pt] 
		    table [col sep=semicolon, x index=0, y index=1,] {csv/4vegasvegasrtt.csv};
		    \addplot [color=red, mark=none, line width=1.2pt] 
		    table [col sep=semicolon, x index=2, y index=3,] {csv/4vegasvegasrtt.csv};
		    \addplot [color=blue, mark=none, line width=1.2pt] 
		    table [col sep=semicolon, x index=4, y index=5,] {csv/4vegasvegasrtt.csv};
		    \addplot [color=yellow, mark=none, line width=1.2pt] 
		    table [col sep=semicolon, x index=6, y index=7,] {csv/4vegasvegasrtt.csv};
		    \legend{\small 100ms, \small 200ms, \small 300ms, \small 400ms}
		\end{axis}
	\end{tikzpicture}
	\label{fig:4vegasrttrtt}
\end{subfigure}%
\begin{subfigure}{0.32\textwidth}
    \centering
	\begin{tikzpicture}
	\begin{axis}[
            title style={at={(0.5,1.1)},anchor=north},
	        width = 5.8cm, height = 4cm,
	        title={4 Vegas flows},
		    xlabel={t $\lbrack s \rbrack$},
            xlabel style={at={(0.5,-0.37)},anchor=south},
		    ylabel={\small{Throughput $\lbrack bps \rbrack$}},
    		legend style={at={(0.99,0.63)},anchor=south east, column sep=-0.08cm, legend cell align=left, row sep=-0.05cm},	
    		legend columns=2,
		    xmin=0, xmax=60, ymin=-0.1, ymax=80000000, grid,]
		    \addplot [color=green, mark=none, line width=1.2pt] 
		    table [col sep=semicolon, x index=0, y index=1,] {csv/4vegasvegasthroughput.csv};
		    \addplot [color=red, mark=none, line width=1.2pt] 
		    table [col sep=semicolon, x index=2, y index=3,] {csv/4vegasvegasthroughput.csv};
		    \addplot [color=blue, mark=none, line width=1.2pt] 
		    table [col sep=semicolon, x index=4, y index=5,] {csv/4vegasvegasthroughput.csv};
		    \addplot [color=yellow, mark=none, line width=1.2pt] 
		    table [col sep=semicolon, x index=6, y index=7,] {csv/4vegasvegasthroughput.csv};
		    \legend{\small 100ms, \small 200ms, \small 300ms, \small 400ms}
		\end{axis}
	\end{tikzpicture}
	\label{fig:4vegasthrtt}
\end{subfigure}%
\begin{subfigure}{0.32\textwidth}
    \centering
	\begin{tikzpicture}
	\begin{axis}[
            title style={at={(0.5,1.1)},anchor=north},
	        width = 5.8cm, height = 4cm,
		    xlabel={t $\lbrack s \rbrack$},
		    title={4 Vegas flows},
            xlabel style={at={(0.5,-0.37)},anchor=south},
		    ylabel={\small{Jain's index}},
    		legend style={at={(0.99,0.76)},anchor=south east,column sep=-0.08cm, legend cell align=left, row sep=-0.05cm},	
		    xmin=0, xmax=60, ymin=0.25, ymax=1, grid,]
		    \addplot [color=blue!50!red, mark=none, line width=1.2pt] 
		    table [col sep=semicolon, x index=0, y index=1,] {csv/4fairnesvegasvegas.csv};
		\end{axis}
	\end{tikzpicture}
	\label{fig:4vegasfairnesrtt}
\end{subfigure}

\begin{subfigure}{0.32\textwidth}
    \centering
	\begin{tikzpicture}
	\begin{axis}[
		    title={4 BBR flows},
            title style={at={(0.5,1.1)},anchor=north},
	        width = 5.8cm, height = 4cm,
		    xlabel={t $\lbrack s \rbrack$},
            xlabel style={at={(0.5,-0.37)},anchor=south},
		    ylabel={\small{RTT $\lbrack ms \rbrack$}},
    		legend style={at={(0.85,0.0)},anchor=south east, column sep=-0.08cm, legend cell align=left, row sep=-0.05cm },	
    		legend columns=2,
		    xmin=0, xmax=60, ymin=-0.1, ymax=1200, grid,]
		    \addplot [color=green, mark=none, line width=1.2pt] 
		    table [col sep=semicolon, x index=0, y index=1,] {csv/4bbrbbrrtt.csv};
		    \addplot [color=red, mark=none, line width=1.2pt] 
		    table [col sep=semicolon, x index=2, y index=3,] {csv/4bbrbbrrtt.csv};	
		    \addplot [color=blue, mark=none, line width=1.2pt] 
		    table [col sep=semicolon, x index=4, y index=5,] {csv/4bbrbbrrtt.csv};		    
		    \addplot [color=yellow, mark=none, line width=1.2pt] 
		    table [col sep=semicolon, x index=6, y index=7,] {csv/4bbrbbrrtt.csv};		    
		    \legend{\small 100ms, \small 200ms, \small 300ms, \small 400ms}
		\end{axis}
	\end{tikzpicture}
	\label{fig:4bbrrttrtt}
\end{subfigure}%
\begin{subfigure}{0.32\textwidth}
    \centering
	\begin{tikzpicture}
	\begin{axis}[
		    title={4 BBR flows},
            title style={at={(0.5,1.1)},anchor=north},
	        width = 5.8cm, height = 4cm,
		    xlabel={t $\lbrack s \rbrack$},
            xlabel style={at={(0.5,-0.37)},anchor=south},
		    ylabel={\small{Throughput $\lbrack bps \rbrack$}},
    		legend style={at={(0.99,0.63)},anchor=south east, column sep=-0.08cm, legend cell align=left, row sep=-0.05cm},	
    		legend columns=2,
		    xmin=0, xmax=60, ymin=-0.1, ymax=90000000, grid,]
		    \addplot [color=green, mark=none, line width=1.2pt] 
		    table [col sep=semicolon, x index=0, y index=1,] {csv/4bbrbbrthroughput.csv};
		    \addplot [color=red, mark=none, line width=1.2pt] 
		    table [col sep=semicolon, x index=2, y index=3,] {csv/4bbrbbrthroughput.csv};
		    \addplot [color=blue, mark=none, line width=1.2pt] 
		    table [col sep=semicolon, x index=4, y index=5,] {csv/4bbrbbrthroughput.csv};	
		    \addplot [color=yellow, mark=none, line width=1.2pt] 
		    table [col sep=semicolon, x index=6, y index=7,] {csv/4bbrbbrthroughput.csv};	  
		    \legend{\small 100ms, \small 200ms, \small 300ms, \small 400ms}
		\end{axis}
	\end{tikzpicture}
	\label{fig:4bbrthrtt}
\end{subfigure}%
\begin{subfigure}{0.32\textwidth}
    \centering
	\begin{tikzpicture}
	\begin{axis}[
		    title={4 BBR flows},
            title style={at={(0.5,1.1)},anchor=north},
	        width = 5.8cm, height = 4cm,
		    xlabel={t $\lbrack s \rbrack$},
            xlabel style={at={(0.5,-0.37)},anchor=south},
		    ylabel={\small{Jain's index}},
    		legend style={at={(0.01,0.76)},anchor=south west,column sep=-0.08cm, legend cell align=left, row sep=-0.05cm},	
    		legend columns=2,
		    xmin=0, xmax=60, ymin=0.25, ymax=1, grid,]
		    \addplot [color=blue!50!red, mark=none, line width=1.2pt] 
		    table [col sep=semicolon, x index=0, y index=1,] {csv/4fairnesbbrbbr.csv};
		\end{axis}
	\end{tikzpicture}
	\label{fig:4bbrfairnessrtt}
\end{subfigure}
		\caption{RTT scenario: Comparison of average RTT, average throughput, and fairness index for representatives of the congestion control algorithm classes in case the link is shared by 4 flows using the same algorithm but different RTTs (time unit 600ms).}\label{fig:dual4}
	\end{center}
\end{figure*}

We observe RTT-fairness issues for all three groups of algorithms, with TCP Hybla being the least sensitive (Tab.~\ref{tab:fairness2}). When the number of competing flows is 2, two Vegas flows achieve a similar throughput on average, both claiming a similar share of the available bandwidth. Their convergence time increases though, as compared to the previous scenario (Sec.~\ref{Sec:dual1}), and is $\approx 5\;s$. However, when the number of Vegas flows at the bottleneck increases, the flow with the lowest RTT claims all the available bandwidth, starving the other flows (Fig. ~\ref{fig:dual4}). The fairness index increases over time, but due to a very conservative congestion avoidance approach of Vegas, even after $60s$, flows do not converge. The observed queuing delay increases for all flows by almost a factor of $10$ (from $\approx 2ms$ to $\approx 20ms$). 

\begin{table*}[!htb]
	\begin{center}
		\caption{RTT scenario with 4 flows: Different metrics for representatives of the congestion control algorithm classes in case the link is shared by four flows using the same algorithm and having different RTTs.}
		\label{tab:dual4}
		\begin{tabular}{l | l | l | c | c | c | c | c | c | c}
			\toprule
			Protocol & Group & Algorithm & Average & Average & Average & Average & Average & Average & Average\\
			& & & goodput & goodput ratio  &  cwnd & RTT  & sending rate & throughput & Jain's index\\ 
			& &           & $\lbrack Mbps \rbrack$ & $\lbrack \% \rbrack$ & $\lbrack \#packets \rbrack$ &  $\lbrack ms \rbrack$ & $\lbrack Mbps \rbrack$ & $\lbrack Mbps \rbrack$ \\            
			\midrule    
			\multirow{12}{0.5cm}{TCP} 
			& \multirow{4}{2cm}{Loss-based}             
			&Cubic(100ms)  &  40.13  & 94.76  & 1988.66 & 550.68 & 46.11 & 41.74 &   \multirow{4}{1cm}{0.64}  \\   
			& &Cubic(200ms)  & 11.38 & 91.15 & 914.639 & 755.17 & 12.83 & 11.86 &   \\     
			& &Cubic(300ms)  & 17.63 & 92.92 & 1626.1 & 749.21 & 20.31 & 18.48 &   \\     
			& &Cubic(400ms)  &  8.38 & 88.87 & 897.348 & 828.25 &  9.83 &  8.73 &   \\     
			\cline{2-10}                                                                                           
			& \multirow{4}{2cm}{Delay-based}
			&Vegas(100ms)  &  41.54 & 93.94 & 503.01 & 125.87 & 47.35 & 43.19 &   \multirow{4}{1cm}{0.57} \\		
			& &Vegas(200ms) &    6.79 & 92.28 & 129.78 & 226.26 &  7.60 &  7.10 &     \\       
			& &Vegas(300ms) &    3.56 & 89.60 & 99.58  & 326.22 &  4.07 &  3.73 &     \\       
			& &Vegas(400ms) &   15.63 & 91.48 & 574.68 & 426.19 & 17.46 & 16.32 &     \\       
			
			\cline{2-10}                                                                                           
			& \multirow{4}{2cm}{Hybrid}                                                             
			&BBR(100ms)      & 32.34 & 92.73 & 2658.96 & 509.66 & 35.702 & 33.76 &   \multirow{4}{1cm}{0.70} \\ 
			& &BBR(200ms)    &  5.92 & 86.90 & 404.054 & 613.11 &  7.19 &  6.23 &  \\      
			& &BBR(300ms)    & 22.34 & 93.26 & 2488.10 & 722.18 & 25.18 & 23.46 &  \\      
			& &BBR(400ms)    & 19.27 & 91.98 & 2365.76 & 816.02 & 21.24 & 20.23 &  \\      
			
			\hline
			
		\end{tabular}
	\end{center}
\end{table*}

All analyzed loss-based algorithms favour the flow with the lower RTT (Tab.~\ref{tab:fairness2}). Even algorithms, such as Cubic, that claim RTT-fairness, were shown to have a similar behavior \cite{6726892}. This is most noticeable when analyzing two Cubic flows in Fig.~\ref{fig:dual3}. Even when the number of flows increases to 4 (Fig. ~\ref{fig:dual4}), the flows with a lower RTT immediately claim all the available bandwidth, leaving a very small share to the other flows in the first $20\;s$. Moreover, even after the flows converge to the same bandwidth, after a loss event occurs, the flows need to be synchronized again, reducing the fairness index (Fig.~\ref{fig:dual4}). Several improvements addressing this problem, such as TCP Libra \cite{marfia2007tcp} have been proposed. However, current kernel implementations do not capture these improvements.

\begin{table*}[!htb]
	\begin{center}
		\caption{RTT scenario with 2 flows: Comparison of Jain's for different congestion control algorithms.}
		\label{tab:fairness2}
		\begin{tabular}{  l | c  c  c  c  c  c  c | c  c | c  c  c  c }
			
			\multirow{2}{1.3cm}{Delay Difference} & \multicolumn{7}{|c|}{Loss-based}    &   \multicolumn{2}{|c|}{Delay-based} & \multicolumn{4}{|c}{Hybrid} \\
			\cline{2-14}
			& Reno & BIC & Cubic & HS-TCP & H-TCP & Hybla & Westwood & Vegas & LoLA & Veno & Illinois & YeAH & BBR \\            
			\hline
			0ms       & 0.94 & 0.96 & 0.93 & 0.95 & 0.96 & 0.90 & 0.96 & 0.98 & 0.86 & 0.89 & 0.94 & 0.95 & 0.87 \\ 
			200ms     & 0.85 & 0.86 & 0.72 & 0.90 & 0.92 & 0.92
			& 0.83 & 0.87 & 0.55 & 0.82 & 0.89 & 0.77 & 0.76 \\
		\end{tabular}
	\end{center}
\end{table*}
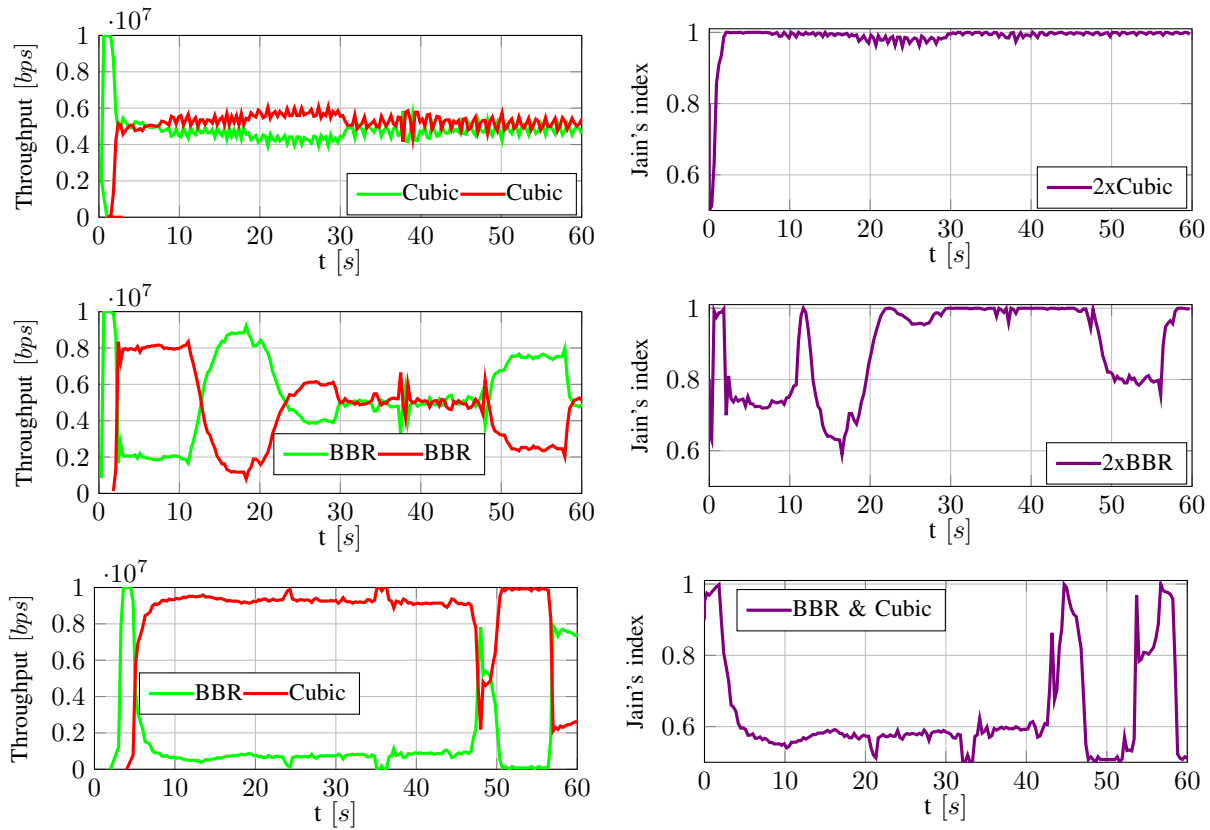
\begin{figure*}[!htb]
	\begin{center}
		\begin{subfigure}{0.45\textwidth}
    \centering
	\begin{tikzpicture}
	\begin{axis}[
            title style={at={(0.5,1.1)},anchor=north},
	        width = 8cm, height = 4cm,
		    xlabel={t $\lbrack s \rbrack$},
            xlabel style={at={(0.5,-0.37)},anchor=south},
		    ylabel={\small{Throughput $\lbrack bps \rbrack$}},
    		legend style={at={(0.99,0.01)},anchor=south east, column sep=-0.08cm, legend cell align=left, row sep=-0.05cm},	
    		legend columns=2,
		    xmin=0, xmax=60, ymin=-0.1, ymax=10000000, grid,]
		    \addplot [color=green, mark=none, line width=1.2pt] 
		    table [col sep=semicolon, x index=2, y index=3,] {csv/QUIC2Cubicthroughput.csv};
		    \addplot [color=red,mark=none, line width=1.2pt] 
		    table [col sep=semicolon, x index=6, y index=7,] {csv/QUIC2Cubicthroughput.csv};
		    \legend{\small Cubic, \small Cubic}
		\end{axis}
	\end{tikzpicture}
	\label{fig:4cubicthrtt}
\end{subfigure}%
\begin{subfigure}{0.45\textwidth}
    \centering
	\begin{tikzpicture}
	\begin{axis}[
            title style={at={(0.5,1.1)},anchor=north},
	        width = 8cm, height = 4cm,
		    xlabel={t $\lbrack s \rbrack$},
            xlabel style={at={(0.5,-0.37)},anchor=south},
		    ylabel={\small{Jain's index}},
    		legend style={at={(0.99,0.01)},anchor=south east,column sep=-0.08cm, legend cell align=left, row sep=-0.05cm},	
		    xmin=0, xmax=60, ymin=0.5, ymax=1.01, grid,]
		    \addplot [color=blue!50!red, mark=none, line width=1.2pt] 
		    table [col sep=semicolon, x index=0, y index=1,] {csv/fairnessCubicCubicQUIC.csv};
		    \legend{\small 2xCubic}
		\end{axis}
	\end{tikzpicture}
	\label{fig:4fairnescubicrtt}
\end{subfigure}

\begin{subfigure}{0.45\textwidth}
	\centering
	\begin{tikzpicture}
	\begin{axis}[
	title style={at={(0.5,1.1)},anchor=north},
	width = 8cm, height = 4cm,
	xlabel={t $\lbrack s \rbrack$},
	xlabel style={at={(0.5,-0.37)},anchor=south},
	ylabel={\small{Throughput $\lbrack bps \rbrack$}},
	legend style={at={(0.80,0.1)},anchor=south east, column sep=-0.08cm, legend cell align=left, row sep=-0.05cm},	
	legend columns=2,
	xmin=0, xmax=60, ymin=-0.1, ymax=10000000, grid,]
	\addplot [color=green, mark=none, line width=1.2pt] 
	table [col sep=semicolon, x index=2, y index=3,] {csv/QUICBBRBBRthroughput.csv};
	\addplot [color=red,mark=none, line width=1.2pt] 
	table [col sep=semicolon, x index=6, y index=7,] {csv/QUICBBRBBRthroughput.csv};
	\legend{\small BBR, \small BBR}
	\end{axis}
	\end{tikzpicture}
	\label{fig:quic2bbrth}
\end{subfigure}%
\begin{subfigure}{0.45\textwidth}
	\centering
	\begin{tikzpicture}
	\begin{axis}[
	title style={at={(0.5,1.1)},anchor=north},
	width = 8cm, height = 4cm,
	xlabel={t $\lbrack s \rbrack$},
	xlabel style={at={(0.5,-0.37)},anchor=south},
	ylabel={\small{Jain's index}},
	legend style={at={(0.99,0.01)},anchor=south east,column sep=-0.08cm, legend cell align=left, row sep=-0.05cm},	
	xmin=0, xmax=60, ymin=0.5, ymax=1.01, grid,]
	\addplot [color=blue!50!red, mark=none, line width=1.2pt] 
	table [col sep=semicolon, x index=0, y index=1,] {csv/fairnessBBRBBRQUIC.csv};
	\legend{\small 2xBBR}
	\end{axis}
	\end{tikzpicture}
	\label{fig:quic2bbrfairness}
\end{subfigure}

\begin{subfigure}{0.45\textwidth}
	\centering
	\begin{tikzpicture}
	\begin{axis}[
	title style={at={(0.5,1.1)},anchor=north},
	width = 8cm, height = 4cm,
	xlabel={t $\lbrack s \rbrack$},
	xlabel style={at={(0.5,-0.37)},anchor=south},
	ylabel={\small{Throughput $\lbrack bps \rbrack$}},
	legend style={at={(0.55,0.3)},anchor=south east, column sep=-0.08cm, legend cell align=left, row sep=-0.05cm},	
	legend columns=2,
	xmin=0, xmax=60, ymin=-0.1, ymax=10000000, grid,]
	\addplot [color=green, mark=none, line width=1.2pt] 
	table [col sep=semicolon, x index=4, y index=5,] {csv/QUICBBRCUBICthroughput.csv};
	\addplot [color=red,mark=none, line width=1.2pt] 
	table [col sep=semicolon, x index=8, y index=9,] {csv/QUICBBRCUBICthroughput.csv};
	\legend{\small BBR, \small Cubic}
	\end{axis}
	\end{tikzpicture}
	\label{fig:quic4bbrcubicth}
\end{subfigure}%
\begin{subfigure}{0.45\textwidth}
	\centering
	\begin{tikzpicture}
	\begin{axis}[
	title style={at={(0.5,1.1)},anchor=north},
	width = 8cm, height = 4cm,
	xlabel={t $\lbrack s \rbrack$},
	xlabel style={at={(0.5,-0.37)},anchor=south},
	ylabel={\small{Jain's index}},
	legend style={at={(0.5,0.72)},anchor=south east,column sep=-0.08cm, legend cell align=left, row sep=-0.05cm},	
	xmin=0, xmax=60, ymin=0.5, ymax=1.01, grid,]
	\addplot [color=blue!50!red, mark=none, line width=1.2pt] 
	table [col sep=semicolon, x index=0, y index=1,] {csv/QUICBBRRCUBICfairness.csv};
	\legend{\small BBR \& Cubic}
	\end{axis}
	\end{tikzpicture}
	\label{fig:quicbbrcubicfairness}
\end{subfigure}
		\caption{QUIC BW scenario: Comparison of average throughput and fairness index for representatives of the congestion control algorithm groups in the case the link is shared by 2 flows (time unit 300ms).}\label{fig:dual7}
	\end{center}
\end{figure*}

Hybrid-based algorithms, such as BBR or YeAH, favour the flow with the higher RTT, while other algorithms, such as Illinois and Veno favour the flow with a lower RTT. BBR flow with a higher RTT overestimates the bottleneck link, claiming all the available resources and increasing the queuing delay present in the network \cite{scholztowards, hock2017experimental}. BBR flows should synchronize when a BBR flow with a larger share of resources enters the ProbeRTT phase (reduces the cwnd to 4 packets). When this occurs, a large portion of the queue at the bottleneck is drained, resulting in all other flows measuring a better $RTT_{prop}$ estimate. This can be observed in Fig.~\ref{fig:dual3} and Fig.~\ref{fig:dual4}. Every 10 seconds, the flow with a smaller share claims more bandwidth, while the throughput of the other flow drops significantly. However, this lasts a very short time and the flow with a lower RTT overestimates the bandwidth again, claiming a bigger share of resources. Even when the number of flows increases to four, the flows with a higher RTT outperform the flows with a lower RTT. In addition, two flows with a higher RTT ($300ms$ and $400ms$) and two flows with a lower RTT (with $RTT=100ms$ and $RTT=200ms$) start competing for resources among themselves, oscillating around the same throughput.

Moreover, when the number of BBR flows increases, the average RTT increases, reaching values comparable to the ones observed by the loss-based algorithms (Fig.~\ref{fig:dual4}).  


\textbf{Summary.} We observed that RTT-fairness is poor for all groups of algorithms. While delay-based algorithms perform best compared to the other two groups, they still take time to converge towards their fair share. Loss-based algorithms such as Cubic perform poorly, contrary to expectations and their own claims, favouring flows with lower RTTs. When loss-based algorithms converge to a fair share, the convergence time is so slow that the fairness index is still low. Finally, hybrid algorithms such as BBR suffer from significant dynamics in the sharing among its own flows, favoring those with higher RTT. This leads to complex dynamics between these flows. 

\subsection{Results: QUIC}

\begin{figure*}[!htb]
	\begin{center}
		\begin{subfigure}{0.45\textwidth}
	\centering
	\begin{tikzpicture}
	\begin{axis}[
	title style={at={(0.5,1.1)},anchor=north},
	width = 8cm, height = 4cm,
	xlabel={t $\lbrack s \rbrack$},
	xlabel style={at={(0.5,-0.37)},anchor=south},
	ylabel={\small{Throughput $\lbrack bps \rbrack$}},
	legend style={at={(0.99,0.01)},anchor=south east, column sep=-0.08cm, legend cell align=left, row sep=-0.05cm},	
	legend columns=2,
	xmin=0, xmax=60, ymin=-0.1, ymax=10000000, grid,]
	\addplot [color=green, mark=none, line width=1.2pt] 
	table [col sep=semicolon, x index=2, y index=3,] {csv/QUICCubicCubic200msthroughput.csv};
	\addplot [color=red,mark=none, line width=1.2pt] 
	table [col sep=semicolon, x index=6, y index=7,] {csv/QUICCubicCubic200msthroughput.csv};
	\legend{\small Cubic(0ms), \small Cubic(200ms)}
	\end{axis}
	\end{tikzpicture}
	\label{fig:quic2cubicthrtt}
\end{subfigure}%
\begin{subfigure}{0.45\textwidth}
	\centering
	\begin{tikzpicture}
	\begin{axis}[
	title style={at={(0.5,1.1)},anchor=north},
	width = 8cm, height = 4cm,
	xlabel={t $\lbrack s \rbrack$},
	xlabel style={at={(0.5,-0.37)},anchor=south},
	ylabel={\small{Jain's index}},
	legend style={at={(0.99,0.01)},anchor=south east,column sep=-0.08cm, legend cell align=left, row sep=-0.05cm},	
	xmin=0, xmax=60, ymin=0.5, ymax=1.01, grid,]
	\addplot [color=blue!50!red, mark=none, line width=1.2pt] 
	table [col sep=semicolon, x index=0, y index=1,] {csv/fairnessCubicCubicQUIC200ms.csv};
	\legend{\small 2xCubic}
	\end{axis}
	\end{tikzpicture}
	\label{fig:quic2cubicfairnessrtt}
\end{subfigure}

\begin{subfigure}{0.45\textwidth}
	\centering
	\begin{tikzpicture}
	\begin{axis}[
	title style={at={(0.5,1.1)},anchor=north},
	width = 8cm, height = 4cm,
	xlabel={t $\lbrack s \rbrack$},
	xlabel style={at={(0.5,-0.37)},anchor=south},
	ylabel={\small{Throughput $\lbrack bps \rbrack$}},
	legend style={at={(0.99,0.4)},anchor=south east, column sep=-0.08cm, legend cell align=left, row sep=-0.05cm},	
	legend columns=2,
	xmin=0, xmax=60, ymin=-0.1, ymax=10000000, grid,]
	\addplot [color=green, mark=none, line width=1.2pt] 
	table [col sep=semicolon, x index=2, y index=3,] {csv/QUICBBRBBR200msthroughput.csv};
	\addplot [color=red,mark=none, line width=1.2pt] 
	table [col sep=semicolon, x index=6, y index=7,] {csv/QUICBBRBBR200msthroughput.csv};
	\legend{\small BBR(0ms), \small BBR(200ms)}
	\end{axis}
	\end{tikzpicture}
	\label{fig:quic2bbrthrtt}
\end{subfigure}%
\begin{subfigure}{0.45\textwidth}
	\centering
	\begin{tikzpicture}
	\begin{axis}[
	title style={at={(0.5,1.1)},anchor=north},
	width = 8cm, height = 4cm,
	xlabel={t $\lbrack s \rbrack$},
	xlabel style={at={(0.5,-0.37)},anchor=south},
	ylabel={\small{Jain's index}},
	legend style={at={(0.5,0.72)},anchor=south east,column sep=-0.08cm, legend cell align=left, row sep=-0.05cm},	
	xmin=0, xmax=60, ymin=0.5, ymax=1.01, grid,]
	\addplot [color=blue!50!red, mark=none, line width=1.2pt] 
	table [col sep=semicolon, x index=0, y index=1,] {csv/fairnessBBRBBRQUIC200ms.csv};
	\legend{\small 2xBBR}
	\end{axis}
	\end{tikzpicture}
	\label{fig:quic2bbrfairnessrtt}
\end{subfigure}
		\caption{QUIC RTT scenario: Comparison of average throughput and fairness index for representatives of the congestion control algorithm groups in the case the link is shared by 2 flows using the same algorithm but different RTTs (time unit 300ms).}\label{fig:dual10}
	\end{center}
\end{figure*}
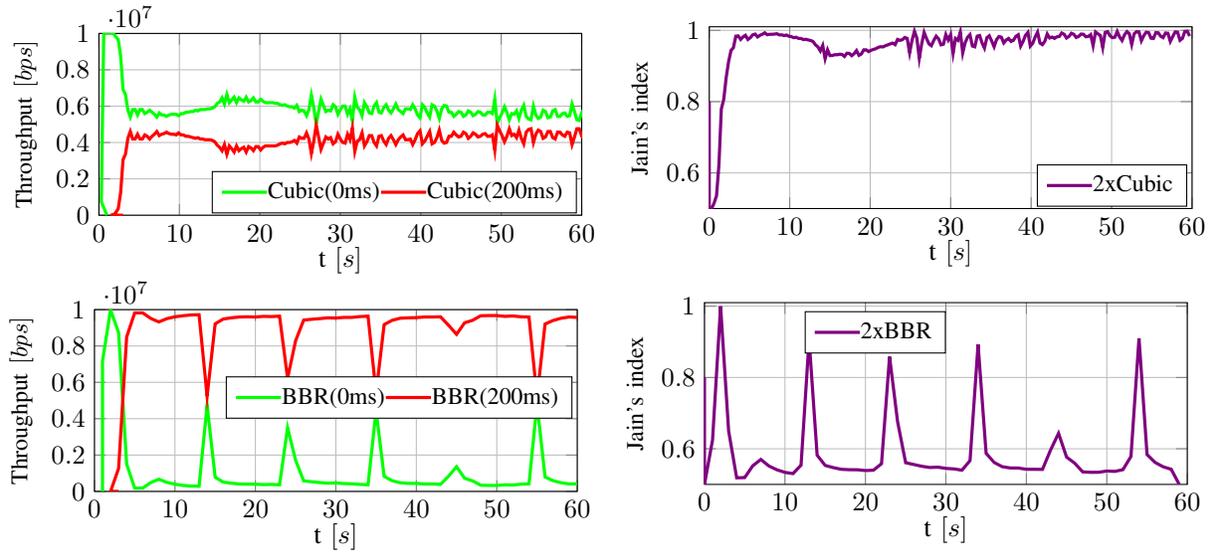

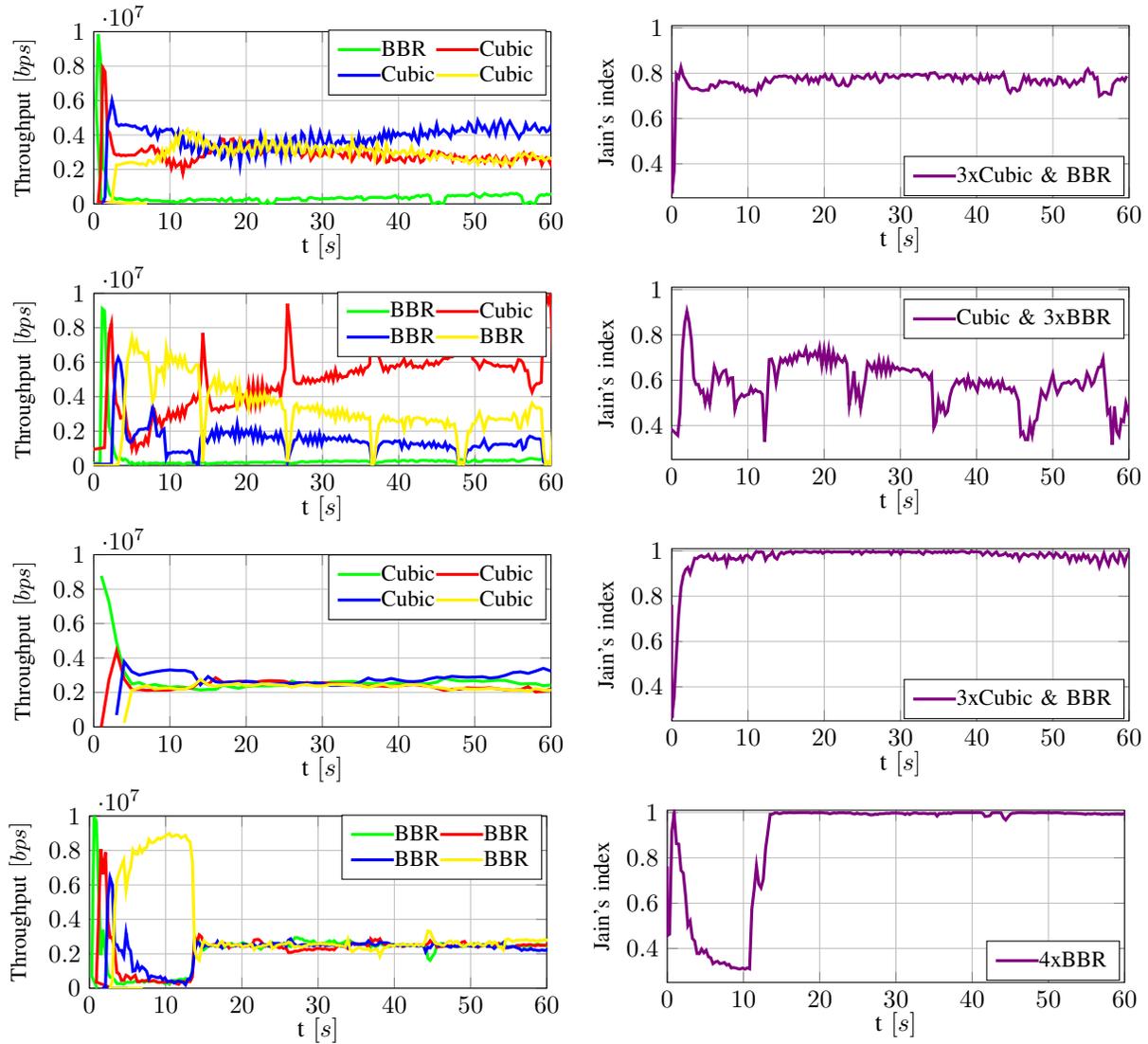
\begin{figure*}[!htb]
	\begin{center}
		\begin{subfigure}{0.45\textwidth}
    \centering
	\begin{tikzpicture}
	\begin{axis}[
            title style={at={(0.5,1.1)},anchor=north},
	        width = 8cm, height = 4cm,
		    xlabel={t $\lbrack s \rbrack$},
            xlabel style={at={(0.5,-0.37)},anchor=south},
		    ylabel={\small{Throughput $\lbrack bps \rbrack$}},
    		legend style={at={(0.99,0.63)},anchor=south east, column sep=-0.08cm, legend cell align=left, row sep=-0.05cm},	
    		legend columns=2,
		    xmin=0, xmax=60, ymin=-0.1, ymax=10000000, grid,]
		    \addplot [color=green, mark=none, line width=1.2pt] 
		    table [col sep=semicolon, x index=2, y index=3,] {csv/QUICthroughput3Cubic1Bbr.csv};
		    \addplot [color=red,mark=none, line width=1.2pt] 
		    table [col sep=semicolon, x index=6, y index=7,] {csv/QUICthroughput3Cubic1Bbr.csv};
		    \addplot [color=blue,mark=none, line width=1.2pt] 
table [col sep=semicolon, x index=10, y index=11,] {csv/QUICthroughput3Cubic1Bbr.csv};
		    \addplot [color=yellow,mark=none, line width=1.2pt] 
table [col sep=semicolon, x index=14, y index=15,] {csv/QUICthroughput3Cubic1Bbr.csv};		    
		    \legend{\small BBR, \small Cubic, \small Cubic, \small Cubic}
		\end{axis}
	\end{tikzpicture}
	\label{fig:quic3cubicbbrth}
\end{subfigure}%
\begin{subfigure}{0.45\textwidth}
    \centering
	\begin{tikzpicture}
	\begin{axis}[
            title style={at={(0.5,1.1)},anchor=north},
	        width = 8cm, height = 4cm,
		    xlabel={t $\lbrack s \rbrack$},
            xlabel style={at={(0.5,-0.37)},anchor=south},
		    ylabel={\small{Jain's index}},
    		legend style={at={(0.99,0.01)},anchor=south east,column sep=-0.08cm, legend cell align=left, row sep=-0.05cm},	
		    xmin=0, xmax=60, ymin=0.25, ymax=1.01, grid,]
		    \addplot [color=blue!50!red, mark=none, line width=1.2pt] 
		    table [col sep=semicolon, x index=0, y index=1,] {csv/fairness3Cubic1BbrQUIC.csv};
		    \legend{\small 3xCubic \& BBR}
		\end{axis}
	\end{tikzpicture}
	\label{fig:quic3cubicbbrfairness}
\end{subfigure}

\begin{subfigure}{0.45\textwidth}
	\centering
	\begin{tikzpicture}
	\begin{axis}[
	title style={at={(0.5,1.1)},anchor=north},
	width = 8cm, height = 4cm,
	xlabel={t $\lbrack s \rbrack$},
	xlabel style={at={(0.5,-0.37)},anchor=south},
	ylabel={\small{Throughput $\lbrack bps \rbrack$}},
	legend style={at={(0.99,0.63)},anchor=south east, column sep=-0.08cm, legend cell align=left, row sep=-0.05cm},	
	legend columns=2,
	xmin=0, xmax=60, ymin=-0.1, ymax=10000000, grid,]
	\addplot [color=green, mark=none, line width=1.2pt] 
	table [col sep=semicolon, x index=0, y index=1,] {csv/QUIC3BBR1CUBICRTTthroughput.csv};
	\addplot [color=red,mark=none, line width=1.2pt] 
	table [col sep=semicolon, x index=2, y index=3,] {csv/QUIC3BBR1CUBICRTTthroughput.csv};
	\addplot [color=blue,mark=none, line width=1.2pt] 
	table [col sep=semicolon, x index=4, y index=5,] {csv/QUIC3BBR1CUBICRTTthroughput.csv};
	\addplot [color=yellow,mark=none, line width=1.2pt] 
	table [col sep=semicolon, x index=6, y index=7,] {csv/QUIC3BBR1CUBICRTTthroughput.csv};		    
	\legend{\small BBR, \small Cubic, \small BBR, \small BBR}
	\end{axis}
	\end{tikzpicture}
	\label{fig:quiccubic3bbrth}
\end{subfigure}%
\begin{subfigure}{0.45\textwidth}
	\centering
	\begin{tikzpicture}
	\begin{axis}[
	title style={at={(0.5,1.1)},anchor=north},
	width = 8cm, height = 4cm,
	xlabel={t $\lbrack s \rbrack$},
	xlabel style={at={(0.5,-0.37)},anchor=south},
	ylabel={\small{Jain's index}},
	legend style={at={(0.99,0.7)},anchor=south east,column sep=-0.08cm, legend cell align=left, row sep=-0.05cm},	
	xmin=0, xmax=60, ymin=0.25, ymax=1.01, grid,]
	\addplot [color=blue!50!red, mark=none, line width=1.2pt] 
	table [col sep=semicolon, x index=0, y index=1,] {csv/fairness3BBR1CUBIC.csv};
	\legend{\small Cubic \& 3xBBR}
	\end{axis}
	\end{tikzpicture}
	\label{fig:quiccubic3bbrfairness}
\end{subfigure}

\begin{subfigure}{0.45\textwidth}
	\centering
	\begin{tikzpicture}
	\begin{axis}[
	title style={at={(0.5,1.1)},anchor=north},
	width = 8cm, height = 4cm,
	xlabel={t $\lbrack s \rbrack$},
	xlabel style={at={(0.5,-0.37)},anchor=south},
	ylabel={\small{Throughput $\lbrack bps \rbrack$}},
	legend style={at={(0.99,0.62)},anchor=south east, column sep=-0.08cm, legend cell align=left, row sep=-0.05cm},	
	legend columns=2,
	xmin=0, xmax=60, ymin=-0.1, ymax=10000000, grid,]
	\addplot [color=green, mark=none, line width=1.2pt] 
	table [col sep=semicolon, x index=2, y index=3,] {csv/QUIC4CUBICthroughput.csv};
	\addplot [color=red,mark=none, line width=1.2pt] 
	table [col sep=semicolon, x index=6, y index=7,] {csv/QUIC4CUBICthroughput.csv};
	\addplot [color=blue,mark=none, line width=1.2pt] 
	table [col sep=semicolon, x index=10, y index=11,] {csv/QUIC4CUBICthroughput.csv};
	\addplot [color=yellow,mark=none, line width=1.2pt] 
	table [col sep=semicolon, x index=14, y index=15,] {csv/QUIC4CUBICthroughput.csv};		    
	\legend{\small Cubic, \small Cubic, \small Cubic, \small Cubic}
	\end{axis}
	\end{tikzpicture}
	\label{fig:quic4cubicth}
\end{subfigure}%
\begin{subfigure}{0.45\textwidth}
	\centering
	\begin{tikzpicture}
	\begin{axis}[
	title style={at={(0.5,1.1)},anchor=north},
	width = 8cm, height = 4cm,
	xlabel={t $\lbrack s \rbrack$},
	xlabel style={at={(0.5,-0.37)},anchor=south},
	ylabel={\small{Jain's index}},
	legend style={at={(0.99,0.01)},anchor=south east,column sep=-0.08cm, legend cell align=left, row sep=-0.05cm},	
	xmin=0, xmax=60, ymin=0.25, ymax=1.01, grid,]
	\addplot [color=blue!50!red, mark=none, line width=1.2pt] 
	table [col sep=semicolon, x index=0, y index=1,] {csv/QUICfairness4CUBIC.csv};
	\legend{\small 3xCubic \& BBR}
	\end{axis}
	\end{tikzpicture}
	\label{fig:quic4cubicfairness}
\end{subfigure}

\begin{subfigure}{0.45\textwidth}
	\centering
	\begin{tikzpicture}
	\begin{axis}[
	title style={at={(0.5,1.1)},anchor=north},
	width = 8cm, height = 4cm,
	xlabel={t $\lbrack s \rbrack$},
	xlabel style={at={(0.5,-0.37)},anchor=south},
	ylabel={\small{Throughput $\lbrack bps \rbrack$}},
	legend style={at={(0.99,0.63)},anchor=south east, column sep=-0.08cm, legend cell align=left, row sep=-0.05cm},	
	legend columns=2,
	xmin=0, xmax=60, ymin=-0.1, ymax=10000000, grid,]
	\addplot [color=green, mark=none, line width=1.2pt] 
	table [col sep=semicolon, x index=2, y index=3,] {csv/QUIC4BBRthroughput.csv};
	\addplot [color=red,mark=none, line width=1.2pt] 
	table [col sep=semicolon, x index=6, y index=7,] {csv/QUIC4BBRthroughput.csv};
	\addplot [color=blue,mark=none, line width=1.2pt] 
	table [col sep=semicolon, x index=10, y index=11,] {csv/QUIC4BBRthroughput.csv};
	\addplot [color=yellow,mark=none, line width=1.2pt] 
	table [col sep=semicolon, x index=14, y index=15,] {csv/QUIC4BBRthroughput.csv};		    
	\legend{\small BBR, \small BBR, \small BBR, \small BBR}
	\end{axis}
	\end{tikzpicture}
	\label{fig:quic4bbrth}
\end{subfigure}%
\begin{subfigure}{0.45\textwidth}
	\centering
	\begin{tikzpicture}
	\begin{axis}[
	title style={at={(0.5,1.1)},anchor=north},
	width = 8cm, height = 4cm,
	xlabel={t $\lbrack s \rbrack$},
	xlabel style={at={(0.5,-0.37)},anchor=south},
	ylabel={\small{Jain's index}},
	legend style={at={(0.99,0.01)},anchor=south east,column sep=-0.08cm, legend cell align=left, row sep=-0.05cm},	
	xmin=0, xmax=60, ymin=0.25, ymax=1.01, grid,]
	\addplot [color=blue!50!red, mark=none, line width=1.2pt] 
	table [col sep=semicolon, x index=0, y index=1,] {csv/QUICfairness4BBR.csv};
	\legend{\small 4xBBR}
	\end{axis}
	\end{tikzpicture}
	\label{fig:quic4bbrfairness}
\end{subfigure}
		\caption{QUIC BW scenario: Comparison of average throughput and fairness index for representatives of the congestion control algorithm classes groups in case the link is shared by 4 flows (time unit 300ms).}\label{fig:dual8}
	\end{center}
\end{figure*}

\begin{figure*}[!htb]
	\begin{center}
		\begin{subfigure}{0.45\textwidth}
    \centering
	\begin{tikzpicture}
	\begin{axis}[
		title = {4 Cubic flows},	
            title style={at={(0.5,1.1)},anchor=north},
	        width = 8cm, height = 4cm,
		    xlabel={t $\lbrack s \rbrack$},
            xlabel style={at={(0.5,-0.37)},anchor=south},
		    ylabel={\small{Throughput $\lbrack bps \rbrack$}},
    		legend style={at={(0.99,0.63)},anchor=south east, column sep=-0.08cm, legend cell align=left, row sep=-0.05cm},	
    		legend columns=2,
		    xmin=0, xmax=60, ymin=-0.1, ymax=10000000, grid,]
		    \addplot [color=green, mark=none, line width=1.2pt] 
		    table [col sep=semicolon, x index=4, y index=5,] {csv/QUIC4CUBICRTTthroughput.csv};
		    \addplot [color=red,mark=none, line width=1.2pt] 
		    table [col sep=semicolon, x index=8, y index=9,] {csv/QUIC4CUBICRTTthroughput.csv};
		    \addplot [color=blue,mark=none, line width=1.2pt] 
table [col sep=semicolon, x index=12, y index=13,] {csv/QUIC4CUBICRTTthroughput.csv};
		    \addplot [color=yellow,mark=none, line width=1.2pt] 
table [col sep=semicolon, x index=16, y index=17,] {csv/QUIC4CUBICRTTthroughput.csv};		    
		    \legend{\small 100ms, \small 200ms, \small 300ms, \small 400ms}
		\end{axis}
	\end{tikzpicture}
	\label{fig:quic4cubicthrtt}
\end{subfigure}%
\begin{subfigure}{0.45\textwidth}
    \centering
	\begin{tikzpicture}
	\begin{axis}[
		title = {4 Cubic flows},
            title style={at={(0.5,1.1)},anchor=north},
	        width = 8cm, height = 4cm,
		    xlabel={t $\lbrack s \rbrack$},
            xlabel style={at={(0.5,-0.37)},anchor=south},
		    ylabel={\small{Jain's index}},
    		legend style={at={(0.99,0.01)},anchor=south east,column sep=-0.08cm, legend cell align=left, row sep=-0.05cm},	
		    xmin=0, xmax=60, ymin=0.25, ymax=1.01, grid,]
		    \addplot [color=blue!50!red, mark=none, line width=1.2pt] 
		    table [col sep=semicolon, x index=0, y index=1,] {csv/fairnessQUIC4CUBICRTT.csv};
		    \legend{\small 4xCubic}
		\end{axis}
	\end{tikzpicture}
	\label{fig:quic4cubicfairnessrtt}
\end{subfigure}

\begin{subfigure}{0.45\textwidth}
	\centering
	\begin{tikzpicture}
	\begin{axis}[
	title = {4 BBR flows},
	title style={at={(0.5,1.1)},anchor=north},
	width = 8cm, height = 4cm,
	xlabel={t $\lbrack s \rbrack$},
	xlabel style={at={(0.5,-0.37)},anchor=south},
	ylabel={\small{Throughput $\lbrack bps \rbrack$}},
	legend style={at={(0.60,0.63)},anchor=south east, column sep=-0.08cm, legend cell align=left, row sep=-0.05cm},	
	legend columns=2,
	xmin=0, xmax=60, ymin=-0.1, ymax=10000000, grid,]
	\addplot [color=yellow,mark=none, line width=1.2pt] 
table [col sep=semicolon, x index=16, y index=17,] {csv/QUIC4BBRRTTthroughput.csv};		    	
	\addplot [color=red,mark=none, line width=1.2pt] 
	table [col sep=semicolon, x index=8, y index=9,] {csv/QUIC4BBRRTTthroughput.csv};
	\addplot [color=blue,mark=none, line width=1.2pt] 
	table [col sep=semicolon, x index=12, y index=13,] {csv/QUIC4BBRRTTthroughput.csv};
	\addplot [color=green, mark=none, line width=1.2pt] 
table [col sep=semicolon, x index=4, y index=5,] {csv/QUIC4BBRRTTthroughput.csv};
	\legend{\small 100ms, \small 200ms, \small 300ms, \small 400ms}
	\end{axis}
	\end{tikzpicture}
	\label{fig:quic4bbrthrtt}
\end{subfigure}%
\begin{subfigure}{0.45\textwidth}
	\centering
	\begin{tikzpicture}
	\begin{axis}[
	title = {4 BBR flows},	
	title style={at={(0.5,1.1)},anchor=north},
	width = 8cm, height = 4cm,
	xlabel={t $\lbrack s \rbrack$},
	xlabel style={at={(0.5,-0.37)},anchor=south},
	ylabel={\small{Jain's index}},
	legend style={at={(0.40,0.01)},anchor=south east,column sep=-0.08cm, legend cell align=left, row sep=-0.05cm},	
	xmin=0, xmax=60, ymin=0.25, ymax=1.01, grid,]
	\addplot [color=blue!50!red, mark=none, line width=1.2pt] 
	table [col sep=semicolon, x index=0, y index=1,] {csv/QUIC4BBRRTTfairness.csv};
	\legend{\small 4xBBR}
	\end{axis}
	\end{tikzpicture}
	\label{fig:quic4bbrfairnessrtt}
\end{subfigure}
		\caption{QUIC RTT scenario: Comparison of average throughput and fairness index for representatives of the congestion control algorithm groups in the case the link is shared by 4 flows using the same algorithm but different RTTs (time unit 300ms).}\label{fig:dual9}
	\end{center}
\end{figure*}
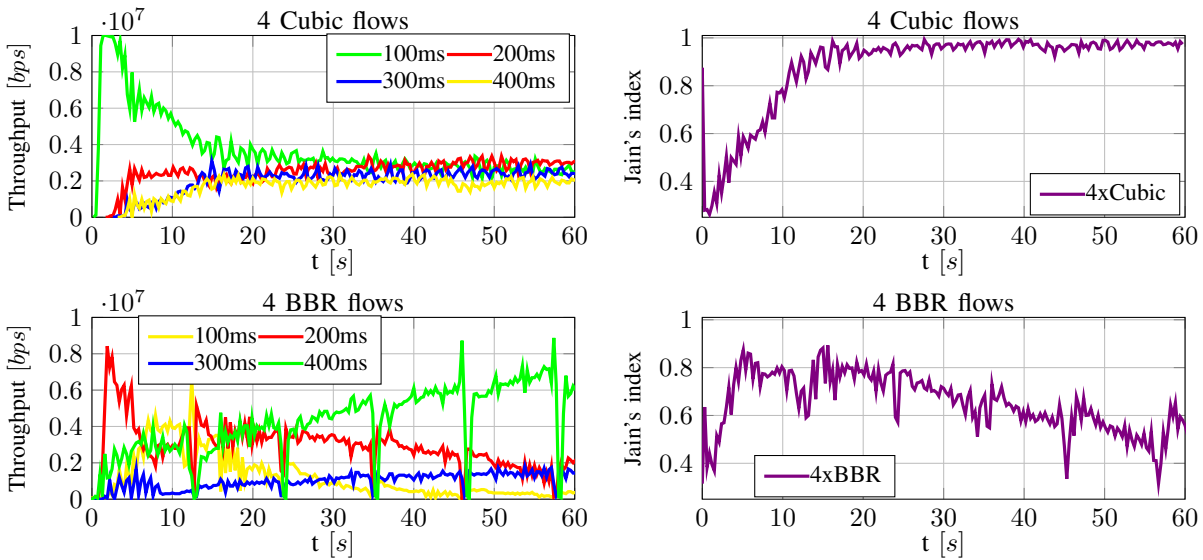

When QUIC is used with different congestion control algorithms, we observed similar interactions as earlier (Fig.~\ref{fig:dual7},~\ref{fig:dual10},~\ref{fig:dual8},~\ref{fig:dual9}). Thus, we can conclude that the choice of the transport protocol has no significant influence on the interaction between the algorithms.

\begin{table*}[!htb]
	\begin{center}
		\caption{QUIC BW \& RTT scenario with 2 flows: Different metrics for representatives of the congestion control algorithm groups.}
		\label{tab:dual7}
		\begin{tabular}{l| l | l | c | c | c | c | c}
			\toprule
			Protocol & Group & Algorithm & Average & Average & Average & Average & Average \\
			
			& & &  goodput & goodput ratio  & sending rate & throughput & Jain's index\\ 
			&   &           & $\lbrack Mbps \rbrack$ & $\lbrack \% \rbrack$ & $\lbrack Mbps \rbrack$ & $\lbrack Mbps \rbrack$ \\            
			\midrule                                                                                         
			\multirow{10}{0.5cm}{QUIC} & 
			
			\multirow{2}{2.7cm}{Hybrid vs Loss-based}                                             
			& BBR    & 0.72 & 95.88 & 0.79 &0.76 & \multirow{2}{1cm}{0.62}\\ 
			& &Cubic  & 8.82 & 95.63 & 9.44 & 9.25 &                    \\ 
			\cline{2-8}                                                                                   
			
			&\multirow{2}{2.7cm}{Loss- vs. Loss-based}                                             
			&Cubic    & 4.50 & 95.39 & 4.83 & 4.72 & \multirow{2}{1cm}{0.98}\\ 
			& &Cubic  & 5.04 & 95.39 & 5.40 & 5.29 &                     \\ 
			\cline{2-8}                                                                                   
			& \multirow{2}{2.7cm}{Hybrid vs. Hybrid}                                                     
			&BBR    & 4.75 & 95.90 & 4.98 & 4.98& \multirow{2}{1cm}{0.89}  \\   
			& &BBR  & 4.79 & 95.90 & 5.02& 5.02 &                          \\   
			\cline{2-8}
			
			&\multirow{2}{2.7cm}{Loss- vs. Loss-based}                                             
			&Cubic(0ms)  & 5.51 & 93.77 & 6.08 & 5.66 & \multirow{2}{1cm}{0.93 }  \\ 
			& &Cubic(200ms)  & 3.89 & 93.46 &4.20 & 4.08&                           \\ 
			\cline{2-8}                                                                                   
			& \multirow{2}{2.7cm}{Hybrid vs. Hybrid}                                                     
			&BBR(0ms)    & 1.12 & 94.14 & 1.18 & 1.17 & \multirow{2}{1cm}{0.59}  \\   
			& &BBR(200ms)    & 8.50 & 95.90 & 8.94 & 8.91 &                           \\   
			\hline				
			
		\end{tabular}
		\vspace{-0.5cm}
	\end{center}
\end{table*}

With BBR, we observe the same RTT-unfairness properties as with the TCP BBR, which always favours the flows with a higher RTT (with an average fairness index of $0.59$). Similarly, QUIC with Cubic always favours the flow with a lower RTT. However, the difference between the throughput of the two QUIC Cubic flows is much smaller than the one observed for the TCP equivalent, with an average fairness index of $0.93$.

\begin{table*}[!htb]
	\begin{center}
		\caption{BW scenario with 4 flows: Different metrics for representatives of the three congestion control algorithm groups.}
		\label{tab:dual8}
		\begin{tabular}{l| l | l | c | c | c | c | c}
			\toprule
			Protocol & Group & Algorithm & Average & Average & Average & Average & Average\\
			& & & goodput & goodput ratio  & sending rate & throughput & Jain's index\\ 
			&   &           & $\lbrack Mbps \rbrack$ & $\lbrack \% \rbrack$ & $\lbrack Mbps \rbrack$ & $\lbrack Mbps \rbrack$ \\            
			\midrule                                                                                         
			\multirow{16}{0.5cm}{QUIC} & 

			\multirow{4}{2.7cm}{Hybrid vs. Loss-based}                                             
&BBR  &0.32 & 95.46 & 0.34 & 0.34 & \multirow{4}{1cm}{0.77}  \\  
& &Cubic  & 2.66 & 94.11 & 2.87 & 2.79 &                       \\ 
& &Cubic  & 3.82 & 95.26 & 4.08 & 4.00 &                       \\ 
& &Cubic  & 2.74 & 95.05 & 2.91 & 2.87 &                       \\ 
\cline{2-8}                                                                                   
& \multirow{4}{2.7cm}{Loss-based vs. Hybrid}                                                     
&Cubic  & 5.25 & 95.88 & 5.62& 5.51& \multirow{4}{1cm}{0.59}  \\  
& &BBR & 2.70 & 95.79 & 2.91 & 2.83&                        \\ 
& &BBR  & 1.26 & 95.71 & 1.32& 1.32&                        \\ 
& &BBR  & 0.25 & 95.46 & 0.26& 0.26&                      \\ 
\cline{2-8}                        			
		
			&\multirow{4}{2.7cm}{Loss- vs. Loss-based}                                             
			&Cubic    & 2.38 & 94.85 & 2.61 &2.49 & \multirow{4}{1cm}{0.95}  \\  
			& &Cubic  & 2.20 & 94.68 & 2.38& 2.30 &                       \\ 
			& &Cubic  & 2.70 & 94.67 & 2.91& 2.83 &                       \\ 
			& &Cubic  & 2.20 & 94.50 & 2.34 & 2.30  &                       \\ 
			\cline{2-8}                                                                                   
			& \multirow{4}{2.7cm}{Hybrid vs. Hybrid}                                                     
			&BBR  & 2.38 & 95.90 & 2.49& 2.49& \multirow{4}{1cm}{0.90}  \\  
			& &BBR & 2.38 & 95.90 & 2.49 & 2.49&                        \\ 
			& &BBR  & 2.30& 95.90 & 2.45& 2.41&                        \\ 
			& &BBR  & 2.45 & 95.90 & 2.57& 2.56&                      \\ 
			\cline{2-8}                                                                                   
			\hline				
			
		\end{tabular}
		\vspace{-0.5cm}
	\end{center}
\end{table*}

\begin{table*}[!htb]
	\begin{center}
		\caption{RTT scenario with 4 flows: Different metrics for representatives of the congestion control algorithm groups in case the link is shared by two flows using the same algorithm and having different RTTs.}
		\label{tab:dual9}
		\begin{tabular}{l| l | l | c | c | c | c | c}
			\toprule
			Protocol & Group & Algorithm & Average & Average & Average & Average & Average\\
			& & & goodput & goodput ratio  & sending rate & throughput & Jain's index\\ 
			&   &           & $\lbrack Mbps \rbrack$ & $\lbrack \% \rbrack$ & $\lbrack Mbps \rbrack$ & $\lbrack Mbps \rbrack$ & \\            
			\midrule                                                                                         

				\multirow{8}{0.5cm}{QUIC} &
			 \multirow{4}{2.7cm}{Loss- vs. Loss-based}             
			& Cubic(100ms)   & 3.72 & 94.69 & 4.01 & 3.90 &   \multirow{4}{1cm}{0.84}  \\   
			& &Cubic(200ms)  & 2.34 & 94.29 & 2.57 & 2.46 &    \\     
			& &Cubic(300ms)  & 1.92 & 94.78 & 2.04 & 2.01 &    \\     
			& &Cubic(400ms)  & 1.63 & 95.03 & 1.76 & 1.71 &    \\     
					\cline{2-8}                                                                                          
			&		\multirow{4}{2.7cm}{Hybrid vs. Hybrid}                                                             
			&BBR(100ms)      & 0.94 & 95.90  & 1.02 & 1.02 &     \multirow{4}{1cm}{0.62} \\ 
			& &BBR(200ms)    & 2.36 & 95.89  & 2.53 & 2.48 &    \\      
			& &BBR(300ms)    & 1.09 & 95.89  & 1.20 & 1.14 &    \\      
			& &BBR(400ms)    & 4.57 & 95.89  & 4.79 & 4.59 &    \\

			\hline				
			
		\end{tabular}
		\vspace{-0.5cm}
	\end{center}
\end{table*}

In all our QUIC scenarios where hybrid (BBR) and loss-based (Cubic) flows compete, Cubic outperforms BBR. Over time, as QUIC BBR flows detect a higher RTT and adopt a more aggressive approach, BBR grabs more bandwidth at the expense of the Cubic flows. However, this process is slow and the throughput of the BBR flow remains low.

\section{Conclusion}\label{Sec_Conclusion}

After dividing existing congestion control algorithms into three groups (e.g., loss-based algorithms, delay-based algorithms, and hybrid algorithms), we studied their interactions. 


We observed multiple fairness issues, among flows of the same group, across groups, as well as when flows having different RTTs were sharing a bottleneck link. We found that delay-based, as well as hybrid algorithms, suffer from a decrease in performance when competing with flows from the loss-based group, making them unusable in a typical network where the majority of flows will rely on a loss-based algorithm. Not only do they get an unfair share of the available bandwidth, but they also suffer from a huge increase in the observed delay when the loss-based algorithms fill the queues. In addition, the observed convergence times were large (up to $60s$), and might be larger than the duration time of a typical flow for many applications. Finally, we found that hybrid algorithms, such as BBR, not only favour the flow with a higher RTT at the expense of the other flows, but they also cannot maintain a low queuing delay as promised.

Our work therefore shows that to support applications that require low latency, a good congestion control algorithm on its own won't be enough. Indeed, guaranteeing that flows of a given group (in terms of type of congestion control) will receive their expected share of resources, requires that resource isolation be provided between the different groups.


\bibliographystyle{IEEEtran}
\bibliography{ref}
\end{document}